\pdfoutput=1
\documentclass[12pt,a4paper]{article}
\usepackage[T1]{fontenc}
\usepackage{times}
\DeclareMathAlphabet{\MATHIT}{OT1}{ptm}{m}{it}
\DeclareSymbolFont{Letters}{OML}{ztmcm}{m}{it}
\DeclareSymbolFontAlphabet{\mathNormal}{Letters}
\usepackage[utf8]{inputenc}
\usepackage[english]{babel}
\usepackage{amsmath}
\usepackage{amsfonts}
\usepackage{amssymb}
\usepackage{amsthm}
\usepackage{enumitem}
\usepackage{dcolumn}
\usepackage{bm}
\usepackage{bbm}
\usepackage{cancel}
\usepackage{tikz}
\usetikzlibrary{calc,shapes,arrows,patterns,decorations,positioning,automata,decorations.markings}
\usepackage{tabularx}
\usepackage{booktabs}
\usepackage{tensor}
\usepackage{scalerel}
\usepackage[small]{caption}
\usepackage{xcolor}
\usepackage{mathtools}
\usepackage{float}
\usepackage{xfrac}
\usepackage[pdfborder={0 0 0}]{hyperref}
\definecolor{darkblue}{rgb}{0,0,.5}
\definecolor{darkgreen}{rgb}{0,0.5,.5}
\definecolor{darkyellow}{rgb}{0.5,0.5,0}
\definecolor{fhl}{rgb}{1,0,0}
\hypersetup{colorlinks=true, breaklinks=true, linkcolor=darkblue, menucolor=darkblue, urlcolor=darkblue, linktocpage=true}
\usepackage{geometry}
\usepackage{textpos}
\usepackage{relsize}
\usepackage[page,toc]{appendix}
\usepackage[section]{placeins}
\usepackage[sorting=none,style=ieee,citestyle=numeric-comp]{biblatex}
\bibliography{biblio}

\newcommand{\mt}[1]{\textrm{\scriptsize #1}}
\def\Nc{N_\mt{c}}

\makeatletter 
\newsavebox\myboxA 
\newsavebox\myboxB 
\newlength\mylenA

\newcommand*\xoverline[2][0.75]{%
    \sbox{\myboxA}{$\m@th#2$}%
    \setbox\myboxB\null
    \ht\myboxB=\ht\myboxA%
    \dp\myboxB=\dp\myboxA%
    \wd\myboxB=#1\wd\myboxA
    \sbox\myboxB{$\m@th\overline{\copy\myboxB}$}
    \setlength\mylenA{\the\wd\myboxA}
    \addtolength\mylenA{-\the\wd\myboxB}%
    \ifdim\wd\myboxB<\wd\myboxA%
       \rlap{\hskip 0.5\mylenA\usebox\myboxB}{\usebox\myboxA}%
    \else 
        \hskip -0.5\mylenA\rlap{\usebox\myboxA}{\hskip 0.5\mylenA\usebox\myboxB}%
    \fi}
\makeatother
\numberwithin{equation}{section}

\let\originalleft\left
\let\originalright\right
\renewcommand{\left}{\mathopen{}\mathclose\bgroup\originalleft}
\renewcommand{\right}{\aftergroup\egroup\originalright}

\newcommand{\e}{\operatorname{e}}

\newcommand{\SU}[1]{\operatorname{SU}(#1)}

\newcommand{\su}[1]{\mathfrak{su}\left(#1\right)}

\newcommand{\of}[1]{\left(#1\right)}

\newcommand{\bof}[1]{\biggl(\bigg.#1\bigg.\biggr)}
\newcommand{\sbof}[1]{\Bigl(\Big.#1\Big.\Bigr)}
\newcommand{\sof}[1]{\bigl(\big.#1\big.\bigr)}
\newcommand{\ssof}[1]{(#1)}
\newcommand{\fof}[1]{\left[#1\right]}

\newcommand{\cof}[1]{\left\{#1\right\}}

\newcommand{\scof}[1]{\bigl\{\big.#1\big.\bigr\}}
\newcommand{\sscof}[1]{\{#1\}}

\newcommand{\avof}[1]{\left\langle #1\right\rangle}

\newcommand{\savof}[1]{\big\langle #1\big\rangle}

\newcommand{\trace}{\operatorname{tr}}
\newcommand{\re}{\operatorname{Re}}
\newcommand{\im}{\operatorname{Im}}
\renewcommand*{\det}{\mathop{\mathrm{det}}\nolimits}

\newcommand{\Det}[1]{\operatorname{Det}\left(#1\right)}

\newcommand{\bra}[1]{\left\langle #1\right|}
\newcommand{\ket}[1]{\left| #1\right\rangle}

\newcommand{\Repart}[1]{\operatorname{Re}\left(#1\right)}

\newcommand{\ii}{\mathrm{i}}
\newcommand{\idd}[2]{\mathrm{d}^{#2}#1}
\newcommand{\dd}{\mathrm{d}}
\newcommand{\DD}[1]{\mathcal{D}\bigl[#1\bigr]}
\newcommand{\cD}{\mathrm{D}}
\newcommand{\totd}[3]{\frac{\dd^{#3} #1}{\dd #2^{#3}}}

\newcommand{\partd}[2]{\frac{\partial #1}{\partial #2}}
\newcommand{\partdf}[3]{\left.\frac{\partial #1}{\partial #2}\right\vert_{#3}}

\newcommand{\order}[1]{\mathcal{O}\left(#1\right)}
\newcommand{\sorder}[1]{\mathcal{O}\big(#1\big)}
\newcommand{\ssorder}[1]{\mathcal{O}(#1)}

\newcommand{\obs}{\mathcal{O}}

\newcommand{\id}{\mathbbm{1}}

\newcommand{\abs}[1]{\left| #1\right|}
\newcommand{\ssabs}[1]{| #1|}
\newcommand{\sabs}[1]{\big| #1\big|}

\renewcommand*\[{\begin{equation}}
\renewcommand*\]{\end{equation}}
\renewcommand*\hat[1]{\widehat{#1}}
\let\oldstackrel\stackrel
\renewcommand*\stackrel[2]{{\scriptstyle\oldstackrel{#1}{#2}}}

\definecolor{emphcol}{RGB}{0,0,0}
\let\oldemph\emph
\renewcommand*\emph[1]{\oldemph{\textcolor{emphcol}{#1}}}
\let\oldstackrel\stackrel
\renewcommand*\stackrel[2]{{\scriptstyle\oldstackrel{#1}{#2}}}

\newcommand{\ucases}[1]{\begin{cases}#1\end{cases}}

\DeclarePairedDelimiter\floor{\lfloor}{\rfloor}

\newcommand*\getscale[1]{%
  \begingroup
    \pgfgettransformentries{\scaleA}{\scaleB}{\scaleC}{\scaleD}{\whatevs}{\whatevs}%
    \pgfmathsetmacro{#1}{sqrt(abs(\scaleA*\scaleD-\scaleB*\scaleC))}%
    \expandafter
  \endgroup
  \expandafter\edef\expandafter#1\expandafter{#1}%
}

\newlength{\slinewidth}

\tikzset{fontscale/.style={font=\relsize{#1}}}
\tikzset{->-/.style={decoration={
  markings,
  mark=at position #1 with {\arrow{>}}},postaction={decorate}}}
\tikzset{-<-/.style={decoration={
  markings,
  mark=at position #1 with {\arrow{<}}},postaction={decorate}}}

\tikzset{cross/.style={cross out,draw,minimum size=2*(#1-\pgflinewidth),inner sep=0pt, outer sep=0pt}}

\DeclareRobustCommand\sketchlink{\vcenter{\hbox{\tikz[scale=0.4,nodes={inner sep=0}]{\pgfpointtransformed{\pgfpointxy{1}{1}};
  \pgfgetlastxy{\vx}{\vy}
  \begin{scope}[node distance=\vx and \vy,fontscale=-1]
    \draw node[black,below] at (0,-0.2) {$x\phantom{\hat{\mu}}$};
    \draw[-<-=.5,thick,black] (0,0) -- (4,0);
    \draw node[black,above] at (2,0.3) {$U_{\mu}\of{x}$};
    \draw node[black,below] at (4.2,-0.2) {$x+\hat{\mu}$};
    \draw[black,fill=white] (0,0) circle (5pt); 
    \draw[black,fill=white] (4,0) circle (5pt); 
  \end{scope}}}}\,}
  
\DeclareRobustCommand\sketchplaq{\vcenter{\hbox{\tikz[scale=0.4,nodes={inner sep=0}]{\pgfpointtransformed{\pgfpointxy{1}{1}};
  \pgfgetlastxy{\vx}{\vy}
  \begin{scope}[node distance=\vx and \vy,fontscale=-1]
    \draw[-<-=.5,thick,black] (0,0) -- (4,0);
    \draw[-<-=.5,thick,black] (4,0) -- (4,4);
    \draw[-<-=.5,thick,black] (4,4) -- (0,4);
    \draw[-<-=.5,thick,black] (0,4) -- (0,0);
    
    \draw node[black,below] at (0,-0.2) {$x\phantom{\hat{\mu}}$};
    \draw node[black,below] at (4.2,-0.2) {$x+\hat{\mu}$};
    \draw node[black,above] at (4.4,4.3) {$x+\hat{\mu}+\hat{\nu}$};
    \draw node[black,above] at (-0.2,4.3) {$x+\hat{\nu}$};
    \draw node[black] at (2,1.9) {$U_{\mu\nu}\of{x}$};
    \draw[black,fill=white] (0,0) circle (5pt); 
    \draw[black,fill=white] (4,0) circle (5pt);
    \draw[black,fill=white] (4,4) circle (5pt);
    \draw[black,fill=white] (0,4) circle (5pt); 
  \end{scope}}}}\,}
  
\DeclareRobustCommand\sketchdensme{\vcenter{\hbox{\tikz[scale=0.5,nodes={inner sep=0}]{%
\pgfpointtransformed{\pgfpointxy{1}{1}};
  \pgfgetlastxy{\vx}{\vy}
  \def\xs{5};
  \def\ys{4};
  \def\ls{2};
  \begin{scope}[node distance=\vx and \vy]
    \foreach \i in {0,...,\xs} {
        \draw [very thin,gray] (\i,0) -- ($({\i},{0.5*\ys})$)  node[below] at (\i,-0.3) {};
        \draw [very thin,dashed,gray] ($({\i},{0.5*\ys})$) -- ($({\i},{0.5*\ys+1})$);
        \draw [very thin,gray] ($({\i},{0.5*\ys+1})$) -- ($({\i},{\ys+1})$);
    }
    \foreach \i in {0,...,\ys} {
        \draw [very thin,gray] (0,\i) -- (\xs+1,\i) node[left] at (-0.2,\i) {};
    }
    \node[anchor=west,xshift=2pt,scale=1] at (\xs+1,\ys+1) {$N_t$};
    \node[anchor=west,xshift=4pt,scale=1] at (\xs+1,0) {$0$};

    \draw[thick,dashed,red!100] ($({0},{(\ys+1)*1.0+0.025})$) -- ($({(\xs+1)},{(\ys+1)*1.0+0.025})$) node[pos=0.5,anchor=north,yshift=-2pt,scale=1] {$\psi_{2}$};
    \draw[thick,red!100] ($({0},{(\ys+1)*0.0-0.025})$) -- ($({(\xs+1)},{(\ys+1)*0.0-0.025})$) node[pos=0.5,anchor=south,yshift=2pt,scale=1] {$\psi_{1}$};
    
  \end{scope}%
  }}}\,}
  
\DeclareRobustCommand\sketchreddensme{\vcenter{\hbox{\tikz[scale=0.5,nodes={inner sep=0}]{%
  \pgfpointtransformed{\pgfpointxy{1}{1}};
  \pgfgetlastxy{\vx}{\vy}
  \def\xs{5};
  \def\ys{4};
  \def\ls{2};
  \begin{scope}[node distance=\vx and \vy]
    \foreach \i in {0,...,\xs} {
        \draw [very thin,gray] (\i,0) -- ($({\i},{0.5*\ys})$)  node[below] at (\i,-0.3) {};
        \draw [very thin,dashed,gray] ($({\i},{0.5*\ys})$) -- ($({\i},{0.5*\ys+1})$);
        \draw [very thin,gray] ($({\i},{0.5*\ys+1})$) -- ($({\i},{\ys+1})$);
    }
    \foreach \i in {0,...,\ys} {
        \draw [very thin,gray] (0,\i) -- (\xs+1,\i) node[left] at (-0.2,\i) {};
    }
    \draw[thick,dashed,blue!100] ($(0,{(\ys+1)*1.0})$) -- ($({(\xs+1-\ls)},{(\ys+1)*1.0})$) node[pos=0.5,anchor=north,yshift=-2pt,scale=1] {$r_{B}$};
    \draw[thick,blue!100] ($(0,{(\ys+1)*0.0})$) -- ($({(\xs+1-\ls)},{(\ys+1)*0.0})$) node[pos=0.5,anchor=south,yshift=2pt,scale=1] {$r_{B}$};

    \draw[thick,dashed,red!100] ($({(\xs+1-\ls)},{(\ys+1)*1.0+0.025})$) -- ($({(\xs+1)},{(\ys+1)*1.0+0.025})$) node[pos=0.5,anchor=north,yshift=-2pt,scale=1] {$\psi_{A,2}$};
    \draw[thick,red!100] ($({(\xs+1-\ls)},{(\ys+1)*0.0-0.025})$) -- ($({(\xs+1)},{(\ys+1)*0.0-0.025})$) node[pos=0.5,anchor=south,yshift=2pt,scale=1] {$\psi_{A,1}$};
    
    \node[red] at ($({(\xs-0.5*\ls+1)},{(\ys+1)*0.5})$) {$\mathbf{A}$};
    \node[blue] at ($({0.5*(\xs-\ls+1)},{(\ys+1)*0.5})$) {$\mathbf{B}$};
    
  \end{scope}%
  }}}\,}

\newcommand{\be}{\begin{equation}} \newcommand{\ee}{\end{equation}}
\newcommand{\bea}{\begin{eqnarray}} \newcommand{\eea}{\end{eqnarray}}

\begin{document}\selectlanguage{english}
\title{\begin{textblock*}{100pt}(395pt,-95pt)
\textnormal{\small \texttt{HIP-2023-4/TH}}
\end{textblock*}
Disentangling the gravity dual of Yang-Mills theory
}
\iftrue
{\smaller{\author{
Niko Jokela,\thanks{niko.jokela@helsinki.fi}\quad Kari Rummukainen,\thanks{kari.rummukainen@helsinki.fi}\quad Ahmed Salami\thanks{ahmed.salami@helsinki.fi}\\
Helsinki Institute of Physics and Department of Physics,\\ P.O. Box 64, FI-00014 University of Helsinki, Finland\\ \\
Arttu P\"onni\thanks{arttuponni@gmail.com}\\
Micro and Quantum Systems Group,\\
Department of Electronics and Nanoengineering,\\
Aalto University, Otakaari 1B, FI-00076 Aalto, Finland \\ \\
Tobias Rindlisbacher\thanks{trindlis@itp.unibe.ch}\\
Albert Einstein Center for Fundamental Physics and\\
Institute for Theoretical Physics, University of Bern,\\
Sidlerstrasse 5, CH-3012 Bern, Switzerland 
}}}
\fi
\maketitle

\iftrue
\begin{abstract}
A construction of a gravity dual to a physical gauge theory requires confronting data. We establish a proof-of-concept for precision holography, {\emph{i.e.}}, the explicit reconstruction of the dual background metric functions directly from the entanglement entropy (EE) of strip subregions that we extract from pure  glue Yang-Mills theory discretized on a lattice. Our main focus is on a three-dimensional Euclidean $\SU{2}$ theory in the deconfining phase. 
Holographic EE suggests, and we find evidence for, that the scaling of the thermal entropy with temperature is to power 7/3 and that it approaches smoothly the critical point, consistent with black hole thermodynamics.
In addition, we provide frugal results on the potential between quenched quarks by the computation of the Polyakov loop correlators on the lattice. Holographic arguments pique curiosity in the substratum of Debye screening at strong coupling.
\end{abstract}
\fi

\clearpage
\tableofcontents

\newpage

\section{Introduction and summary}\label{sec:introduction}

Our current understanding of fundamental particle physics rests on non-Abelian gauge field theories. Simplest examples among these are pure Yang-Mills (YM) theories, the studies of which constantly contribute to our ever-improving understanding of nonperturbative aspects of Standard Model of particle physics. The lucrative simplicity of YM theories void of fermionic matter is deceptive, however. The prime example is the inability to currently extract a satisfying narrative of the deconfinement phase transition. Attempts to extrapolate obtained lessons to situations in the presence of fundamental quarks described within the theory of strong interactions, quantum chromodynamics (QCD), falter. A description for the deconfinement phase transition is inherently nonperturbative, which means that standard diagrammatic methods based on weak coupling expansions may not be optimal starting points. On the other hand, lattice approach hits the challenge of simulating fermions in the fundamental representation of the gauge group. New approaches to shed light on this immensely important and highly convoluted issue are certainly welcome.

Here we do not directly address the phenomenon of deconfinement phase transition. Rather, we wish to point out another outstanding issue in pure YM theories, the lack of understanding of the behavior of entanglement entropy (EE), $S_{EE}$. Entanglement entropy may though play a role in our quest to understand the deconfinement phase transition \cite{Nishioka:2006gr,Klebanov:2007ws} as it is a probe of the underlying energy scale or the finite correlation length \cite{Jokela:2020wgs}, an intrinsic dynamical one in the cases of interest to us in this work. In somewhat similar fashion to the case of deconfinement phase transition, computations of EE in non-Abelian gauge field theories are highly complicated. Only existing results directly from the field theory for slab subregions have been obtained at zero temperature from lattice simulations of (3+1)-dimensional pure YM theories with $\SU{\Nc}$, $\Nc=2,3,4$ \cite{Buividovich:2008kq,Velytsky:2008sv,Nakagawa:2009jk,Nakagawa:2011su,Itou:2015cyu,Rabenstein:2018bri} as well as at around critical temperature in $\SU{3}$ \cite{Nakagawa:2011su}.\footnote{Other echoing gauge theory computations of the entanglement entropy include \cite{Soni:2015yga,Agarwal:2016cir,Donnelly:2019zde,Anegawa:2021osi,Panizza:2022gvd,Liu:2022qqf,Bulgarelli:2023ofi}.}
In addition, note that the (1+1)-dimensional case can be solved exactly since the Migdal-Kadanoff decimation approach \cite{Migdal:1975zg,Kadanoff:1976jb} is exact. This leads to trivial entanglement entropy \cite{Velytsky:2008rs,Velytsky:2008sv}, {\emph{i.e.}}, in the sense of not depending on the spatial extent of the subregion; in Sec.~\ref{sec:latticeEE} we provide a complementary exposition for this fact. In the context of gauge/gravity duality, the Ryu-Takayanagi proposal \cite{Ryu:2006bv} allows extracting the entanglement entropy for spatial subregions in the large-$\Nc$ limit in a variety of supersymmetric field theories in diverse dimensions at large coupling.

\begin{figure}
    \centering
    \includegraphics[width=0.9\textwidth]{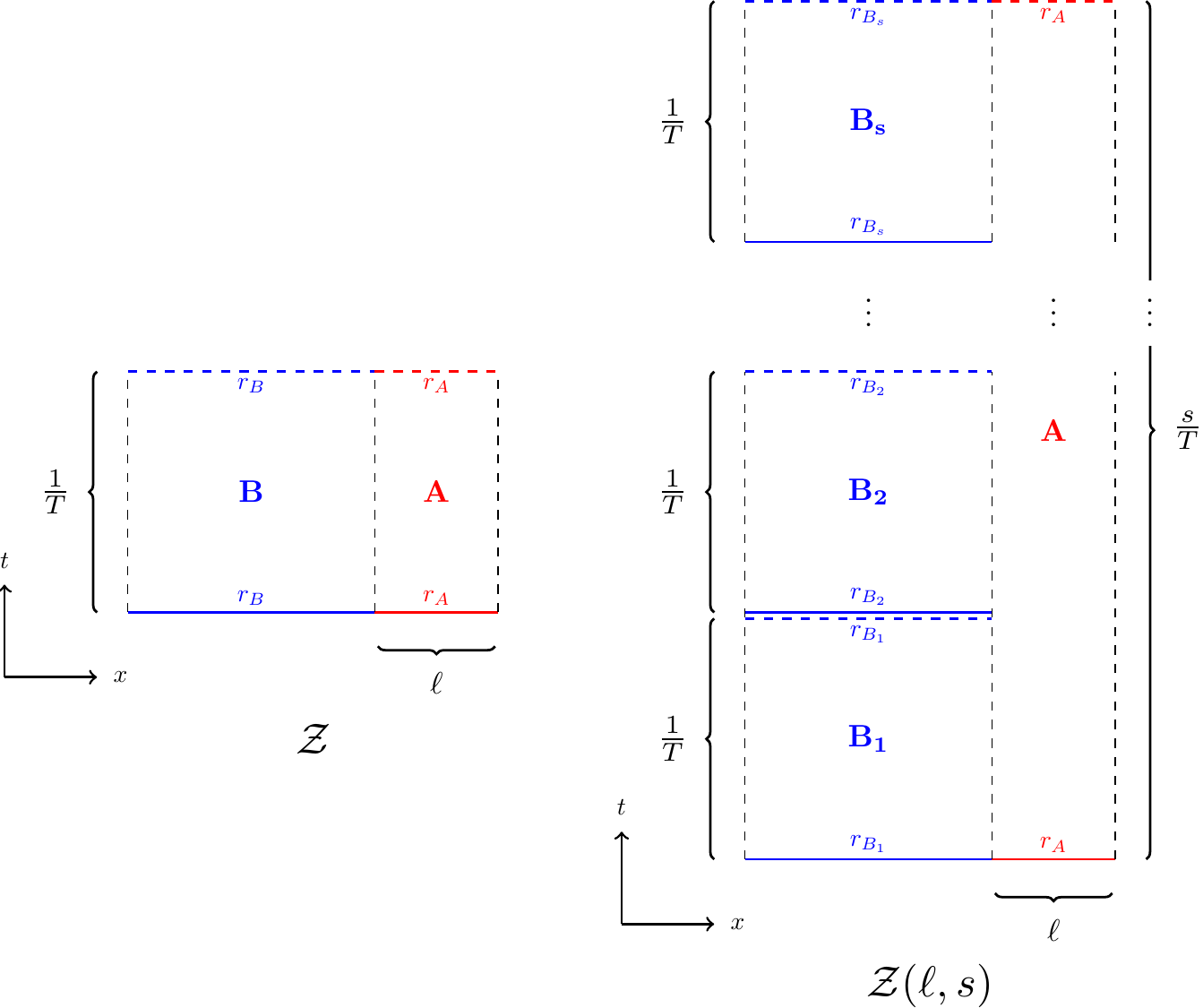}\caption{\textbf{Left:} We divide the system in spatial direction $x$ by drawing an imaginary line at distance $\ell$, which defines regions $A$ and $B$. The ${\cal Z}$ is the associated partition function for field configurations which are $1/T$ periodic in the Euclidean time direction. \textbf{Right:} We introduce $s$ replicas. The fields are $s/T$-periodic in the right part (region $A$), while only $1/T$-periodic in the left part (region B). In the sketch, the periodicity of each box, $A$, $B_i$, is represented by a pair of horizontal lines (a solid one at the bottom and a dashed one at the top) which both carry the same label $r_A$, resp. $r_{B_i}$, indicating that these two boundaries of the box are identified.}\label{fig:replica}
\end{figure}

In this paper, we will report on extracting the lattice derivatives of the entanglement entropy in pure $\SU{\Nc}$ Yang-Mills theories using a powerful new method. Our treatment is general, but we gear our attention to Euclidean three- and four-dimensional cases and consider entanglement entropies for single strip and slab geometries, respectively. Similarly to bulk thermodynamic quantities, as the entanglement entropy itself is not an observable, we will measure the {\emph{derivative}} of the entanglement entropy with respect to the width $\ell$ of the strip (slab) for a given $\ell$: $\partial S_{EE}/\partial\ell$. More precisely, we utilize the replica method \cite{Callan:1994py,Calabrese:2004eu,Calabrese:2005zw}, which on the lattice can be schematically summarized as follows. If we are interested in region $A$, we construct a reduced density matrix by tracing over the degrees of freedom in the complement, region $B$, $\rho_A=\trace_{B}\of{\rho}$, where $\rho$ is the density matrix of the whole system, $A\cup B$. The entanglement entropy,
\be
S_{EE}\equiv -\trace\of{ \rho_A\log \rho_A}\ ,\label{eq:defaultSEE}
\ee
associated with region $A$ can then be written with the help of $s$ replicas, 
\be\label{eq:SEE}
 S_{EE} = -\lim_{s\to 1}\partial_s \log\trace_A\of{\rho_A^s} = -\lim_{s\to 1}\partial_s \log\frac{ {\cal{Z}}(\ell,s)}{{\cal Z}^s} \ .
\ee
Here the first equality is an identity and the second one can be understood through the connection to path integrals. ${\cal Z}$ is the partition function of the original system, where the division to $A$ and $B$ is only imaginary, fields are periodic in $1/T$, and hence ${\cal Z}$ does {\emph{not}} depend on $\ell$ (nor on $s$). The partition function ${\cal{Z}}(\ell,s)$ is for the field configurations which have more complicated boundary conditions: for spatial coordinate values in the $B$ region the fields are $1/T$ periodic and for the spatial coordinate in the region $A$, the fields are $s/T$-periodic; see Fig.~\ref{fig:replica} for a sketch. 

As alluded to before, we are not interested in $S_{EE}$, however, but its derivative with respect to $\ell$. To this end, by acting with this derivative on (\ref{eq:SEE}) we get
\be\label{eq:delSEE}
 \partial_\ell S_{EE} = \lim_{s\to 1}\partial_\ell\partial_s F(\ell,s) \approx \frac{F(\ell+a,2)-F(\ell,2)}{a} \ ,
\ee
where we replaced for the free energy $F(\ell,s)=-\log({\cal Z}(\ell,s))$. In the latter step we approximated the spatial derivative with a discrete derivative on the lattice, with lattice spacing $a$, and for the derivative with respect to replica number $s$ we only consider the dominant contribution. Higher replica numbers give rise to contributions whose sizes are expected to be small \cite{Rabenstein:2018bri}; in free field theories the $\ell$ dependence of EE and the second R\'enyi entropy can even be shown to be the same \cite{Calabrese:2004eu,Casini:2005zv}. 

The right-hand-side of (\ref{eq:delSEE}) is a free energy difference, a finite quantity that we extract from the lattice. This will be detailed in Sec.~\ref{sec:latticeEE}. Notice that there are no issues of gauge invariance as long as one chooses not to pierce the cut through the sides of plaquettes \cite{Aoki:2015bsa}. Therefore, this lattice formulation does not suffer from the muddle associated with the tensor decomposition of the physical Hilbert space that one promptly runs across in gauge theories when formulating the tracing over the complement regions \cite{Buividovich:2008gq,Lin:2018bud}.

Our results for extracting precise data for entanglement entropies are based on a novel computational method. The advantage of our method is that it enables us to outperform previous simulations, based on a clever interpolation ansatz for the free energies \cite{Buividovich:2008gq} needed in (\ref{eq:delSEE}) that we improve upon. This advancement mitigates the severe signal-to-noise ratio, thus allowing more accurate simulations with still moderate computational resources. We can push down statistical error margin considerably relative to previous works in the field. Our method is, in principle, applicable for any number of colors (and dimensions), and so although our main focus in this paper is three-dimensional field theory at non-zero temperature, we will provide original results for the zero temperature $\SU{5}$ in four dimensions elsewhere; see also \cite{Rindlisbacher:2022bhe} for the associated entanglement c-function that we have amused on already.

Armed with the success in four dimensions, we continue the discussion in three-di\-men\-si\-onal case. Although we are limiting our focus on $\SU{2}$ theory, since the simulations are costly, we will consider a novel scenario and consider entanglement entropies at non-zero temperature. In addition to for the first time to extract entanglement entropies in three dimensions, there are interesting features that we get to test. First of all, contrary to four dimensions, there is no UV fixed point, so one does not get to exploit the underlying conformal symmetry to claim success of the comparison to holographic result. 
Interestingly, we will find support that the gravity dual to supersymmetric Yang-Mills theory at strong coupling and in the limit of large-$\Nc$ elucidates many features that would be otherwise unexpected from field theory point of view. 

So, what is the gravity dual of the Euclidean three-dimensional pure Yang-Mills theory that we consider? In an optimal scenario we should really be considering a supersymmetric Yang-Mills theory on the lattice, {\emph{e.g.}}, as in \cite{Catterall:2020nmn}. Since the results that we obtain are pretty encouraging, we believe that our work is a first step towards this direction and similar methods that we construct could be considered in vastly more demanding cases with fermions. We will review salient details of the holographic approach in Sec.~\ref{sec:holo}. 

The key observation made in \cite{Jokela:2020auu} is that the derivative (\ref{eq:delSEE}) can be directly and very effectively used to reconstruct selected components of the metric in the dual geometry to an accuracy given by the statistical error in the measurements of (\ref{eq:delSEE}). This boils down to a few facts, that $\partial S_{EE}/\partial\ell$ is independent of a regularization scheme and that it can be written without an explicit integral that would otherwise be present in the evaluation of the entanglement functional $S_{EE}$. The derivative is instead written in terms of components of the background metric, evaluated on the position of the RT surface. This quantity can be understood as the conjugate momentum that is conserved under translations of the position of the strip (slab) \cite{DiNunno:2021eyf} and it becomes a very simple expression when one chooses to evaluate $\partial S_{EE}/\partial\ell$ at the tip of the RT surface. In particular, in the limit $\ell\to\infty$ one expects to probe the IR physics of the ambient field theory. On general grounds in the deconfining phase, the finite part $S_{EE}/\ell$ should pick only the contribution from the degrees of freedom on the black hole horizon and indeed the expression $\partial S_{EE}/\partial\ell|_{\ell\to\infty}$ reduces to the Bekenstein-Hawking entropy $S_{BH}\propto \Nc^2\,T^{7/3}/\lambda^{1/3}$, where $\lambda=g_{YM}^2 \Nc$ indicates the 't Hooft coupling \cite{Itzhaki:1998dd,Peet:1998wn}. Thereby, per holographic dictionary, extracting the RHS of (\ref{eq:delSEE}) will match onto the thermal entropy of Yang-Mills theory at strong coupling and large-$\Nc$ when $\ell\to\infty$ is taken. We will show numerical evidence for this from our lattice simulations in Sec.~\ref{sec:results}.
That the derivative of the entanglement entropy of some region $A$ becomes in the limit $\ell\to\infty$ proportional to the thermal entropy, $S_{\mathrm{th}}$, can be understood also directly from the definition of the entanglement entropy, Eq.~\eqref{eq:defaultSEE}, by noting that in the limiting case, where $A$ represents the whole system, and its complement, $B$, shrinks to zero volume, one has $\rho_{A}\to \rho$, and therefore:
\[
\lim_{B\to\emptyset} S_{EE}\of{A} = -\trace\of{\rho\,\log\of{\rho}} = S_{\mathrm{th},A}\ .
\]
Now, as the thermal entropy is an extensive quantity, $S_{\mathrm{th},A}$ grows linearly with the volume of $A$, which in turn depends linearly on $\ell$. One would therefore expect that 
\[
\lim_{B\to\emptyset} \partd{S_{EE}\of{A}}{\ell} = \partd{\operatorname{vol}\of{A}}{\ell}\,s_{\mathrm{th},A}\ ,\label{eq:seederivandsthrel}
\]
with $s_{\mathrm{th},A}$ being the density of thermal entropy in $A$, and $\partial_{\ell}\operatorname{vol}\of{A}=\abs{A_{\perp \ell}}$ is simply the area of the cross section of $A$ that is perpendicular to the direction in which $\ell$ grows. We will discuss this in more detail in Sec.~\ref{sec:eeandterelation}. The relation Eq.~\eqref{eq:seederivandsthrel} could also be of interest for future thermodynamic studies, as it allows for a non-perturbative determination of the thermal entropy of a system without having to fix integration constants.

At intermediate energy scales (or strip widths) we find evidence for the scaling  $S_{EE}\sim \ell^{-4/3}$ that is expected from (supersymmetric) YM theory at zero temperature in three dimensions \cite{vanNiekerk:2011yi}. 
We present the results that we have obtained from lattice studies in Sec.~\ref{sec:results}. We also show that the corresponding geometry can be reconstructed using this data.
Although the reconstructed metric might give further information than the entropy density on some field theory quantities, and possibly even make useful predictions, we defer these investigations in the future.

Sec.~\ref{sec:conclusions} then contains our conclusions and a discussion on future outgrowths of our work, and we supplement the article with an Appendix~\ref{sec:Wilson} where we give a peek to other novel results that we extracted from the lattice in the case of 3d YM at finite temperature. We present data for the Polyakov loop and extract the potential $V(L)$ between quarks and anti-quarks, separated by a distance $L$. Among other things, in the intermediate energy scales, we show evidence for the scaling $V(L)\sim L^{-2/3}$ \cite{Maldacena:1998im} stemming from the same D2-brane background in the UV as we utilized for the entanglement entropy. At the deep IR, {\emph{i.e.}}, for very large separation, the D2-brane background suggests that the real part of the potential ${\rm{Re}} V\sim L^{-10/3}$, while the imaginary part would grow as ${\rm{Im}}V\sim L$. Our results are consistent with the former behavior though we cannot dissect it from the expectation of the Debye screening behavior below $T_c$: ${\rm{Re}}V\sim e^{-m_D L}/L^{1/2}$ where $m_D$ would be the Debye mass \cite{Dumitru:2002cf}.\footnote{Notice that in the perturbative regime at zero temperature one finds the Coulomb-type logarithmic potential \cite{Pineda:2010mb}, plus higher order corrections.} To this end, a systematic scan in the parameters is required, in particular to analytically continue the Euclidean correlators to real time \cite{Burnier:2014ssa,Burnier:2016mxc} to see how the imaginary part of the potential behaves.

\section{Entanglement entropy on the lattice}\label{sec:latticeEE}

In this section we describe in some more detail how the aforementioned replica method~\cite{Callan:1994py,Calabrese:2004eu,Calabrese:2005zw} can be applied to determine entanglement measures in $\SU{\Nc}$ Yang-Mills theories using non-perturbative lattice Monte Carlo methods. For the reader who is unfamiliar with the lattice formulation of Yang-Mills theory and Monte Carlo techniques, we refer to the text books~\cite{Creutz:1983njd,Rothe:1992nt,smit_2002,Gattringer:2010zz}.

The first implementation of the replica method for $\SU{\Nc}$ lattice gauge theory was introduced almost 15 years ago~\cite{Buividovich:2008kq} and used ever since~\cite{Nakagawa:2009jk,Nakagawa:2011su,Itou:2015cyu,Rabenstein:2018bri}. We will here briefly review the working principle of the original method and sketch the considerations that lead to our new approach. 

In Appendix~\ref{sec:changeoftempbc} we provide some more details on our update algorithm from which also an argument arises for why in $\of{1+1}$ dimensions the entanglement entropy must be independent of the width, $\ell$, of the slab region $A$.

\subsection{Established method}\label{ssec:estlatEEmethod}
The replica method is based on a path-integral representation of matrix elements, $\bra{\psi_1}\rho\ket{\psi_2}$, of the density matrix $\rho$, which can be thought of as the (Euclidean) time-evolution operator from some initial time $x^d_1$ to some final time $x^d_2$, and $\psi_1$ and $\psi_2$ represent instantaneous states of the system at these times. 

Let us now consider a pure $\SU{\Nc}$ gauge theory on a $d$-dimensional, periodic, finite lattice of temporal extent $N_t$. The ordinary Euclidean lattice partition function for this system can be written as,
\[
Z=\int\DD{U}\,\e^{-S_G\fof{U}}\label{eq:wgpartf}
\]
where we choose $S_G$ to be the Wilson gauge action~\cite{Wilson:1974sk},
\[
S_{G}=\frac{\beta_{g}}{\Nc}\,\sum_{x}\sum_{\mu<\nu} \re\trace\of{\id-U_{\mu\nu}\of{x}}\ ,\label{eq:wilsonSG}
\]
with $\beta_g=\frac{2\,\Nc}{g_0^2}$ being the inverse bare lattice gauge coupling and the $U_{\mu\nu}\of{x}$ are the so-called plaquette variables,
\[
U_{\mu\nu}\of{x}=U_{\mu}\of{x}\,U_{\nu}\of{x+\hat{\mu}}\,U^{\dagger}_{\mu}\of{x+\hat{\nu}}\,U^{\dagger}_{\nu}\of{x}= \sketchplaq\ ,\label{eq:plaqvar}
\]
which are formed from the elementary gauge degrees of freedom, the link variables: 
\[
U_{\mu}\of{x}=\exp\of{\ii\,a\,A_{\mu}\of{a\,x}}=\sketchlink\quad\forall x\in\mathbb{Z}^d,\,\mu\in\cof{1,\ldots,d}\ .\label{eq:linkvar}
\]
We note that the $g^2_0$ in the above expression for $\beta_g$ is the dimensionless lattice gauge coupling, which in terms of the continuum theory gauge coupling, $g^2_c$, and the lattice spacing, $a$, can be written as $g^2_0=g^2_c\,a^{4-d}$. Furthermore, we note that a finite temporal extent, $N_t$, of the lattice corresponds to a finite temperature $T=1/(a\,N_t)$. 

Now, to define a matrix element $\bra{\psi_{1}}\rho\ket{\psi_{2}}$ of the density matrix $\rho$ for this theory, we drop the temporal periodicity of the lattice and think of the initial and final states, $\psi_1$ and $\psi_2$, as defining (up to gauge transformations) values for the gauge links that touch or are within the time slices $x^d=0$ and $x^d=N_t$:
\[
\bra{\psi_{1}}\rho\ket{\psi_{2}}=\frac{1}{Z}\int\limits_{\mathclap{\substack{U\ssof{\bar{x},0}=U_{\psi_1}\ssof{\bar{x}}\\ U\ssof{\bar{x},N_t}=U_{\psi_2}\ssof{\bar{x}}}}}\DD{U}\,\e^{-S_G\fof{U}}=\frac{1}{Z}\sketchdensme\ ,\label{eq:densmatpirep}
\]
where the $U_\psi$ are made up from link-variables (\ref{eq:linkvar}).
The normalization factor $Z^{-1}$ is obtained by noting that 
taking the trace over $\rho$  means to identify the two boundaries at $x^d=0$ and $x^d=N_t$ with each other and summing over all possible values. From the normalization condition
\[
\trace\of{\rho}=1
\]
it then follows that:
\[
\trace\of{\rho}=\int\dd{\psi}\bra{\psi}\rho\ket{\psi}=\frac{1}{Z}\int\DD{U}\,\e^{-S_{G}\fof{U}}=1\ .
\]

\begin{figure}[h]
  \centering
  \begin{tikzpicture}[scale=0.7,nodes={inner sep=0}]
    \node[anchor=south east,xshift=-10pt] at (0,0) {\includegraphics[width=0.41\linewidth]{imgresults/boundary_conditions3}};
    \node[anchor=south west,xshift=10pt] at (0,0) {\includegraphics[width=0.41\linewidth]{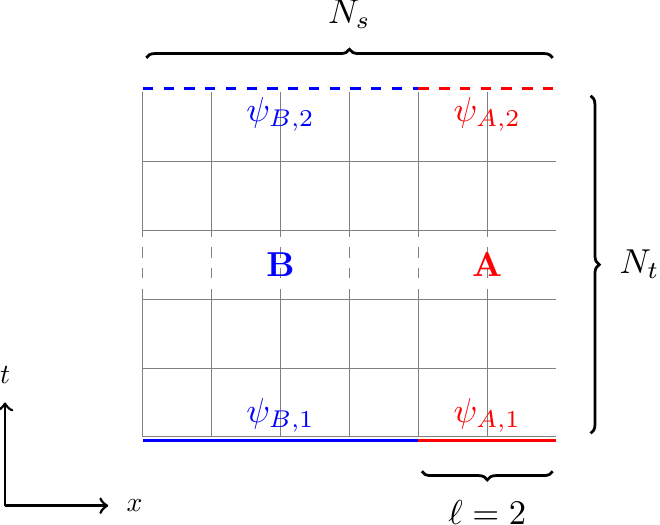}};
  \end{tikzpicture}
  \caption{Illustration in $\of{2+1}$ dimensions of how we divide the system with fixed temporal boundary states, $\psi_1$ and $\psi_2$, into two parts, $A$ and $B$. After the partitioning, we have $\psi_1=\psi_{B,1}\otimes\psi_{A,1}$ and $\psi_2=\psi_{B,2}\otimes\psi_{A,2}$, where $\psi_{B,1}$, $\psi_{B,2}$ represent the boundary states for part $B$ and $\psi_{A,1}$, $\psi_{A,2}$ the boundary states for part $A$.}
  \label{fig:dividedlat}
\end{figure}

By splitting the lattice system into two parts, $A$ and $B$, as depicted in Fig.~\ref{fig:dividedlat}, matrix elements for the reduced density matrix of part $A$ can then be represented as
\[
\bra{\psi_{A,1}}\rho_A\ket{\psi_{A,2}}=\int\dd{\psi_B}\bra{\psi_{B}\otimes\psi_{A,1}}\rho\ket{\psi_{B}\otimes\psi_{A,2}}=\frac{1}{Z}\sketchreddensme\ ,\label{eq:reddensmatpirep}
\]
where only the temporal boundary states, $\psi_{B,1}$ and $\psi_{B,2}$ for part $B$ get identified with each other and summed over all possible values, indicated in the drawing after the last equality sign by assigning the label $r_B$ to both temporal boundaries. The temporal boundaries for part $A$ are specified by the boundary states $\psi_{A,1}$, $\psi_{A,2}$.

\begin{figure}[H]
  \centering
  \begin{tikzpicture}[scale=0.7,nodes={inner sep=0}]
    \node[anchor=south east,xshift=-10pt] at (0,0) {\includegraphics[width=0.41\linewidth]{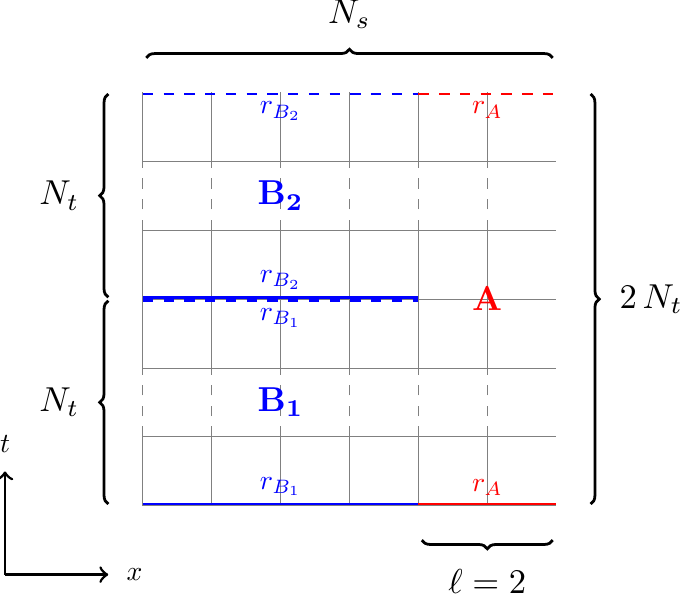}};
    \node[anchor=north east,xshift=-10pt,yshift=-10pt] at (0,0) {\includegraphics[width=0.41\linewidth]{imgresults/boundary_conditions3}};
    \node[anchor=south west,xshift=10pt] at (0,0) {\includegraphics[width=0.41\linewidth]{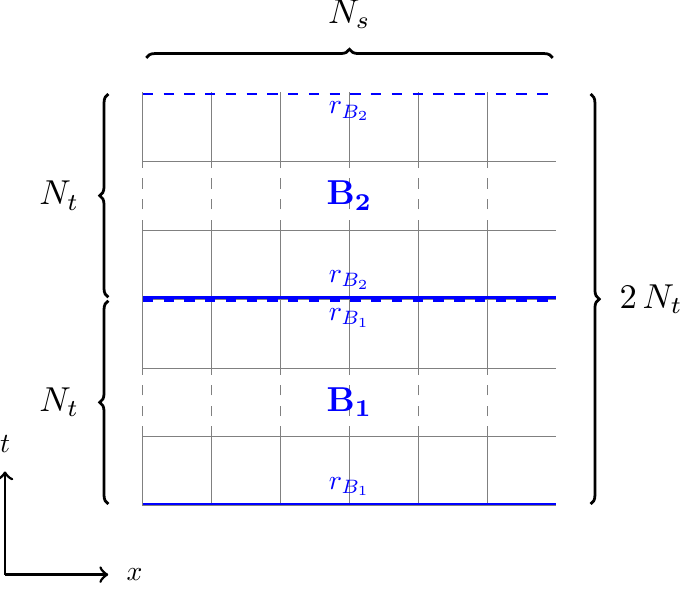}};
    \node[anchor=north west,xshift=10pt,yshift=-10pt] at (0,0) {\includegraphics[width=0.41\linewidth]{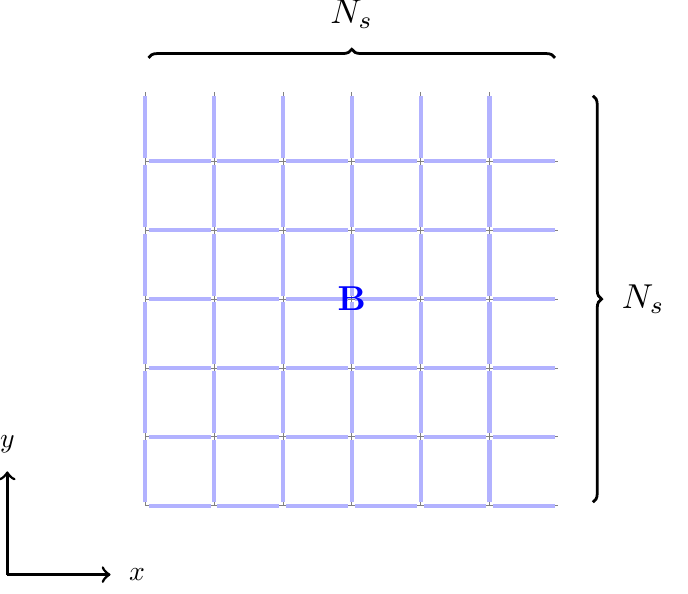}};
  \end{tikzpicture}
  \caption{Illustration of two different topologies for a $\of{2+1}$ dimensional lattice of size $V=N_s^2\times s\,N_t$ with $s=2$. The left-hand side shows the topology for the system described by the partition function $Z_c\of{\ell,s,N_t,N_s}$ in Eq.~\eqref{eq:renyientropyords}, and the right-hand side the system described by $Z^s\of{N_t,N_s}$, {\emph{i.e.}}, the product of $s$ copies of the system described by the ordinary lattice partition function $Z\of{N_t,N_s}$.}
  \label{fig:renyientropylat}
\end{figure}

With these definitions, it is now straightforward to obtain an expression for the so-called \emph{R{\'e}nyi entropy} of order $s\in\mathbb{N}$ in terms of a ratio of two lattice partitions that represent $\trace\of{\rho_A^s}$:
\begin{multline}
H_{s}\of{\ell,N_t,N_s}=\frac{1}{1-s}\log\trace\of{\rho_{A}^{s}}=\frac{1}{1-s}\log\frac{Z_{c}\of{\ell,s,N_{t},N_{s}}}{Z^{s}\of{N_{t},N_{s}}}\\
=\frac{-1}{1-s}\of{F_c\of{\ell,s,N_t,N_s}-s\,F\of{N_t,N_s}}\ .\label{eq:renyientropyords}
\end{multline}
Here $Z_c\of{\ell,s,N_t,N_s}$ is the partition function for a system with the topology depicted on the left-hand side of Fig.~\ref{fig:renyientropylat}. The example shows the case of $s=2$ in (2+1) dimensions with region $A$ having width $\ell=2$. The right hand side of Fig.~\ref{fig:renyientropylat} shows the same lattice but with a different topology, so that it decomposes into a product of two copies of the system described by the normal partition function $Z\of{N_t,N_s}$.

Using the expression for entanglement entropy given in Eq.~\eqref{eq:SEE}, one finds with these definitions:
\begin{multline}
S_{EE}\of{\ell,N_t,N_s}=-\lim_{s\to 1} \partd{\log\trace\of{\rho_A^s}}{s}=\lim_{s\to 1}\partd{F_c\of{\ell,s,N_t,N_s}}{s}-F\of{N_t,N_s}\\
\approx F_c\of{\ell,2,N_t,N_s}-2\,F\of{N_t,N_s}\ ,\label{eq:eentropylat}
\end{multline}
where on the last line we approximated 
\[
\partd{F_c\of{\ell,s,N_t,N_s}}{s}\approx \frac{F_c\of{\ell,s+\Delta s,N_t,N_s}-F_c\of{\ell,s,N_t,N_s}}{\Delta s}\ 
\]
with $\Delta s=1$, and used that for $s=1$ one has $F_c\of{\ell,1,N_t,N_s}=F\of{N_t,N_s}$. With this approximation, the entanglement entropy turns out to be just the R{\'e}nyi entropy of order two:
\[
S_{EE}\of{\ell,N_t,N_s}\approx H_{2}\of{\ell,N_t,N_s}\ .\label{eq:seeapproxbyh2}
\]

As the entanglement entropy has a $\ell$-independent UV-divergence, it is convenient to look at the derivative of $S_{EE}$ with respect to $\ell$ instead of $S_{EE}$ itself. On the lattice, this corresponds again to a discrete lattice derivative,
\[
\left.\partd{S_{EE}\of{\ell',N_t,N_s}}{\ell'}\right\vert_{\ell'=\ell+1/2}\approx F_c\of{\ell+1,2,N_t,N_s}-F_c\of{\ell,2,N_t,N_s}\ .\label{eq:eentropylatderiv}
\]
Apart from getting rid of UV divergences, Eq.~\eqref{eq:eentropylatderiv} has from lattice Monte Carlo perspective also the advantage over Eq.~\eqref{eq:eentropylat}, that the change that has to be done to the system to increase the width of region $A$ from $\ell$ to $\of{\ell+1}$ is in general much smaller than the required change to get from width $\ell=0$ to $\ell$. However, also the change from $\ell$ to $\of{\ell+1}$ represents a highly non-local change that affects a large number of degrees of freedom. As a consequence there is a bad overlap problem, meaning that link-variable configurations that contribute to the partition function of the $\ell$ system play essentially no role for the partition function of the $\of{\ell+1}$ system, and vice versa. 

\begin{figure}[htbp]
  \centering
  \begin{tikzpicture}[scale=0.68,nodes={inner sep=0}]
    \node[anchor=south east,xshift=-20pt] at (0,0) {\includegraphics[height=0.31\linewidth,keepaspectratio]{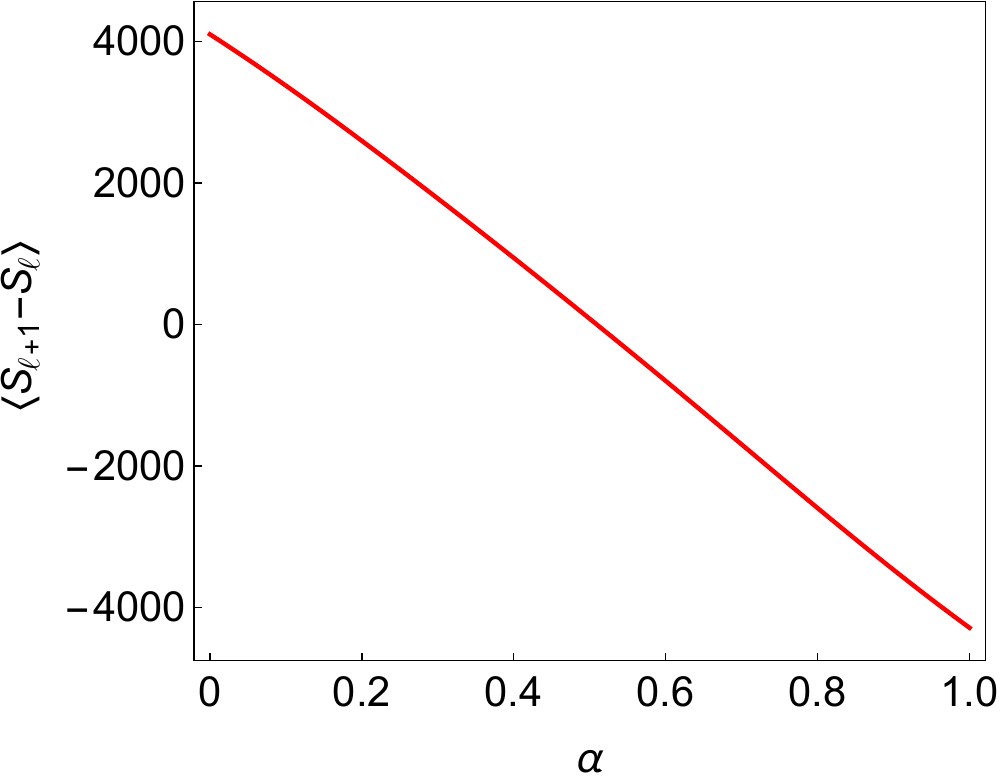}};
    \node[anchor=south west,xshift=20pt] at (0,0) {\includegraphics[height=0.31\linewidth,keepaspectratio]{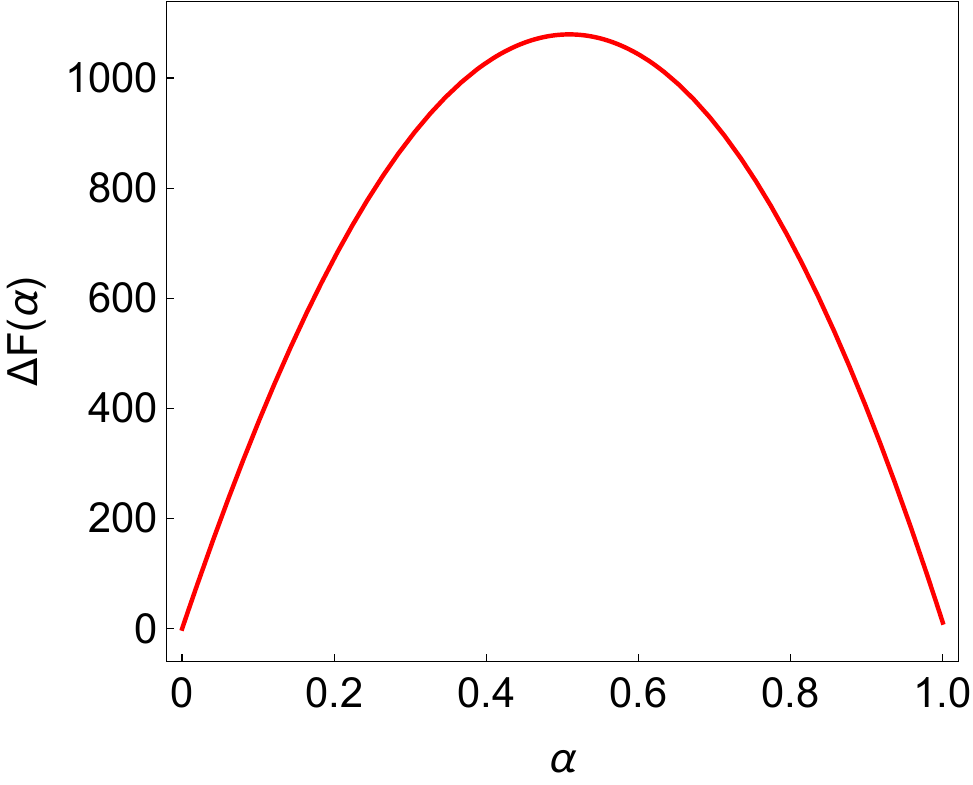}};
  \end{tikzpicture}
  \caption{Interpolation function for Eq.~\eqref{eq:dzldalpha} as function (left) and the corresponding running free energy difference Eq.~\eqref{eq:dseedalpharunningint} (right) as functions of $\alpha$. The data is taken from~\cite{Nakagawa:2009jk} and corresponds to a $\SU{3}$ gauge theory on a (3+1)-dimensional lattice with $N_s=16$, $N_t=16$, $s=2$, $\ell=2$ at $\beta_g=6.0$. The displayed data does not include errors.}
  \label{fig:freeenergyvsalpha}
\end{figure}

To overcome the overlap problem, the authors of~\cite{Buividovich:2008kq} proposed to define a one-parameter family of interpolating partition functions, $Z_{\ell}^{*}\of{\alpha}$, with $\alpha\in\fof{0,1}$, so that 
\[
Z_{\ell}^{*}\of{0}=Z_c\of{\ell,s,N_t,N_s}\quad\text{and}\quad Z_{\ell}^{*}\of{1}=Z_c\of{\ell+1,s,N_t,N_s}\ ,\label{eq:inpartfbv}
\]
where we leave the dependency of $Z_{\ell}^{*}$ on $N_t$ and $N_s$ implicit in order to avoid too lengthy expressions. The interpolating partition function is chosen to be
\[
Z^{*}_{\ell}\of{\alpha}=\int\DD{U}\,\exp\of{-\of{1-\alpha}\,S_{G,\ell}\fof{U}-\alpha\,S_{G,\ell+1}\fof{U}}\ ,\label{eq:partftradip}
\]
where $S_{G,\ell}\fof{U}$ and $S_{G,\ell+1}\fof{U}$ represent for a given configuration of link-variables, $U=\cof{U_{x,\nu}}_{x,\nu}$, the two values of the gauge action where the subsystem $A$ has width $\ell$ and $\of{\ell+1}$, respectively. The derivative of the entanglement entropy with respect to $\ell$ from Eq.~\eqref{eq:eentropylatderiv} is then obtained as
\[
\left.\partd{S_{EE}\of{\ell',N_t,N_s}}{\ell'}\right\vert_{\ell'=\ell+1/2}\approx -\int\limits_0^1\dd\alpha \partd{\log Z^{*}_{\ell}\of{\alpha}}{\alpha}=\int\limits_0^1\dd\alpha \avof{S_{G,\ell+1}-S_{G,\ell}}_{\alpha}\ ,\label{eq:dseedl}
\]
where the integration over $\alpha$ is performed by interpolating lattice results for the expectation values
\[
\partd{\log Z^{*}_{\ell}\of{\alpha}}{\alpha}= \avof{S_{G,\ell+1}-S_{G,\ell}}_{\alpha}\ ,\label{eq:dzldalpha}
\]
for a sufficiently dense set of $\alpha$-values in the integration domain $\fof{0,1}$. An example for a possible outcome of this procedure is shown in Fig.~\ref{fig:freeenergyvsalpha}, using data for a $\SU{3}$ gauge theory on a $\of{3+1}$-dimensional lattice with $N_s=16$, $N_t=16$, $s=2$ and $\ell=2$ at $\beta_g=6.0$, which has been extracted from~\cite{Nakagawa:2009jk}. The left-hand panel of Fig.~\ref{fig:freeenergyvsalpha} shows the interpolating function for Eq.~\eqref{eq:dzldalpha} as function of $\alpha$ and the right-hand panel shows the corresponding running free energy differences,
\[
\Delta F_{\ell}\of{\alpha}=\int\limits_0^{\alpha}\dd\alpha{'}\,\avof{S_{G,\ell+1}-S_{G,\ell}}_{\alpha'}\ ,\label{eq:dseedalpharunningint}
\]
where $\Delta F_{\ell}\of{1}$ corresponds to the value of Eq.~\eqref{eq:dseedl} we are interested in.

The example depicted in Fig.~\ref{fig:freeenergyvsalpha} illustrates the problem that arises when using the just described method for interpolating between the partition functions \eqref{eq:inpartfbv}. During the interpolation process, the free energy has to pass through a huge maximum. This is presumably due to the fact that the configurations that contribute to the interpolating partition function from Eq.~\eqref{eq:partftradip} at $0<\alpha<1$ are simultaneously subject to both actions, the one corresponding to a width $\ell$ of region $A$, and the one for region $A$ having width $\of{\ell+1}$. As the configuration distributions sampled with either of two actions separately have very little overlap (overlap problem), the number of configurations that can give a significant contribution to the interpolation partition function, Eq.~\eqref{eq:partftradip}, must be highly reduced when $\alpha\sim 1/2$, resulting in a significantly increased free energy. This is problematic because the values of Eq.~\eqref{eq:dzldalpha} for different $\alpha$ values are determined with Monte Carlo methods and therefore come with a statistical error. Upon integration over $\alpha$, the positive and negative values of Eq.~\eqref{eq:dzldalpha} that occur as $\alpha$ runs through the interval $\fof{0,1}$ lead to large cancelations, so that the total free energy difference, $\Delta F_{\ell}\of{1}$, is again relatively small, the errors in the measured values of Eq.~\eqref{eq:dzldalpha} cannot cancel but simply accumulate, giving rise to a bad signal-to-noise ratio.

\subsection{Improved method}\label{ssec:imprlatEEmethod}

\begin{figure}[htbp]
\centering
\begin{minipage}[t]{0.49\linewidth}
\centering
\includegraphics[width=0.8\linewidth]{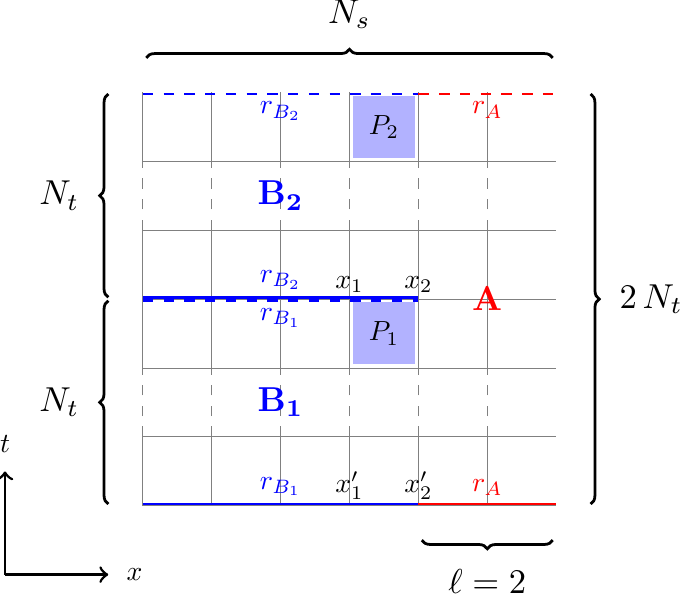}
\end{minipage}\hfill
\begin{minipage}[t]{0.49\linewidth}
\centering
\includegraphics[width=0.8\linewidth]{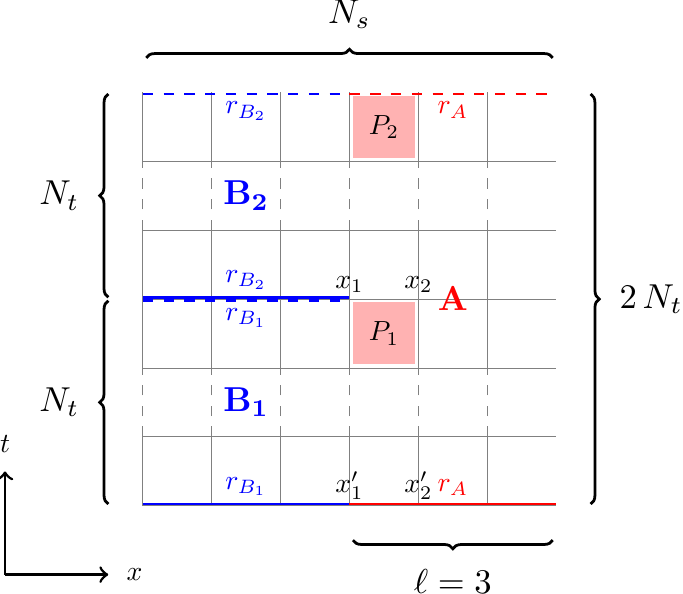}
\end{minipage}\\[10pt]
\begin{minipage}[t]{0.49\linewidth}
\centering
\includegraphics[width=0.8\linewidth]{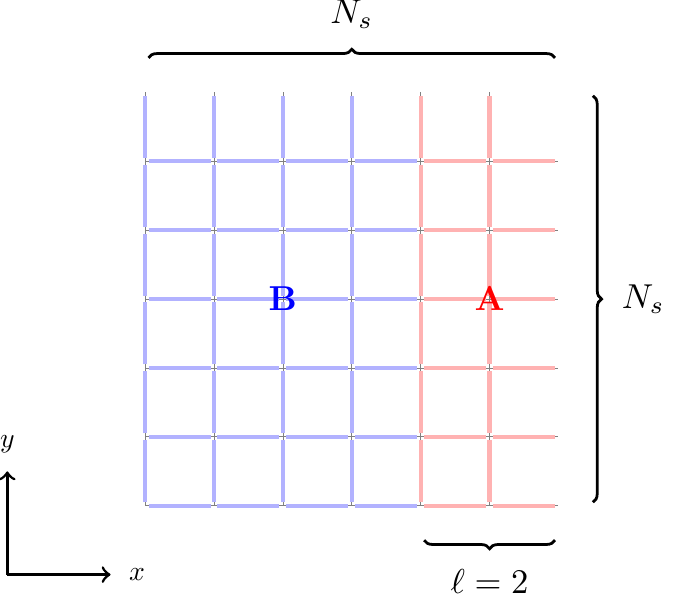}
\end{minipage}\hfill
\begin{minipage}[t]{0.49\linewidth}
\centering
\includegraphics[width=0.8\linewidth]{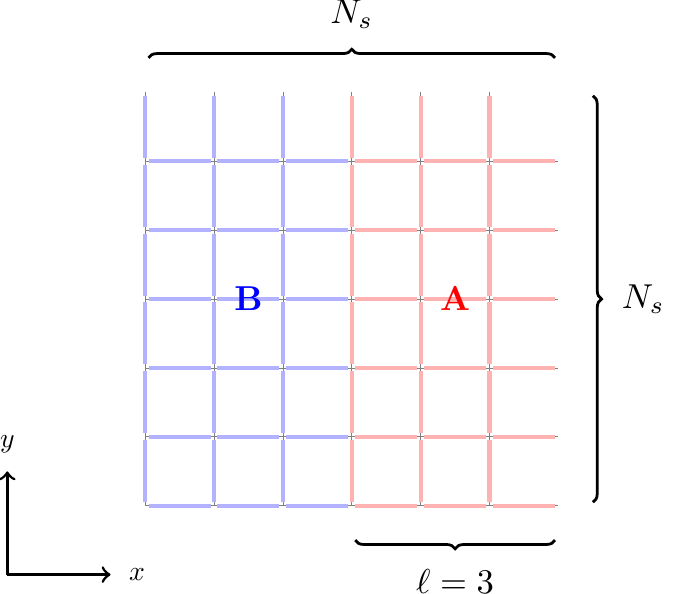}
\end{minipage}
\caption{The figure illustrates at the example of a (2+1)-dimensional lattice of size $N_{s}^2\times s\,N_{t}$ with $s=2$, how a change of temporal boundary conditions, associated to a change of the width $\ell$ of region A from $\ell=2$ to $\ell=3$, affects the way in which plaquettes are computed.\\
The plaquette $P_{1}$ in the top-left panel touches from below the dashed line $r_{B_1}$, meaning that the top-edge of $P_{1}$ does not consist of the link variable that connects $x_{1}$ and $x_{2}$, but rather the link variable that connects $x'_{1}$ and $x'_{2}$. Similarly, the upper edge of plaquette $P_{2}$ does not consist of the link between $x'_{1}$ and $x'_{2}$ but of the link between $x_{1}$ and $x_{2}$. In the top-right panel, the same temporal plaquettes are marked in red and belong now to region $A$: this means that the link that closes the plaquette $P_{1}$ along its top-edge is now indeed the link variable that connects $x_{1}$ and $x_{2}$ and the link that closes $P_{2}$ along its top-edge is the one that connects $x'_{1}$ to $x'_{2}$.\\
The two lower panels show for $\ell=2$ (left) and $\ell=3$ (right) the spatial links over which the temporal plaquettes are subject to the boundary conditions of region $A$ (red) and region $B$ (blue).}
\label{fig:bdcond}
\end{figure}

To avoid the formation of huge free energy barriers and resulting bad signal-to-noise ratios, we can use an alternative interpolation method. To this end, let us denote by $C$ the set of all plaquettes in a $V=s\,N_{t}\times N_{s}^{d-1}$ lattice, which are not affected by a change of temporal boundary conditions. These are all plaquettes except for the temporal ones that touch from below the time-slices with $x^{d}=r\cdot N_{t}$ for $r=1,\ldots,s$ (cf. Fig.~\ref{fig:bdcond}). 

A generalized partition function can then be written as:
\begin{multline}
Z\sof{\beta,s,N_{t},N_{s},\cof{n}}=\\
\int\DD{U}\exp\sbof{\frac{\beta}{\Nc}\sbof{\sum\limits_{\Box \in C}\re\trace\sof{U_{\vphantom{\bar{\ell}}\Box}}+\sum_{\bar{x}}\sum_{\nu=1}^{d-1}\sum_{r=1}^{s}\re\trace\sof{U^{\of{n_{\bar{x},\nu}}}_{\nu d}\of{\bar{x},r\cdot N_{t}-1}}}}\ ,\label{eq:genpartf}
\end{multline}
where $\bar{x}={x^{1},\ldots,x^{d-1}}\in\mathbb{Z}^{d-1}$ labels the spatial positions on the lattice and $U^{\of{0}}_{\nu d}\of{x}$ and $U^{\of{1}}_{\nu d}\of{x}$ are the two different values for the temporal plaquette
\[U_{\nu d}\of{x}=U^{\vphantom{\dagger}}_{\nu\vphantom{\hat{\nu}}}\of{x}\,U^{\vphantom{\dagger}}_{d}\of{x+\hat{\nu}}\,U^{\dagger}_{\nu}\of{x+\smash{\hat{d}}\vphantom{\hat{\nu}}}\,U^{\dagger}_{d}\of{x\vphantom{\hat{\nu}}}\ ,
\] 
depending on whether the plaquette is subject to the boundary conditions from outside or inside region A (see Fig.~\ref{fig:bdcond}: (0) corresponds to blue, (1) to red). Which of these two cases applies is for each spatial link controlled by the value of a corresponding discrete variable $n_{\bar{x},\nu}\in\cof{0,1}$ $\forall \bar{x}\in\mathbb{Z}^{d-1}, \nu\in\cof{1,\ldots,d-1}$. If we set for example
\[
n_{\bar{x},\nu}=\ucases{0,\ x^{1}<N_{s}-\ell\\1,\ x^{1}\geq N_{s}-\ell}\ ,
\]
the generalized partition function \eqref{eq:genpartf} reduced to $Z_{c}\of{\ell,s,N_{t},N_{s}}$ as introduced in the previous section.

To describe how the interpolation between $Z_{c}\of{\ell,s,N_{t},N_{s}}$ and $Z_{c}\of{\ell+1,s,N_{t},N_{s}}$ can be carried, using the generalized partition function Eq.~\eqref{eq:genpartf}, let us denote by
\[
K=\cof{\ssof{\bar{x}_{i},\nu_{i}}}_{i=1,\ldots,N_{K}}
\]
the (somehow) ordered set of spatial links in the boundary region where $N_{s}-\ell-1\leq x_{1}<N_{s}-\ell$, and let $K_{j}=\cof{\ssof{\bar{x}_{i},\nu_{i}}}_{i=1,\ldots,j}$ be the subset of the first $j$ elements of $K$. Let us then further define:
\[
n^{i}_{\bar{x},\nu}=\ucases{0,\,\text{if}\,x_{1}<N_{s}-\ell-1\,\lor\,\of{\bar{x},\nu}\in K\setminus K_{i}\\ 1,\,\text{if}\,x_{1}\geq N_{s}-\ell\,\lor\,\of{\bar{x},\nu}\in K_{i}}\ ,
\]
and abbreviate $Z_{i}=Z\ssof{\beta,s,N_{t},N_{s},\cof{n^{i}}}$ and $F_{i}=-\log\of{Z_{i}}$.
The expression for the derivative of the entanglement entropy \eqref{eq:eentropylatderiv} then becomes:
\[
\left.\partd{S_{EE}\of{\ell',N_t,N_s}}{\ell'}\right\vert_{\ell'=\ell+1/2}\approx-\log\of{Z_{N_{K}}}+\log\of{Z_{0}}=F_{N_{K}}-F_{0}\ .\label{eq:imprfreeenergydiff}
\]
This free energy difference can be measured from a single simulation, using a form of multi-canonical method~\cite{Berg:1992qua} for discrete sets of states, known as Wang-Landau (WL) sampling~\cite{Wang:2000fzi}. With this method one samples the modified partition function,
\begin{multline}
Z_{\text{WL}}\sof{\beta,s,N_{t},N_{s},\cof{f}}=\\
\sum\limits_{i=0}^{N_{K}} e^{f_{i}}\int\DD{U}\exp\sbof{\frac{\beta}{\Nc}\sbof{\sum\limits_{\Box \in C}\re\trace\sof{U_{\vphantom{\bar{\ell}}\Box}}+\sum_{\bar{x}}\sum_{\nu=1}^{d-1}\sum_{r=1}^{s}\re\trace\sof{U^{\of{n^{i}_{\bar{x},\nu}}}_{\vphantom{\bar{x}}\smash{\bar{x}+\ssof{r\cdot N_{t}-1}\cdot\hat{d},\nu d}}}}}\ ,\label{eq:wlpartf}
\end{multline}
and adjusts the set of parameters $\cof{f}=\cof{f_{i}}_{i=0,\ldots,N_{K}}$ till the histogram $\cof{H}=\cof{H_{j}}_{j=0,\ldots,N_{K}}$, with
\[
H_{j}=\totd{\log\of{Z_{\text{WL}}\sof{\beta,s,N_{t},N_{s},\cof{f}}}}{f_{j}}{}\ ,\label{eq:wlhistentry}
\]
is approximately flat. After appropriate values for the $\cof{f}$ have been found with the WL method, one can start to accumulate high statistics for the histogram $\cof{H}$ while keeping the $\cof{f}$ fixed, and obtain an improved estimator for the free energy difference by setting:
\[
F_{N_{K}}-F_{0}=f_{N_{K}}-f_{0}+\log\of{H_{N_{K}}}-\log\of{H_{0}}\ .\label{eq:imprfreeenergydiff2}
\] 
Error bars for Eq.~\eqref{eq:imprfreeenergydiff2} can be obtained from the measurements of the histograms in Eq.~\eqref{eq:wlhistentry} using the jackknife method.

\begin{figure}[htbp]
  \centering
  \begin{tikzpicture}[scale=0.8,nodes={inner sep=0}]
    \getscale{\scalef}
    \setlength{\slinewidth}{\scalef\linewidth}
    \node[anchor=south east,xshift=-30pt] (na) at (0,0) {\includegraphics[height=0.32\slinewidth,keepaspectratio]{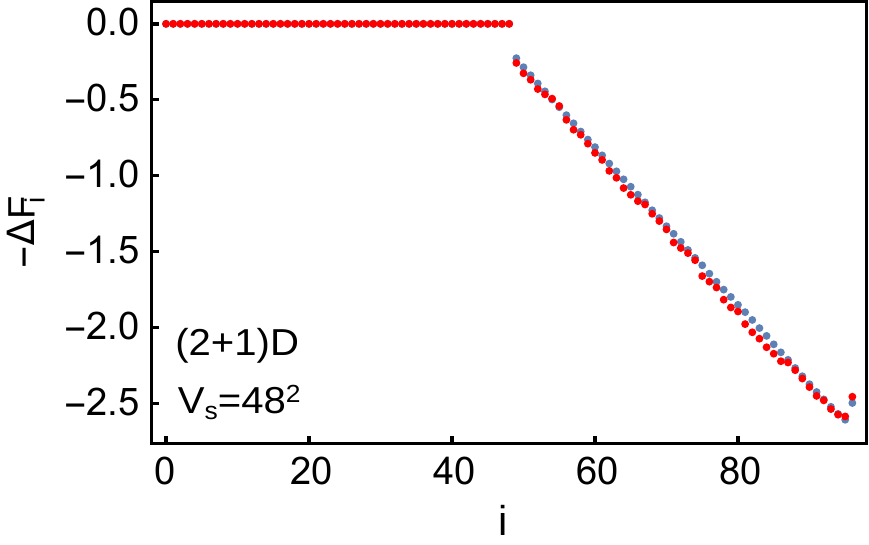}};
    \draw[fill=red!30,opacity=0.3,draw=none] ([xshift=39pt,yshift=26pt]na.south west) rectangle ([xshift=69pt,yshift=140pt]na.south west);
    \node[anchor=south west,xshift=10pt] at (0,0) {\includegraphics[height=0.33\slinewidth,keepaspectratio]{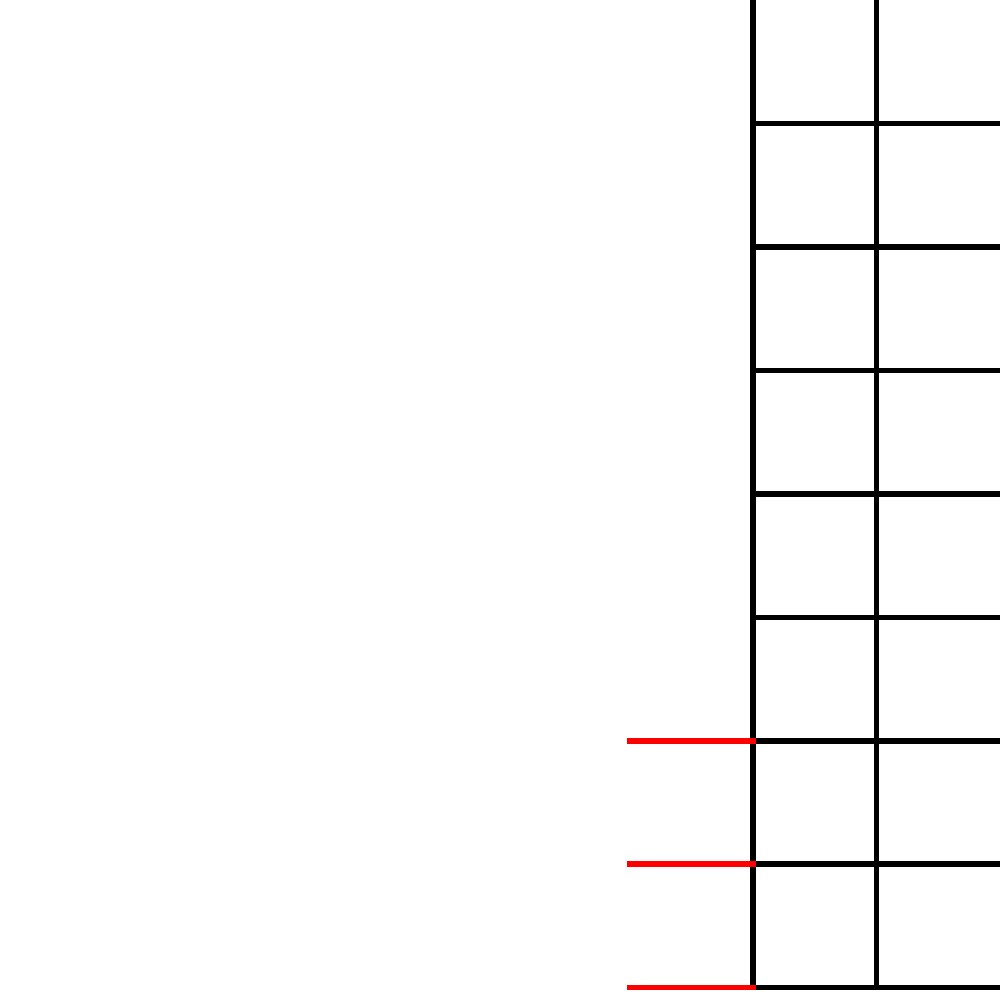}};
  \end{tikzpicture}\\[10pt]
  \begin{tikzpicture}[scale=0.8,nodes={inner sep=0}]
    \getscale{\scalef}
    \setlength{\slinewidth}{\scalef\linewidth}
    \node[anchor=south east,xshift=-30pt] (na) at (0,0) {\includegraphics[height=0.32\slinewidth]{imgresults/free_energy_diff_3d_b}};
    \draw[fill=red!30,opacity=0.3,draw=none] ([xshift=39pt,yshift=26pt]na.south west) rectangle ([xshift=133pt,yshift=140pt]na.south west);
    \node[anchor=south west,xshift=10pt] at (0,0) {\includegraphics[height=0.33\slinewidth,keepaspectratio]{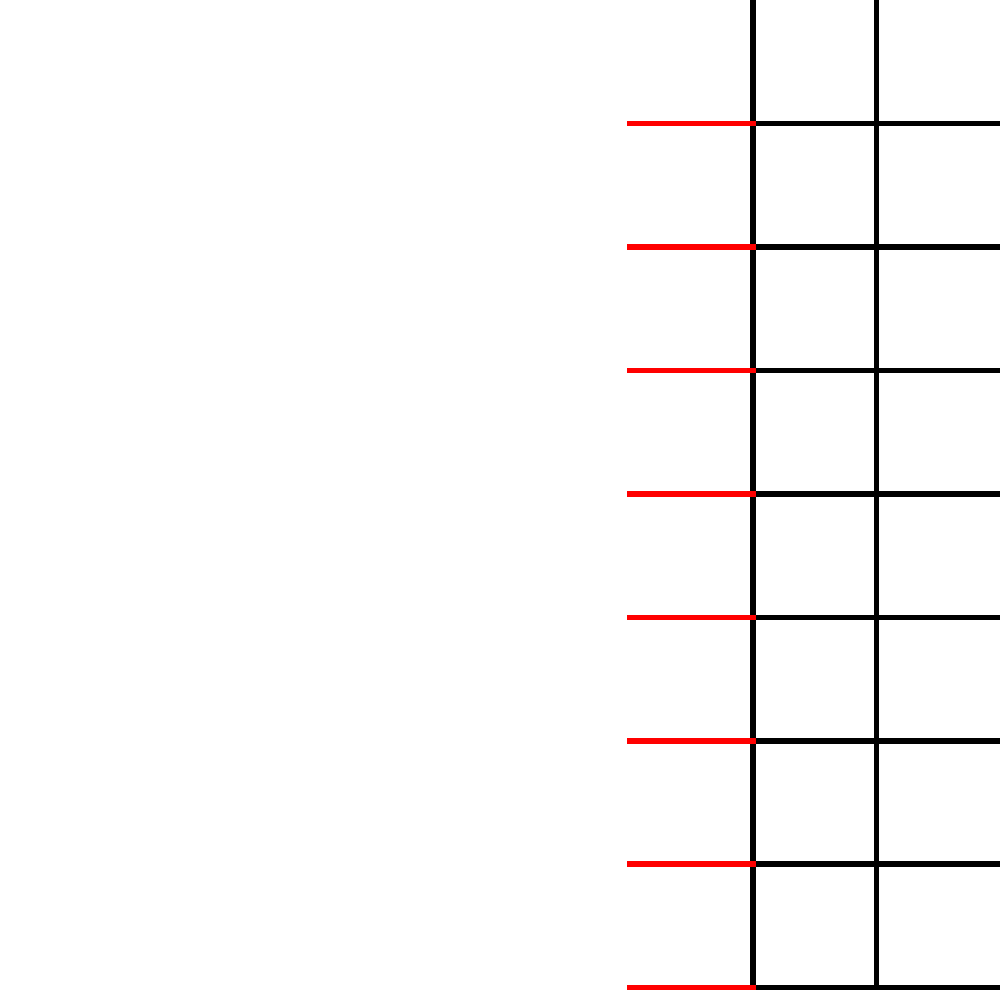}};
  \end{tikzpicture}\\[10pt]
  \begin{tikzpicture}[scale=0.8,nodes={inner sep=0}]
    \getscale{\scalef}
    \setlength{\slinewidth}{\scalef\linewidth}
    \node[anchor=south east,xshift=-30pt] (na) at (0,0) {\includegraphics[height=0.32\slinewidth]{imgresults/free_energy_diff_3d_b}};
    \draw[fill=red!30,opacity=0.3,draw=none] ([xshift=39pt,yshift=26pt]na.south west) rectangle ([xshift=172pt,yshift=140pt]na.south west);
    \node[anchor=south west,xshift=10pt] at (0,0) {\includegraphics[height=0.33\slinewidth,keepaspectratio]{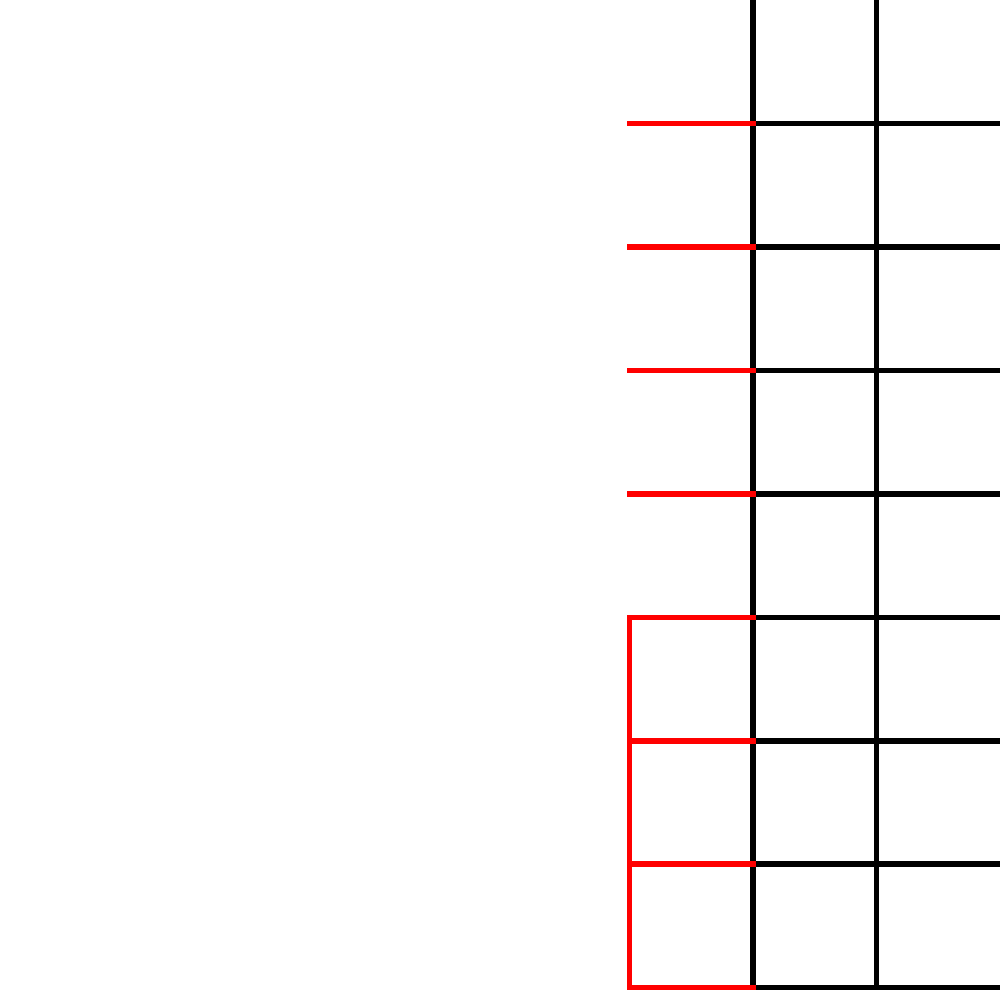}};
  \end{tikzpicture}\\[10pt]
  \begin{tikzpicture}[scale=0.8,nodes={inner sep=0}]
    \getscale{\scalef}
    \setlength{\slinewidth}{\scalef\linewidth}
    \node[anchor=south east,xshift=-30pt] (na) at (0,0) {\includegraphics[height=0.32\slinewidth]{imgresults/free_energy_diff_3d_b}};
    \draw[fill=red!30,opacity=0.3,draw=none] ([xshift=39pt,yshift=26pt]na.south west) rectangle ([xshift=221pt,yshift=140pt]na.south west);
    \node[anchor=south west,xshift=10pt] at (0,0) {\includegraphics[height=0.33\slinewidth,keepaspectratio]{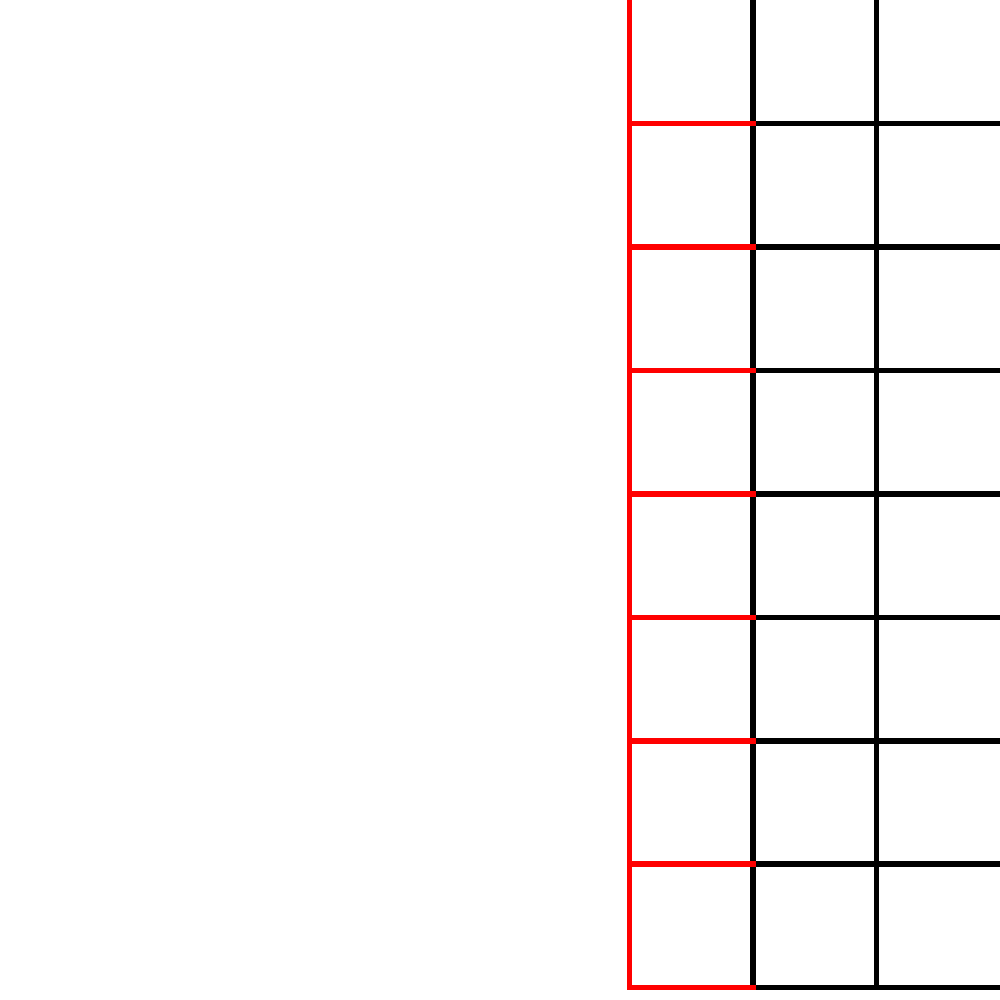}};
  \end{tikzpicture}
  \caption{Illustration of how in $\SU{2}$ gauge theory on a (2+1)-dimensional lattice with topology as in the left-hand panel of Fig.~\ref{fig:bdcond}, with $N_s=48$, $N_t=8$, $s=2$, and $\beta_g=96$, the free energy changes as spatial links are sequentially added to the boundary of region $A$ to grow its width from $\ell=2$ to $\ell=3$. The left-hand panels show the total change in free energy as function of the number $i$ of spatial links added to region $A$. The red data points correspond to the initial Wang-Landau estimate for $\Delta F_i$, while the blue data points are the histogram-improved values. In each panel the red-shaded area marks the change in free energy corresponding to an intermediate shape of region $A$ as sketched to the right. Note that the sketches show a much smaller lattice than the one used for the simulations.}
  \label{fig:freeenergyvsn3d}
\end{figure}

\begin{figure}[htbp]
  \centering
  \begin{tikzpicture}[scale=0.8,nodes={inner sep=0}]
    \getscale{\scalef}
    \setlength{\slinewidth}{\scalef\linewidth}
    \node[anchor=south east,xshift=-30pt] (na) at (0,0) {\includegraphics[height=0.32\slinewidth,keepaspectratio]{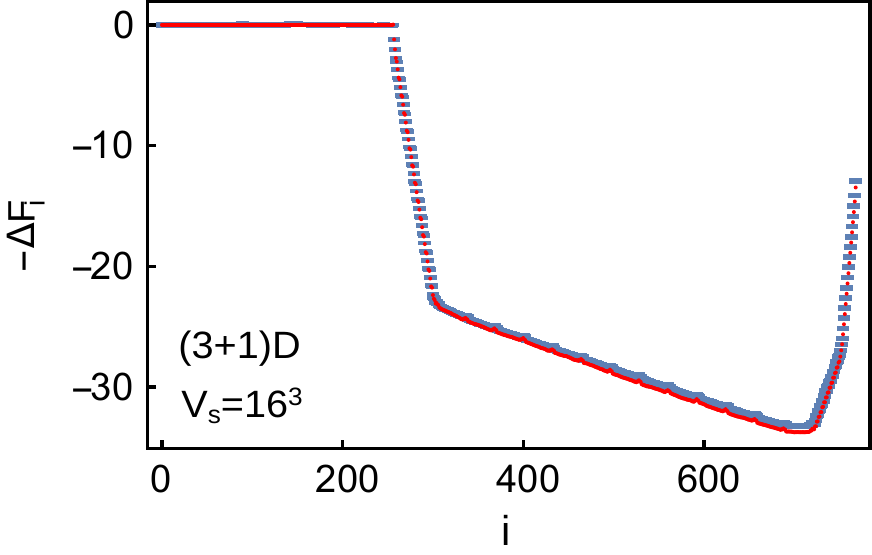}};
    \draw[fill=red!30,opacity=0.3,draw=none] ([xshift=38pt,yshift=27pt]na.south west) rectangle ([xshift=101pt,yshift=140pt]na.south west);
    \node[anchor=south west,xshift=30pt] at (0,0) {\includegraphics[height=0.37\slinewidth,keepaspectratio]{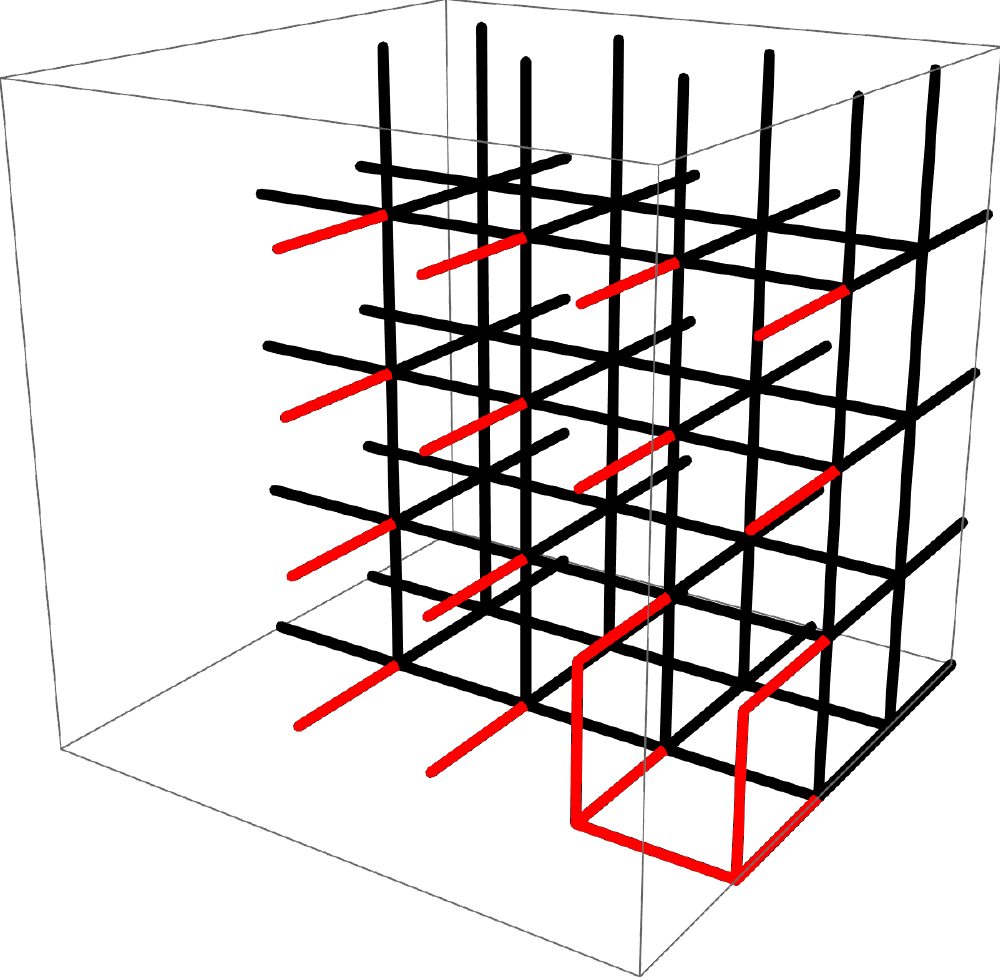}};
  \end{tikzpicture}\\[5pt]
  \begin{tikzpicture}[scale=0.8,nodes={inner sep=0}]
    \getscale{\scalef}
    \setlength{\slinewidth}{\scalef\linewidth}
    \node[anchor=south east,xshift=-30pt] (na) at (0,0) {\includegraphics[height=0.32\slinewidth]{imgresults/free_energy_diff_4d_b}};
    \draw[fill=red!30,opacity=0.3,draw=none] ([xshift=38pt,yshift=27pt]na.south west) rectangle ([xshift=113pt,yshift=140pt]na.south west);
    \node[anchor=south west,xshift=30pt] at (0,0) {\includegraphics[height=0.37\slinewidth,keepaspectratio]{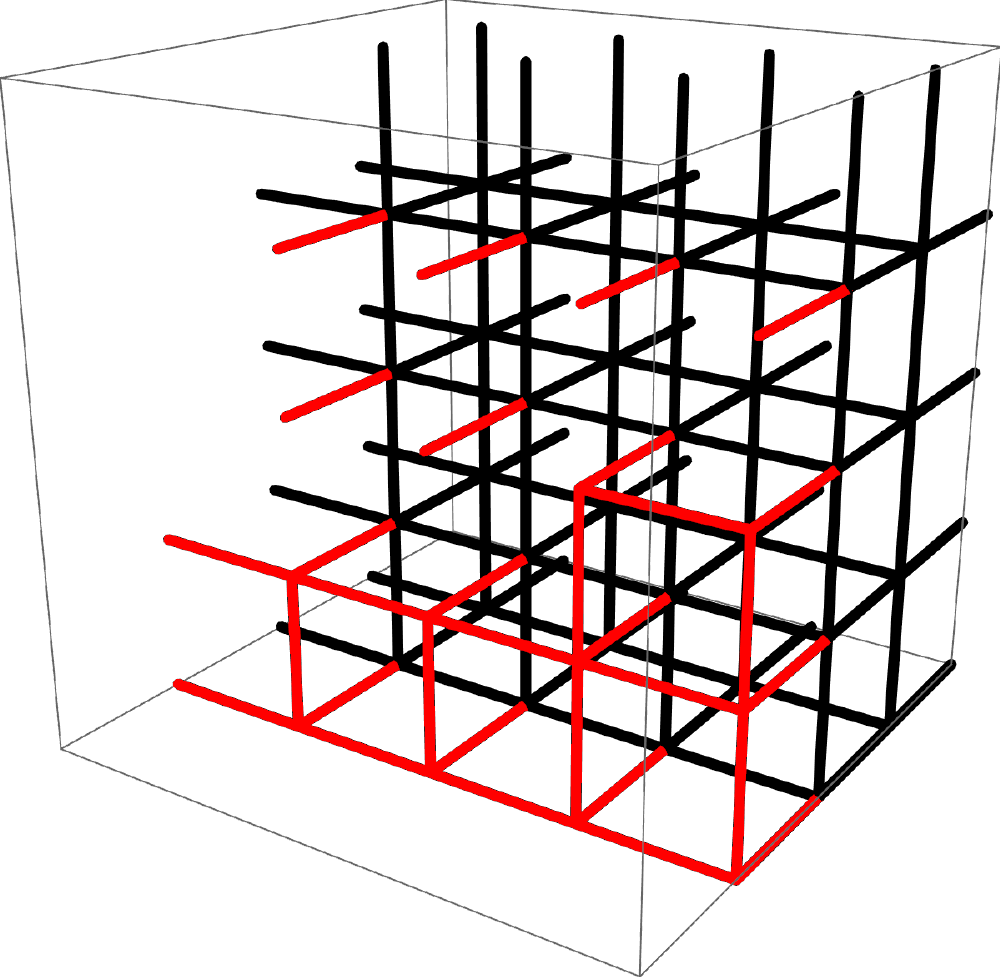}};
  \end{tikzpicture}\\[5pt]
  \begin{tikzpicture}[scale=0.8,nodes={inner sep=0}]
    \getscale{\scalef}
    \setlength{\slinewidth}{\scalef\linewidth}
    \node[anchor=south east,xshift=-30pt] (na) at (0,0) {\includegraphics[height=0.32\slinewidth]{imgresults/free_energy_diff_4d_b}};
    \draw[fill=red!30,opacity=0.3,draw=none] ([xshift=38pt,yshift=27pt]na.south west) rectangle ([xshift=213pt,yshift=140pt]na.south west);
    \node[anchor=south west,xshift=30pt] at (0,0) {\includegraphics[height=0.37\slinewidth,keepaspectratio]{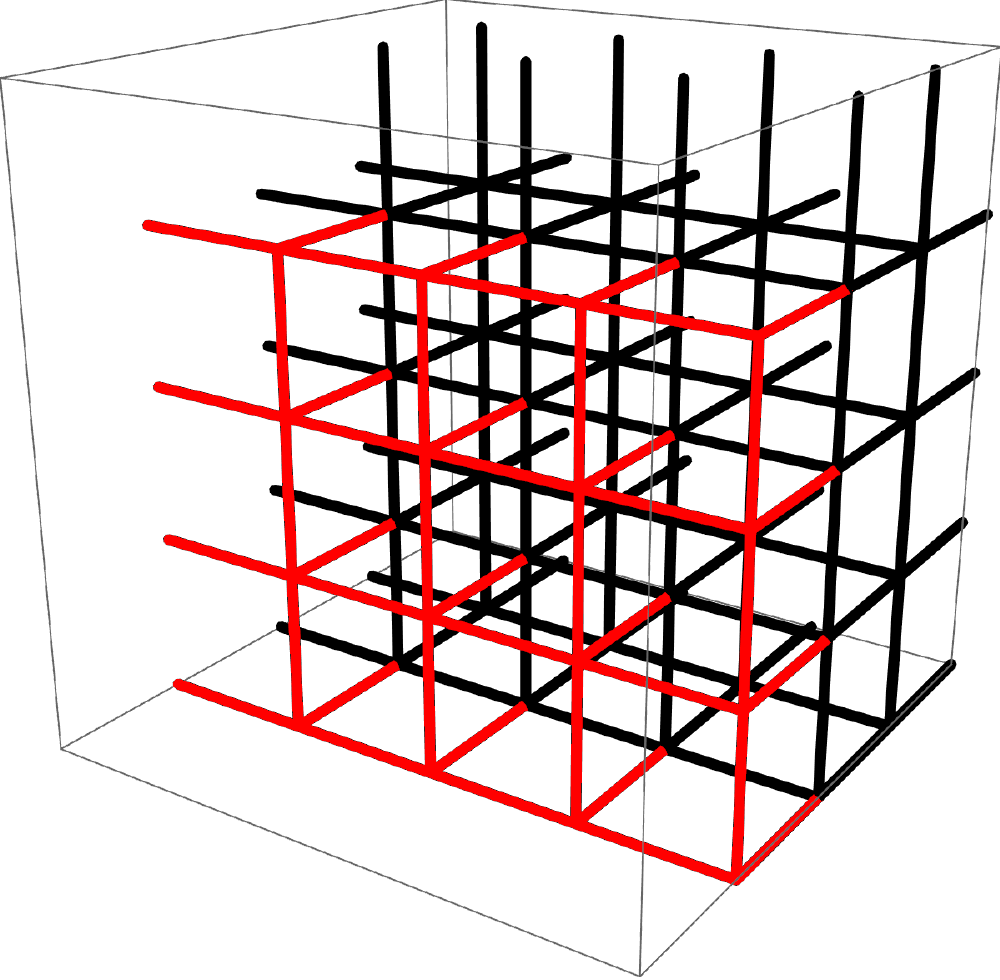}};
  \end{tikzpicture}\\[5pt]
  \begin{tikzpicture}[scale=0.8,nodes={inner sep=0}]
    \getscale{\scalef}
    \setlength{\slinewidth}{\scalef\linewidth}
    \node[anchor=south east,xshift=-30pt] (na) at (0,0) {\includegraphics[height=0.32\slinewidth]{imgresults/free_energy_diff_4d_b}};
    \draw[fill=red!30,opacity=0.3,draw=none] ([xshift=38pt,yshift=27pt]na.south west) rectangle ([xshift=218pt,yshift=140pt]na.south west);
    \node[anchor=south west,xshift=30pt] at (0,0) {\includegraphics[height=0.37\slinewidth,keepaspectratio]{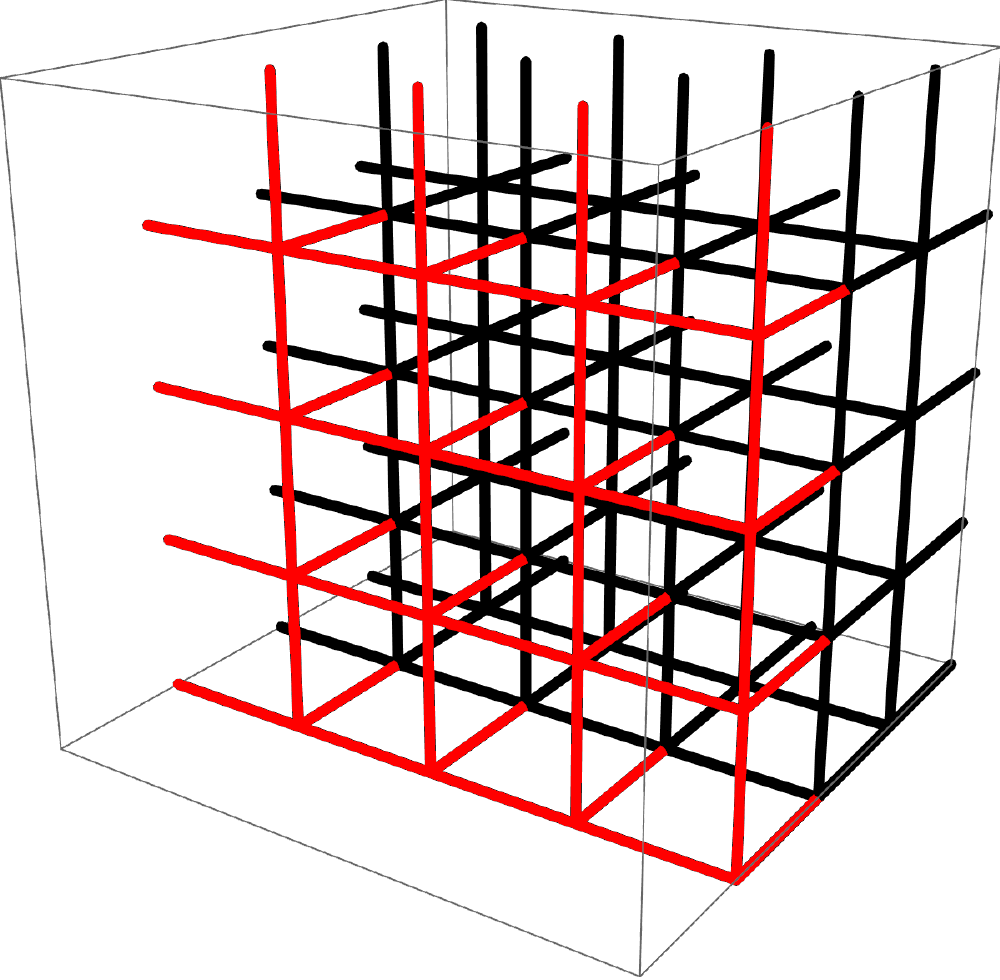}};
  \end{tikzpicture}
  \caption{Same as in Fig.~\ref{fig:freeenergyvsn3d} but for $\SU{3}$ on a (3+1)-dimensional lattice with $N_s=16$, $N_t=16$, $s=2$, and $\beta_g=6.0$. After all spatial links perpendicular to the boundary of region $A$ have been added, one continues to add spatial so as to form complete cubes that fill one row after the other. Note that the sketches on the right-hand side show a much smaller lattice than the one used to obtain the data in the left-hand plots. The sketches are meant to illustrate why the free-energy difference as a function of $n$ changes slope for a certain points. The total free energy difference for the here discussed example is the same as in Fig.~\ref{fig:freeenergyvsalpha} but is obtained without having to overcome a equally huge free energy barrier.}
  \label{fig:freeenergyvsn4d}
\end{figure}

Possible choices for ordering the spatial boundary link variables in $K$ are illustrated in Figs.~\ref{fig:freeenergyvsn3d} and \ref{fig:freeenergyvsn4d} for the (2+1)-dimensional and the (3+1)-dimensional case, respectively. These examples exploit the fact that gauge invariance ensures that a change of temporal boundary conditions over spatial links for which at least one end is always (before and after the update) either completely in region $A$ or completely in region $B$, does not change the free energy (cf. Appendix~\ref{sec:changeoftempbc}). This is the reason why the free energy graphs in Figs.~\ref{fig:freeenergyvsn3d} and \ref{fig:freeenergyvsn4d} are flat at small $i$, where $i$ is used to enumerate the boundary link variables. To determine the total free energy difference in Eq.~\eqref{eq:imprfreeenergydiff2}, one therefore has to consider only the boundary states for $i>N_s^{d-2}$. Note also, that with this choice of ordering of the spatial links in $K$, the change in free energy as function of $i$ is piecewise linear. This will be particularly useful for future studies on larger systems, as it means that one does not have to measure the change in free energy for all $N_K$ boundary states, but only has to determine the slope of $\Delta F_i$ as a function of $i$ in each of the piecewise linear interval.

\section{Holographic approach}\label{sec:holo}

Entanglement entropy is a difficult quantity to compute in general QFTs and even if the theory admits a lattice prescription the computation proceeds nontrivially as we have seen in the previous section. However, if a QFT admits a holographic description, its computation becomes easy using the Ryu-Takayanagi (RT) formula \cite{Ryu:2006bv,Ryu:2006ef,Nishioka:2009un}. The RT-formula states that the entanglement entropy is given by the area of a certain minimal bulk surface in the holographic dual of the QFT. 

In this section we will review salient details on the computation of the derivative of entanglement entropy. In particular, we will show how the derivative can be used to solve the inverse problem, to infer the dual background metric, which can only be obtained up to statistical error inherent in the data of the derivative.

\subsection{Derivative of the entanglement entropy}

\begin{figure}
    \centering
    \includegraphics[width=0.5\textwidth]{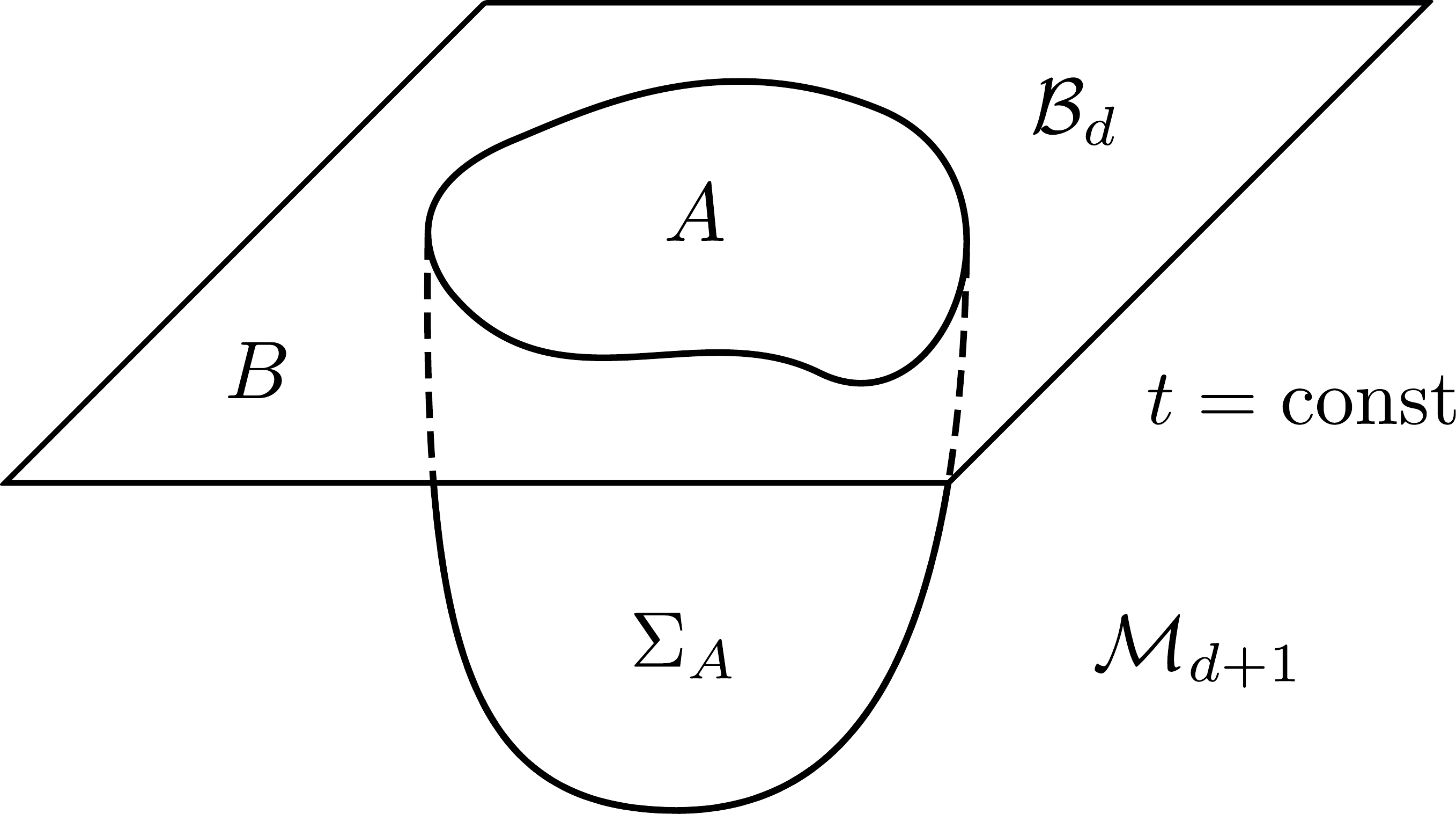}
    \caption{The holographic bulk surface used in computing $S_{EE}(A)$ using the RT-prescription. The QFT lives on the boundary $\mathcal B_d = A \cup B$, and the dual gravity theory lives in the bulk $\mathcal M_{d+1}$ (times the internal geometry). The bulk codimension-2 surface $\Sigma_A$ is anchored to the boundary of $A$ so that $\partial \Sigma_A = \partial A$. The bulk metric in $\mathcal M_{d+1}$ is such that it causes $\Sigma_A$ to hang in the bulk when its area is being minimized.}
    \label{fig:rt_setup}
\end{figure}

In a holographic theory the subsystems $A$ and $B$ are regions on the conformal boundary of the bulk dual theory. The QFT is said to live on this conformal boundary while the dual gravity theory lives in the bulk enclosed by the boundary. The boundary QFT and bulk gravity descriptions are dual in the sense that they encode the same physical information. The two descriptions are equivalent but some quantities might be much easier to compute on one side of the correspondence than on the other. A great example is the entanglement entropy which is given by
\begin{align}
    S_{EE}(A) = \frac{1}{4 G_{\Nc}^{(10)}} \text{Area}(\Sigma_A) \ ,
\end{align}
where $\Sigma_A$ is the minimal area surface associated to the boundary region $A$ and $G_{\Nc}^{(10)}$ is the ten-dimensional bulk Newton constant which can be written explicitly in terms of gauge theory quantities if the dual theory is known. We consider the dual to be Yang-Mills theory and imagine $1/G^{(10)}_{\Nc}\propto \Nc^2$. Here, we are assuming that the external bulk geometry is $(d+1)$-dimensional, {\emph{i.e.}}, that the boundary QFT is $d$-dimensional. The 8-dimensional bulk surface $\Sigma_A$ has to be anchored to the boundary region $A$ whose entanglement entropy we wish to compute, that is, $\partial \Sigma_A = \partial A$. Additionally, $\Sigma_A$ has to be homologous to $A$ in the bulk meaning that one has to be able to retract $\Sigma_A$ smoothly close to $A$ through the bulk. There is an infinite family of bulk surfaces that satisfy these conditions. The RT-formula then instructs to pick the surface with minimal area as measured by the bulk metric tensor. The area of this minimal surface then gives $S_{EE}(A)$. Note that since the region $A$ is defined on a boundary time slice, the bulk surface $\Sigma_A$ is of codimension two. Also note that we are working with an asymptotically $AdS_{d+1}$ bulk theory and therefore the area of $\Sigma_A$ is divergent because it has to reach all the way to the boundary of the spacetime. This divergence dominates the entanglement entropy, and this implies that the holographic formula follows the area law. This is the RT-prescription for static cases which is the case of interest to us in the present work, but there are extensions of the prescription applicable to more general situations \cite{Dong:2016hjy}.

In this paper, we consider slab-like entangling regions $A$. That is, regions defined by
\begin{align}
    A = \{ \ \vec x \ | -\ell/2 \leq x_1 \leq \ell/2 \} \ ,
\end{align}
where $\ell$ is the width of the slab. We will always orient the slab orthogonal to the first spatial boundary dimension $x_1$. The slab is assumed to be large in directions other than $x_1$. This slab shape combined with the translation invariance of the theory implies that the profile of the bulk surface can be represented by a single function $x_1 = x_1(z)$, where $z$ is the holographic coordinate orthogonal to the boundary.

We are interested in the entanglement of such slabs in $(3+1)$ and $(2+1)$ dimensions. We will start with the former case. Consider a bulk metric in the Einstein frame of the form

\begin{align}\label{eq:4dmetric}
    ds^2 = \frac{R^2}{z^2} \left( -\frac{b(z)}{a(z)^2} \dd t^2 + \frac{a(z)^2}{b(z)} \dd z + \dd \vec x^2 \right)+R^2d\Omega_5^2 \ ,
\end{align}
where $b(z) = 1-z^4/z_h^4$ and $R$ is the radius of curvature.  This metric describes a family of asymptotically $AdS_5$ spacetimes if $a(0) = 1$. The holographic coordinate $z$ is such that $z=0$ corresponds to the conformal boundary. The function $b(z)$ is the usual blackening factor of the AdS planar black hole, which puts our theory to a finite temperature. Further, we will only consider functions $a(z)$ such that $a(z_h)=1$, where $z_h$ is the black hole horizon position.
The choice of this family of metrics is mainly motivated by simplicity and the assumption that metrics generalized from the simple AdS planar black hole are useful in reproducing properties of finite temperature entanglement entropy results obtained from lattice simulations in $(3+1)$ dimensions. We add a single parameterized function $a(z)$ to the temporal and holographic directions of the metric, thus preserving symmetries of the transverse and internal spaces. The function $a(z)$ is introduced squared in the metric so that the integrals for slab width and entanglement entropy we shortly derive will have more convenient forms.
It follows that the dual QFT is at temperature
\begin{align}
    T = \frac{1}{\pi z_h} \ .
\end{align}
It is worth pointing out that since we consider a static situation, the entanglement entropy actually does not depend on the metric component $g_{tt}$ at all. We chose to fix the relationship between $g_{tt}$ and $g_{zz}$ so that later we will be able to infer the complete bulk metric.\footnote{Note that we are omitting any compact internal spaces from our bulk which would be present if the metric was derived top-down from a full 10-dimensional string theory.} The entanglement entropy is then given by the integral
\begin{align}
    S_{EE}(\ell) = \frac{V R^3}{2 G_{\Nc}^{(5)}} \int_0^{z_*} \frac{\dd z}{z^3} \sqrt{\frac{a(z)^2}{b(z)} + x_1'(z)^2} \ , \label{eq:S_int}
\end{align}
where $x'_1\of{z}=\partial_z\,x_1\of{z}$, $V$ is the regularized volume of the slab in directions $x_2$ and $x_3$, and we have simply integrated over the internal space so that its volume is absorbed into the definition of the five-dimensional Newton constant $G^{(10)}_{\Nc}=G^{(5)}_{\Nc}R^5{\rm{Vol}}(\Omega_5)$. The point $z = z_*$ is called the turning point of the bulk surface, which is the deepest point in the bulk reached by the minimal surface. The profile of the surface $x_1(z)$ has to be such that its area is minimized. This happens when $x_1(z)$ satisfies the Euler-Lagrange equation with the Lagrangian $L$ given by the integrand of \eqref{eq:S_int}. Since $x_1(z)$ itself does not appear in the integrand, we immediately find a conserved quantity $\partial_{x_1'}L$. We can fix the value of this constant at the turning point $z = z_*$ where $x_1'(z) \to \pm\infty$ and find an equation for the profile which minimizes the area:
\begin{gather}
    \frac{x_1(z)}{z^3 \sqrt{a(z)^2/b(z) + x_1(z)^2}} = \pm\frac{1}{z_*^3} \\
    \rightarrow x_1(z) = \pm \frac{z^3}{z_*^3} \frac{a(z)}{\sqrt{b(z)}} \frac{1}{\sqrt{1-(z/z_*)^6}} \ ,
\end{gather}
where the $\pm$ refers to the two different branches of $\Sigma_A$ as a function of $z$. After we now know the profile, we can give explicit integrals for the slab width $\ell$ and the entanglement entropy $S_{EE}$:
\begin{align}
    T \ell(z_*) &= \frac{2}{\pi z_h} \int_0^{z_*} \frac{z^3}{z_*^3} \frac{a(z)}{\sqrt{b(z)}} \frac{1}{\sqrt{1-(z/z_*)^6}} \dd z \label{eq:l_int_4d} \\
    \frac{4 G_{\Nc}^{(5)}}{R^3 V T^2} S_{EE}(z_*) &= 2 \pi^2 z_h^2 \int_\epsilon^{z_*} \frac{1}{z^3} \frac{a(z)}{\sqrt{b(z)}} \frac{1}{\sqrt{1-(z/z_*)^6}} \dd z \\
    &= \pi^2 \Bigg[ \frac{z_h^2}{\epsilon^2} - \frac{z_h^2}{z_*^2} + 2 z_h^2 \int_0^{z_*} \frac{1}{z^3} \left( \frac{a(z)}{\sqrt{b(z)}} \frac{1}{\sqrt{1-(z/z_*)^6}} - 1 \right) \dd z \Bigg]\ . \label{eq:S_int_4d}
\end{align}
On the last line we have separated the UV-divergence from the $z=\epsilon$ limit of the integral.

It is instructive to study the behavior of $S_{EE}(\ell)$ for small $\ell$, that is, in the UV-limit. This limit is easily obtained from the integrals \eqref{eq:l_int_4d} and \eqref{eq:S_int_4d}. The UV-limit is equivalent to assigning $a(z) = 1$ and $b(z) = 1$, which corresponds to the zero-temperature ${\mathcal{N}} = 4$ supersymmetric Yang-Mills theory. In this limit the integrals can be computed and one finds
\begin{align}\label{eq:HEE}
    S_{EE}(\ell) = \frac{\Nc^2V}{2\pi\epsilon^2} - 2\sqrt\pi\frac{\Gamma\left(\frac{2}{3}\right)^3}{\Gamma\left(\frac{1}{6}\right)^3} \frac{\Nc^2 V}{\ell^2} \ , \quad T \ell \ll 1 \ ,
\end{align}
where $\epsilon$ is the UV-cutoff we used in the integral \eqref{eq:S_int_4d}. Note that here we displayed the canonical field theory example of ${\cal{N}}=4$ $\SU{\Nc}$ super Yang-Mills theory on flat space, dual to $AdS_5\times S^5$, wherein we plugged in known expressions for the parameters in terms of field theory quantities using ${\rm{Vol}}(S^5)=\pi^3$, $R^4=4\pi g_s\alpha'^2\Nc$ and $G^{(10)}_{\Nc}=8\pi^3\alpha'^4g_s^2$. The first term reflects the expected UV-divergence of $S_{EE}(\ell)$. On the QFT side, it originates from short-range correlations across the entangling surface. On the gravity side, the divergence has a geometric interpretation, as mentioned already above: since the surface $\Sigma_A$ has to reach all the way to the boundary, its area is infinite because the boundary is infinitely far away. Note that $S_{EE}$ is indeed proportional to $V$, the volume of directions parallel to the slab: this is the area law of entanglement entropy.

The diverging term is not particularly interesting since it is independent of the slab width $\ell$. The occurence of this divergence is actually related to the fact that the entanglement entropy itself is not an observable, similarly to say the total free energy of a system, which is typically infinite. In order to bypass this problem, one can focus on the finite piece in (\ref{eq:HEE}) or on its change. To this end, it is easy to show that the derivative of $S_{EE}(\ell)$ takes the following simple form \cite{Jokela:2020auu}:
\begin{align}
    \frac{4 G_{\Nc}^{(5)}}{V} \frac{\dd S_{EE}(\ell)}{\dd \ell} = \frac{R^3}{z_*^3} \ , \label{eq:dS_4d}
\end{align}
where we recall that the width $\ell$ is obtained by inverting $z_*=z_*(\ell)$. The expression (\ref{eq:dS_4d}) is particularly nice since all integrals cancel each other out and the right-hand side reduces to a simple function of the metric coefficients. If we were to reconstruct the bulk geometry using lattice measurements of entanglement entropy in (3+1) dimensions, we could use the formula~\eqref{eq:dS_4d} very effectively. Notice further that when considering the infinite width limit $\ell\to\infty$, the tip of the RT surface approaches $z_*\to z_h$, and the only contribution to the finite piece stems from (\ref{eq:dS_4d}), which gives the thermal entropy
\be
 S_{\rm{th}} = \frac{V R^3\ell}{4G_{\Nc}^{(5)}z_h^3} = \frac{\pi^2\Nc^2}{2}V \ell\,T^3\ , 
\ee
where in the latter expression we again wrote the result for ${\cal{N}}=4$ $\SU{\Nc}$ super Yang-Mills theory on $\mathbb{R}^4$ at temperature $T$.

Having reviewed a few key holographic ideas we will next introduce the $(2+1)$-dimensional model we utilize. We use the following bulk geometry in the string frame:
\begin{align}
    \dd s^2 &= \left( \frac{r_p}{z} \right)^\frac{5}{2} \left( - \frac{b(z)}{a(z)^2} \dd t^2 + \dd \vec x^2 \right) + \left(\frac{z}{r_p}\right)^\frac{3}{2} \frac{a(z)^2}{b(z)} \dd z^2 + r_p^\frac{3}{2} \sqrt z \dd \Omega_6^2 \label{eq:3dmetric}\\
    e^\phi &= \left(\frac{z}{z_p}\right)^\frac{5}{4} \ ,
\end{align}
where $z_p$ is a length scale playing the role of the radius of curvature and $e^\phi$ is the dilaton.
Notice that the our ansatz for the dilaton is not crucial; using Weyl rescaling we could aim for reconstructing in the Einstein frame instead. We stick to the string frame to allow for easier comparison with results in the string literature. The blackening factor is $b(z) = 1-(z/z_h)^5$, where $z_h$ is the horizon position. This background for $a(z) = 1$ can be derived from type IIB string theory by considering a stack of D2-branes \cite{Aharony:1999ti} and we assume that the generalized metric with $a(z) \neq 1$ could result in a useful model for finite-temperature lattice results in $(2+1)$ dimensions. Furthermore, we want to keep the metric ansatz simple so we keep the transverse and internal space symmetries, only modifying temporal and holographic directions.
This is a top-down holographic model, so the metric includes a compact internal space; we keep the ten-dimensional Newton constant explicit below. The temperature in this family of geometries, assuming again $a(z_h) = 1$, is
\begin{align}\label{eq:Tzh}
    T = \frac{5}{4\pi} \frac{\sqrt{z_p}}{z_h^{3/2}} \ .
\end{align}
The integrals for the strip width and entanglement entropy can be derived in the same way as in the $(3+1)$-dimensional case. The relevant integrals are
\begin{align}
    T \ell(z_*) &= \frac{5}{2\pi} \frac{1}{z_h^{3/2} z_*^{7/2}} \int_0^{z_*} \frac{a(z)}{\sqrt{b(z)}} \frac{z^4}{\sqrt{1-(z/z_*)^7}} \dd z \label{eq:l_int_3d} \\
    \frac{1}{T V} S_{EE}(z_*) &= \frac{16 \pi^4 z_p^{17/2}}{75 G_{\Nc}^{(10)} \sqrt{z_h}} \Bigg[ \frac{z_h^2}{\epsilon^2} - \frac{z_h^2}{z_*^2} \nonumber \\
    & \hspace{2cm} + 2 z_h^2 \int_0^{z_*} \frac{1}{z^3} \left( \frac{a(z)}{\sqrt{b(z)}} \frac{1}{\sqrt{1-(z/z_*)^7}} - 1 \right) \dd z \Bigg] \label{eq:S_int_3d} \ .
\end{align}
Like before, the derivative of $S_{EE}(\ell)$ is a simple function, free of integral expressions:
\begin{align}
    \frac{4G_{\Nc}^{(4)}}{V} \frac{\dd S_{EE}}{\dd \ell} = \frac{z_p^{7/2}}{ z_*^{7/2}} \ , \label{eq:dS_3d}
\end{align}
where if interested in this relation in terms of ten-dimensional Newton constant one uses ${\rm{Vol}}(S^6)=16\pi^3/15$ for the volume of the internal space.

We can work out the UV limit of $S_{EE}$ by studying \eqref{eq:l_int_3d} and \eqref{eq:S_int_3d} in the limit $z_* \ll z_0$. The result is \cite{vanNiekerk:2011yi},
\begin{align}
    \frac{1}{T V} S_{EE}(\ell) = \frac{16 \pi^4 z_p^{17/2}}{75 G_{\Nc}^{(10)} \sqrt{z_h}} \left[ \frac{z_h^2}{\epsilon^2} - \frac{3 \pi^{7/6}}{7^{1/3} 98} \frac{\Gamma\left(\frac{5}{7}\right)^{7/3}}{\Gamma\left(\frac{17}{14}\right)^{7/3}} \frac{z_h^2}{z_p^{2/3} \ell^{4/3}} \right] \ , \quad T \ell \ll 1 \ .\label{eq:holseeuvlimit}
\end{align}
We note that in order to avoid flowing to the weakly coupled Yang-Mills regime, we need to work with finite UV cutoff $\epsilon$, however, we are only interested in terms depending on $\ell$ in this work.

On the other hand, in the large width limit
\begin{align}
    \frac{1}{T V} S_{EE}(\ell) = \frac{16 \pi^4 z_p^{17/2}}{75 G_{\Nc}^{(10)} \sqrt{z_h}} \left( \frac{z_h^2}{\epsilon^2} + \frac{\sqrt{z_p}}{z_h^{3/2}} \ell + \frac{2 \sqrt{2} \Gamma\left(-\frac{2}{5}\right)}{5 \Gamma\left(\frac{1}{10}\right)} \right) \ , \quad T \ell \gg 1 \ .\label{eq:holseeirlimit}
\end{align}
Here one should also be wary that the strict IR limit is probably not trustworthy at face value, but better captured using 3D superconformal field theory via M2-branes \cite{Itzhaki:1998dd,Ryu:2006ef}. Our focus is on the middle term in the bracket in \eqref{eq:holseeirlimit} (or actually that in (\ref{eq:dS_3d})) that arises solely from black hole horizon and corresponds to the thermal behavior. We believe that it universally describes the thermal entropy of the deconfining phase in the field theory. Indeed, below we show evidence for the leading $S_{EE}\sim T^{7/3}\ell$ (as follows from plugging (\ref{eq:Tzh}) in (\ref{eq:holseeirlimit})) behavior as extracted from lattice computations.

\subsection{Bulk reconstruction}
We now have expressions for the holographic entanglement entropy in both $(3+1)$ and $(2+1)$ dimensions. In this work, however, we only consider the reconstruction in $(2+1)$ dimensions since this is the case for which we have sufficient data from the lattice. The formulas require as input the bulk metric and output entropies. Since we have also computed the derivative of $S_{EE}(\ell)$ with lattice simulations, it is interesting to ask whether it is possible to find such a bulk geometry for which the holographic predictions would agree with the lattice results. Schemes like this where the holographic bulk geometry is inferred from QFT quantities is called bulk reconstruction. We will use statistical methods for reconstructing the bulk from lattice data \cite{Jokela:2020auu}. The lattice data consists of a set of $M$ measurements of $\dd S_{EE}/\dd \ell$ and corresponding uncertainties, $\sigma$, for different widths $\ell$: 
\begin{align}
    \left\{ \left( \frac{1}{T^2 V} \frac{\dd S_{EE}}{\dd \ell} \right)_i \ , \ \ell_i \ , \ \sigma_i \right\} \ , \quad i \in \cof{1, \ldots, M} \ .
\end{align}

The bulk geometry contains only one free function, $a(z)$, which we parametrize as
\begin{align}
    a(z) = 1 + \sum_{i=1}^{N_\text{basis}} a_i \left[ \left(\frac{z}{z_h} \right)^i - \left(\frac{z}{z_h}\right)^{N_\text{basis}+1} \right] \ ,
\end{align}
with $a_i \in \mathbb R$ $\forall i \in \cof{1, \ldots, M}$. The expressivity of the ansatz is controlled by the number of free parameters, $N_\text{basis}$. Furthermore, the ansatz is built so that $a(0) = a(z_h) = 1$ by construction, which keeps the geometry asymptotically conformally $AdS$ and the horizon at a constant temperature. We note that if we were interested in very low temperatures and the confining phases of the theory, a metric ansatz with a collapsing cycle at the IR end of the geometry (a `cigar') would seem more appropriate. 

Our statistical model then has the parameters
\begin{align}
    \left\{ a_{1},a_{2},\ldots, a_{N_\text{basis}}, c \right\} \ ,
\end{align}
where $c\equiv (64 \pi^5 z_p^{17/2})/(375 \sqrt{z_h} G_{\Nc}^{(10)})$ is the multiplicative constant on the right hand side of \eqref{eq:dS_3d}. We will link strip widths $\ell_i$ to entropy derivatives by assuming a likelihood
\begin{align}
    \left( \frac{1}{T^2 V} \frac{\dd S_{EE}}{\dd \ell} \right)_i \Bigg| \left\{ \vec a, c \right\} \sim \mathcal N \left( c \left( \frac{z_h}{z_*(\vec a, \ell_i)} \right)^{7/2}, \sigma_i \right) \ , \label{eq:likelihood}
\end{align}
with $\mathcal{N}\of{\mu,\sigma}$ being the normal distribution with mean $\mu$ and standard deviation $\sigma$.
We will use weakly informative normal priors for $\vec a$ and $c$ with standard deviation 5 and 1, respectively, around their maximum likelihood estimates.
The priors represent our beliefs about the bulk geometries that might reproduce the experimental results obtained from the lattice simulations: the standard deviations are wide so that we prefer values closer to values which maximize \eqref{eq:likelihood} but still want the posterior parameter estimates to be primarily informed by experimental data instead of our priors. We combine the priors with the information from our experimental data by bringing in their likelihood \eqref{eq:likelihood} to form a posterior distribution for the model parameters. The posterior distribution is of the form
\begin{align}
    & p\left( \vec{a}, c \Bigg| \left\{ \left( \frac{1}{T^2 V}\frac{\dd S_{EE}}{\dd \ell}\right)_i, \ell_i, \sigma_i \right\} \right) \nonumber \\
    & \propto \prod_{i=1}^M \frac{1}{\sigma_i} \exp\left\{-\frac{1}{2\sigma_i^2}\left(\left(\frac{1}{T^2 V} \frac{\dd S_{EE}}{\dd \ell}\right)_i - c \left(\frac{z_h}{z_*(\vec{a}, \ell_i)}\right)^{7/2}\right)^2\right\} \nonumber \\
    & \times \exp\left\{ -\frac{1}{2 \cdot 1^2} \left(c - c_{MLE}\right)^2 \right\} \times \prod_{i=1}^{N_\text{basis}} \exp\left\{ - \frac{1}{2 \cdot 5^2} \left(a_i - a_{i,MLE}\right)^2 \right\} \label{eq:posterior} \ .
\end{align}
Finally, we sample this posterior distribution of $\vec a$ and $c$ with Hamiltonian Monte Carlo (HMC) which is a type of Markov chain Monte Carlo method. More specifically we employ a HMC variant called No U-Turn Sampler which reduces the number of sampling related parameters that need hand-tuning~\cite{JSSv076i01}, making it a good method for sampling high-dimensional distributions like \eqref{eq:posterior}. Furthermore, this sampling approach reduces the autocorrelation between samples which reduces the number of samples needed to obtain a good estimate of the posterior distribution. It is important to note that for a given strip width $\ell_i$ there might be multiple turning points $z_*$ corresponding to it. If this is the case, we select the turning point corresponding to the true global minimum of entanglement entropy in accordance with the Ryu-Takayanagi prescription~\cite{Ryu:2006bv}.

The posterior parameter distributions can then be used to find the distribution of bulk metrics consistent with the lattice entanglement entropy data.
In a sense this is like using the AdS/CFT duality ``in reverse''. Then, having access to the bulk gravity model, we could use the AdS/CFT dictionary in the usual direction to derive other quantities of interest, for example, quark-antiquark potentials.

\section{Entanglement and thermal entropy}\label{sec:eeandterelation}

In this section we would like to show that
\[
S_{EE}\of{\ell,T} \approx S_{\mathrm{th},A\of{\ell}}\of{T}\qquad\text{if}\quad \ell\,T \gg 1\ ,\label{eq:eevsterel}
\]
which tell us that at sufficiently large width $\ell$ of the entangling region $A$, the entanglement entropy, $S_{EE}\of{\ell,T}$, of $A$ reduces to the thermal entropy, $S_{\mathrm{th},A}\of{T}$, in $A$. Note that $S_{\mathrm{th},A}\of{T}$ is an extensive quantity and grows therefore linearly with the volume of $A\of{\ell}$, i.e. linearly with $\ell$. A more straightforward way to write this is, to show that the derivative of $S_{EE}$ with respect to $\ell$, becomes $\ell$-independent for large $\ell$:
\[
\lim_{\ell\gg T^{-1}}\frac{1}{\abs{A_{\perp \ell}}}\partd{S_{EE}\of{\ell,T}}{\ell} = s_{\mathrm{th},A}\of{T}\ ,\label{eq:eevstereldens}
\]
where $\abs{A_{\perp \ell}}$ is the $\ell$-independent area of a cross section of the entangling region $A$ perpendicular to $\ell$, and $s_{\mathrm{th},A}$ is the thermal entropy density for region $A$.

As mentioned in the introduction, in terms of density matrices, one can see that the relation in Eq.~\eqref{eq:eevsterel} should hold, by noting that the entanglement entropy in region $A$ is defined by:
\[
S_{EE}\of{A,T} = -\trace_{A}\of{\rho_{A}\of{T}\,\log\of{\rho_{A}\of{T}}}\ ,
\]
where $\rho_{A}\of{T}=\trace_{B}\of{\rho\of{T}}$ is the reduced density matrix for region $A$, obtained by taking the partial trace of the full density matrix, $\rho\of{T}$, with respect to a basis for the complement, $B=A^{c}$, of region $A$. In the limiting case, where $A$ represents the whole system, and $B$ is empty, one has $\rho_{A}\of{T}\to \rho\of{T}$ and therefore
\[
\lim_{B\to\emptyset} S_{EE}\of{A,T} = -\trace\of{\rho\of{T}\,\log\of{\rho\of{T}}} = S_{\mathrm{th},A}\of{T}\ ,
\]
i.e. the expression for the entanglement entropy reduces to the expression for the thermal entropy if the region $A$ is entangled with nothing. This makes it plausible that Eq.~\eqref{eq:eevsterel} holds as soon as $A$ is much bigger than the thermal screening radius, $r_{T}\sim T^{-1}$, as then most of $A$ will not ``feel'' that $B$ has been traced out; only the part of $A$ that is within a distance $r_{T}$ from the boundary to $B$ can experience entanglement with region $B$. 

In the following, we will provide an additional argument for why Eq.~\eqref{eq:eevsterel} resp. Eq.~\eqref{eq:eevstereldens} should hold, based on a comparison of the replica trick expression for the entanglement entropy and the expression for the thermal entropy in terms of a Euclidean path integral. In order to have well-defined path integral expressions, we will consider directly lattice regularized theories. However, the provided relations should apply also when using other valid regularization schemes.      

\subsection{Thermodynamics in terms of a Euclidean (lattice) field theory partition function}\label{ssec:thermodynamics}

Let us define the \emph{lattice free energy}, $F_L\of{N_t,V,N}$ as
\[
F_L\of{N_t,V,N}=-\log\of{Z\of{N_t,V,N}}\ ,
\]
where $Z\of{N_t,V,N}$ is a canonical, Euclidean lattice partition function, depending on the temporal lattice size, $a\,N_t$, the spatial lattice volume, $a^3\,V$, and a conserved charge $N$. We keep the dependency on the lattice spacing, $a$, implicit as we are interested in the properties of a lattice system at fixed $a$. If the theory under consideration does not contain any conserved charges, one can drop the dependency on $N$; similarly, if there are multiple conserved charges, one will have to extend the formalism to an appropriate set of charges, $\cof{N_i}_{i=1,2\ldots}$.

The pure gauge system we study in this paper does not contain any conserved charges in which are interested, but we keep the $N$-dependency for the moment, in order to emphasize the relation between $F_L$ and the usual Helmholtz free energy $F\of{T,V,N}$, which is given by
\[
 F_L\of{N_t,V,N}=N_t\,F\of{T\of{N_t},V,N}\ .\label{eq:fldef}
\]

From Eq.~\eqref{eq:fldef} it follows that the differential of $F_L$ in therms of the usual thermodynamic quantities, $T=1/N_t$ (temperature), $p$ (pressure), $\mu$ (chemical potential), $S$ (entropy), $V$ (spatial volume) and $N$ (conserved charge number), is given by:
\[
\dd F_L = F\,\dd N_t + N_t\,\dd F = U\,\dd N_t - p\,N_t\,\dd V + \mu\,N_t\,\dd N\ ,\label{eq:fldiff}
\]
where after the second equality sign we have used that the internal energy $U$ is related to $F$ by
\[
U=F+\underbrace{T\,S}_{\frac{S}{N_t}}=F+\frac{S}{N_t}\ ,\label{eq:ufsrel}
\]
and that
\[
\dd F=\underbrace{-S\,\dd T}_{\frac{S}{N_t^2}\,\dd N_t} - p\,\dd V + \mu\,\dd N = \frac{S}{N_t^2}\,\dd N_t - p\,\dd V + \mu\,\dd N\ .
\]
In order to express the thermal entropy $S$ in terms of $F_L$, we can use Eqn.~\eqref{eq:ufsrel} and \eqref{eq:fldiff} to write:
\[
S=N_t\,\of{U-F}=N_t\,U-F_L=N_t\,\partdf{F_L}{N_t}{V,N} - F_L\ .\label{eq:entropylat}
\]

\subsection{Relation between entanglement and thermal entropy}\label{ssec:eeandthermalentropy}

\begin{figure}[htbp]
\centering
\includegraphics[width=0.6\linewidth,keepaspectratio]{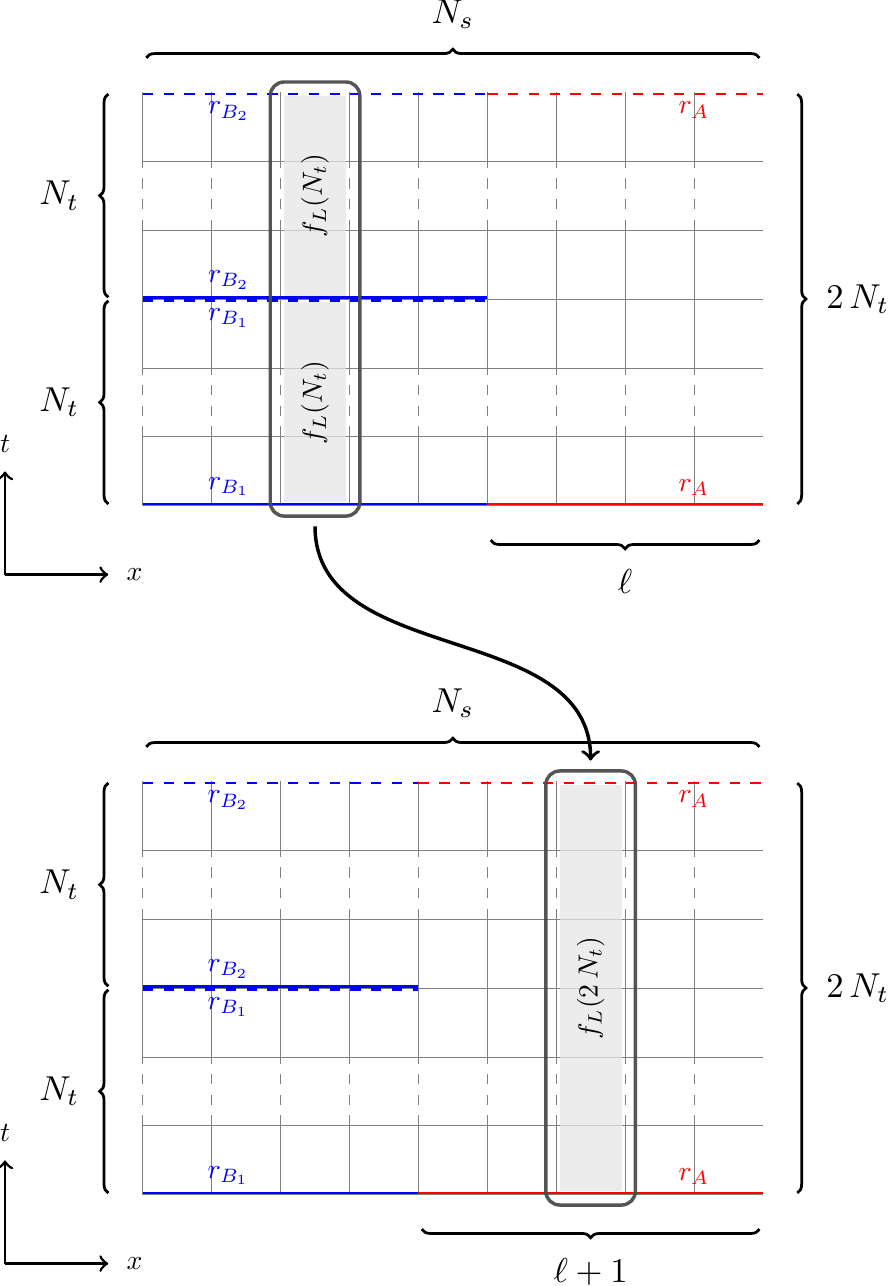}
\caption{The figure illustrates how the width $\ell$ of region $A$ can be increased by removing a slice (which is parallel to the interface between regions $A$ and $B$) from deep within region $B$, changing the temporal boundary conditions of the slice from those of region $B$ to those of region $A$ (cf. Fig.~\ref{fig:bdcond}), and then re-inserting the slice to the system deep inside region $A$.}
\label{fig:eeterel}
\end{figure}

To demonstrate the relation from Eq.~\eqref{eq:eevsterel}, we note that in order to determine Eq.~\eqref{eq:eentropylatderiv}, one does not necessarily have to implement the change $\ell\to\of{\ell+1}$ in the width $\ell$ of the region $A$ by changing the temporal boundary conditions as described in Fig.~\ref{fig:bdcond} along the current boundary of region $A$; one could just as well increase $\ell$ by removing a slice of width $\Delta \ell$ from deep inside region $B$, change the temporal boundary conditions of that slice, and insert it again somewhere in the interior of region $A$, as illustrated in Fig.~\ref{fig:eeterel}. 

Let us now assume that the width $\ell$ of region $A$ and the width $\of{N_s-\ell}$ of region $B$ are both sufficiently large, so that the effect of the interfaces between the two regions, located at $x=0$ and $x=N_s-\ell$, can be neglected for describing the thermodynamic properties of the system deep inside of either of the two regions. Similarly we assume that the change in the volumes of the two regions, $A$ and $B$, when $\ell\to\of{\ell+1}$ can be neglected. 
In this case, deep inside region $B$, the system consists of two copies of a system that can be described by the lattice free energy density
\[
f_L\of{N_t}\ ,
\]
whereas deep inside region $A$, the system consists of just one copy of a system that is described by a lattice free energy density
\[
f_L\of{2\,N_t}\ .
\]
In terms of these lattice free energy densities, the derivative of the entanglement entropy from Eq.~\eqref{eq:eentropylatderiv} is given by
\[
\frac{1}{N_s^{\of{d-2}}}\partdf{S_{EE}\of{\ell',N_t,N_s}}{\ell'}{\ell'=\ell+1/2}\,\approx\,f_L\of{2\,N_t}-2\,f_L\of{N_t}\ ,\label{eq:eentropylatderivfed}
\]
where $N_s^{\of{d-2}}$ is the spatial area (in lattice units) perpendicular to the $x$-direction, so that $N_s^{\of{d-2}}\,\Delta\ell$ with $\Delta\ell=1$ is the spatial volume of the slice that is moved from region $B$ to region $A$. 

Now, as discussed around Eq.~\eqref{eq:seeapproxbyh2}, the lattice expression for $S_{EE}$ actually corresponds to the second order R{\'e}nyi entropy, defined in Eq.~\eqref{eq:renyientropyords}. This is a consequence of approximating in Eq.~\eqref{eq:eentropylat} the derivative with respect to the replica number, $s$, in the limit $\of{s\to 1}$ by a discrete forward derivative: 
\begin{multline}
S_{EE}\of{\ell,N_t,N_s}=-\lim_{s\to 1} \partd{\log\trace\of{\rho_A^s}}{s}=\lim_{s\to 1}\partd{F_c\of{\ell,s,N_t,N_s}}{s}-F\of{N_t,N_s}\\
\approx F_c\of{\ell,2,N_t,N_s}-2\,F\of{N_t,N_s}\ .\label{eq:eentropylatrev}
\end{multline}
The expression on the right-hand side of Eq.~\eqref{eq:eentropylatderivfed} is therefore a consequence of this forward derivative approximation in \eqref{eq:eentropylatrev}. If one tries to reverse-engineer what the right-hand side of Eq.~\eqref{eq:eentropylatderivfed} would look like if the discrete derivative approximation in Eq.~\eqref{eq:eentropylatrev} were not necessary, one quickly finds as possible solution that:
\[
f_L\of{2\,N_t}-2\,f_L\of{N_t}\approx \partdf{f_L\of{s\,N_t}}{s}{s=1}-f_L\of{N_t}=N_t\,\partd{f_L\of{N_t}}{N_t}-f_L\of{N_t}\ ,
\]
in which case one would have:
\[
\frac{1}{N_s^{\of{d-2}}}\partdf{S_{EE}\of{\ell',N_t,N_s}}{\ell'}{\ell'=\ell+1/2}\,=\,N_t\,\partd{f_L\of{N_t}}{N_t}-f_L\of{N_t}=s_{\mathrm{th},A}\ ,\label{eq:eentropylatderivfed2}
\]
with $s_{\mathrm{th},A}$ being the density of thermal entropy, $S_{\mathrm{th},A}$, of region $A$, as follows from Eq.~\eqref{eq:entropylat}.

With Eq.~\eqref{eq:eentropylatderivfed2}, one could therefore compute the thermal entropy of a quantum field theory directly from a single lattice Monte Carlo simulation, using the method for determining the derivative of the entanglement entropy, described in Sec.~\ref{ssec:imprlatEEmethod}. This is not possible by using only Eq.~\eqref{eq:entropylat}, as this would involve knowledge of the lattice free energy, $F_L$, itself. Instead one would have to measure for example $\partial_{T} S\of{T}$, where $T=1/N_t$, for multiple values of $T$, that will allow one to numerically approximate the integral in
\[
S = S_0+\int\limits_{0}^{T}\dd T'\,\partd{S\of{T'}}{T'}\ ,
\]
and where the integration constant, $S_0=S\of{T=0}=0$, is fixed by the 3rd law of thermodynamics. This latter procedure can be computationally very expensive, as one has to simulate low temperatures, which means large $N_t$ and therefore large lattices.

The relation Eq.~\eqref{eq:eentropylatderivfed2} might therefore in the future also be useful to study the thermodynamic properties of strongly interacting field theories. 

\section{Results}\label{sec:results}

In this section we will discuss results that we have obtained from running simulations on the lattice as described in Sec.~\ref{ssec:imprlatEEmethod}. A comparison between results obtained with our new method and corresponding literature results for the case of four-dimensional Yang-Mills theory has been briefly exposed in \cite{Rindlisbacher:2022bhe,Jokela:2022fvh}.
In the following we will present our results on the entanglement entropy in (2+1)-dimensional pure $\SU{2}$ gauge theory at high temperatures. Along the way we recall how the holographic approach, as described in Sec.~\ref{sec:holo}, guides us, {\emph{e.g.}}, in picking correct power laws for fitting. 
We provide an example for how the replica trick, used to determine the entanglement entropy, can be utilized to determine also the thermal entropy, resorting to $\lim_{\ell\to\infty}\partial_{\ell}\,S_{EE}\of{\ell}\propto S_{\text{thermal}}$, as discussed in Sec.~\ref{sec:eeandterelation}.
In the final subsection we reconstruct the bulk geometry from the entanglement entropy measurements.

\subsection{Three-dimensional Yang-Mills theory at high temperature}\label{sec:3dYM}

We are interested in lattice results on the behavior of the derivative of the entanglement entropy with respect to the slab-width, $\ell$, at high temperatures ($T>T_c$) and large slab-width ($\ell>T^{-1}$), where $T$ is the temperature and $T_c$ the critical temperature for deconfinement, which serves as physical reference energy scale.

\begin{table}[!h]
\centering
\begin{tabular}{| c | c | c | c | c | c | c | c | c |} 
 \hline
 $\beta_g$ & $N_s$ & $a\,T_c$ & $N_t$ & $T/T_c$ & $\ell_{\text{min}}/a$ & $\ell_{\text{min}}\,T_c$ & $\ell_{\text{max}}/a$ & $\ell_{\text{max}}\,T_c$ \\[2pt] 
 \hline
 16. & 48 & 0.0976593 & 2 & 5.11984 & 2.5 & 0.244148 & 20.5 & 2.00202 \\
  &  &  & 4 & 2.55992 &  &  &  &  \\
  &  &  & 6 & 1.70661 &  &  &  &  \\
  &  &  & 8 & 1.27996 &  &  &  &  \\
 \hline
 24. & 48 & 0.0641104 & 4 & 3.89953 & 2.5 & 0.160276 & 15.5 & 0.99371 \\
  &  &  & 6 & 2.59968 &  &  &  &  \\
  &  &  & 8 & 1.94976 &  &  &  &  \\
 \hline
 32. & 64 & 0.0477628 & 4 & 5.2342 & 2.5 & 0.119407 & 20.5 & 0.979136 \\
  &  &  & 6 & 3.48947 &  &  &  &  \\
  &  &  & 8 & 2.6171 &  &  &  &  \\
  &  &  & 10 & 2.09368 &  &  &  &  \\
 \hline
 48. & 96 & 0.0316115 & 6 & 5.27234 & 2.5 & 0.0790287 & 29.5 & 0.932539 \\
  &  &  & 8 & 3.95426 &  &  &  &  \\
  &  &  & 10 & 3.16341 &  &  &  &  \\
  &  &  & 12 & 2.63617 &  &  &  &  \\
 \hline
 64. & 128 & 0.0236208 & 8 & 5.29194 & 2.5 & 0.0590521 & 40.5 & 0.956643 \\
  &  &  & 12 & 3.52796 &  &  &  &  \\
  &  &  & 16 & 2.64597 &  &  &  &  \\
 \hline
\end{tabular}
\caption{Simulation parameters and corresponding lattice spacing- and $T/T_c$-values for $\SU{2}$ pure gauge theory on a (2+1)-dimensional lattice, as used for the study of the large distance behavior of $\dd S_{EE}/\dd \ell$ at high temperatures. The number of replica is always $s=2$.}
\label{tab:simparamforee}
\end{table}

We have performed simulations at five different values of the lattice spacing, $a$, and various different temperatures. The list of simulated setups is given in Table~\ref{tab:simparamforee}.

\begin{figure}[!h]
\centering
\includegraphics[width=0.55\linewidth]{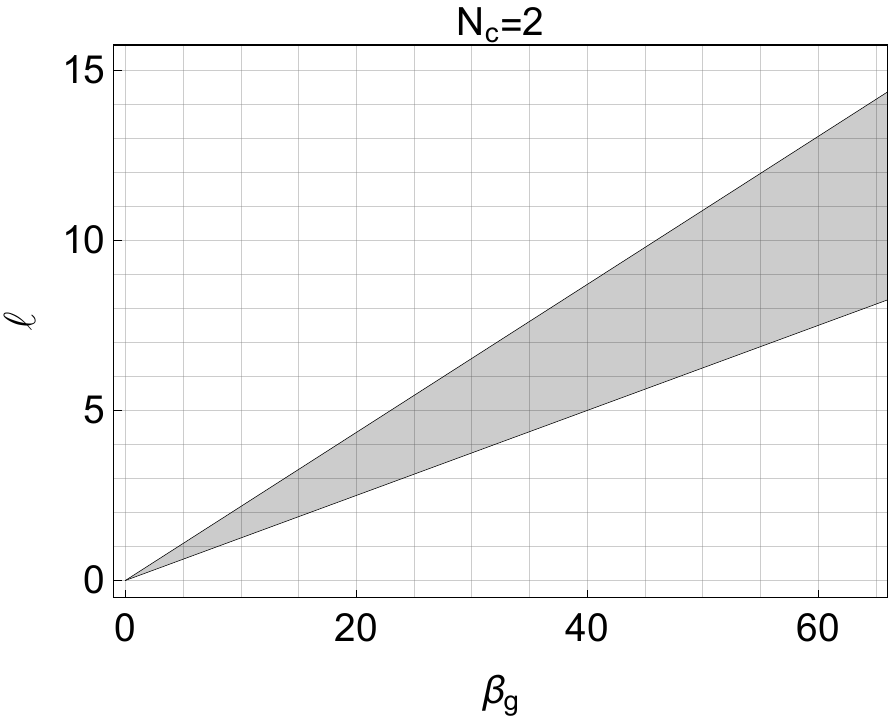}
\caption{The range of validity for the holographic entanglement entropy formula derived in the weak gravity limit as function of the inverse gauge coupling, $\beta_g$ and the slab width, $\ell$ in lattice units, using Eq.~\eqref{eq:holovalidityinequality} for $\Nc=2$.}
\label{fig:validityrange}
\end{figure}

\begin{figure}[h]
\centering
\includegraphics[width=0.8\linewidth]{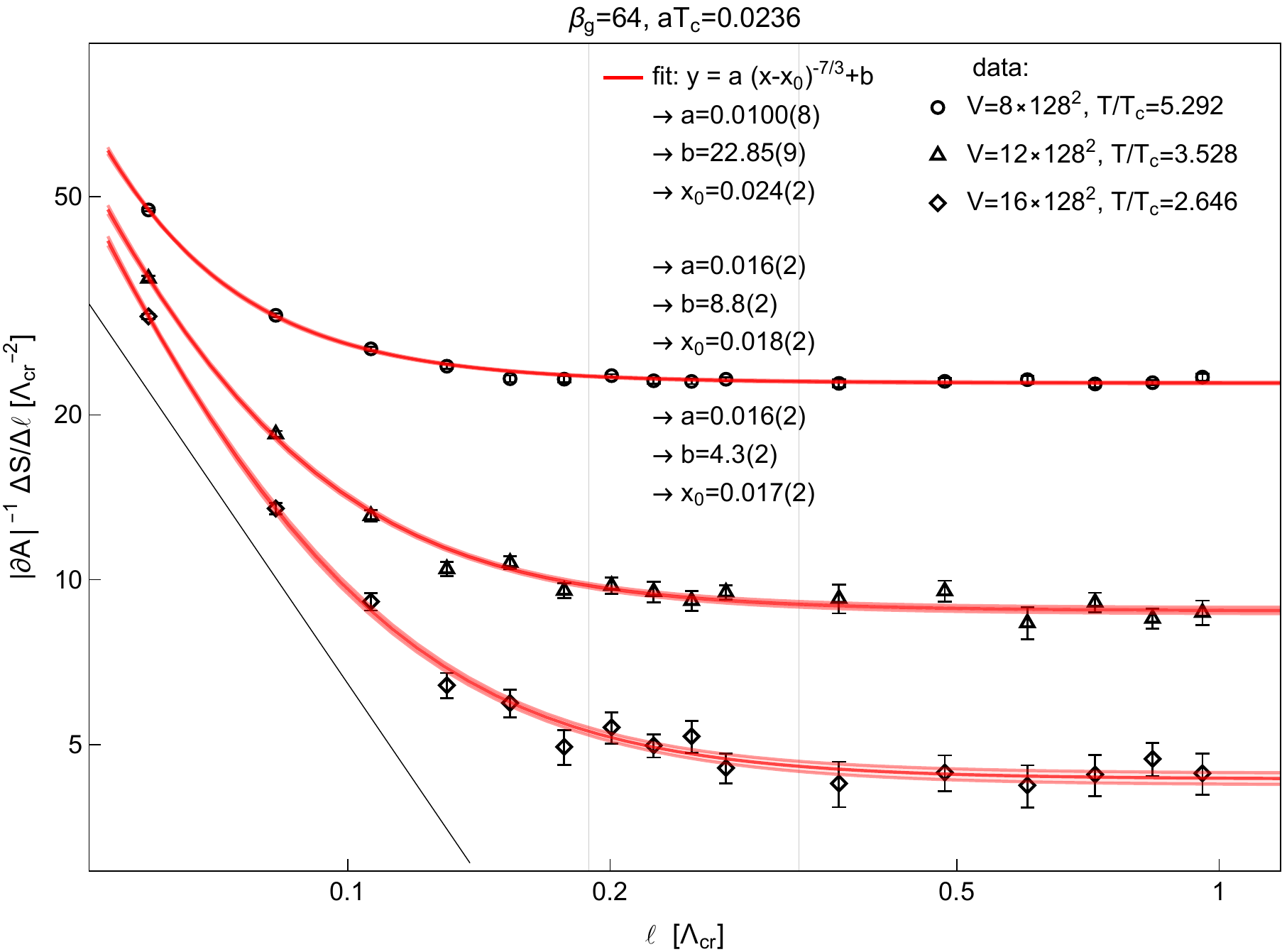}
\caption{Example of fitting the lattice equivalent of $\dd S_{EE}/\dd \ell$ as a function of $\ell$ for $\SU{2}$ in (2+1) dimensions at three different temperatures $T\sim 1/N_t$ with $N_t$=8,12,16 for a spatial lattice size of $V=128^2$. As $T/T_c >1$, the derivative of the entanglement entropy with respect to $\ell$ approaches at large $\ell$ a temperature-dependent plateau-value. The length unit $\Lambda_{cr}$ corresponds to the inverse critical temperature, {\emph{i.e.}}, $\Lambda_{cr}=1/T_c$. The faint vertical lines represent the lower and upper limit of the holography-validity interval from Eq.~\eqref{eq:holovalidityinequality}. Note that the data, corresponding to different $N_t$-values (\emph{i.e.} different temperatures), does not yet lay on a unique curve at short distances, indicating that our lattice data does not resolve a sufficiently high UV-regime at which the temperature would no longer matter. Furthermore, we note that the short distance data lays outside the holography-validity range and does not follow the expected pure power law $\sim x^{-7/3}$ (black line); we had to allow for a shift, $x_0$, in order to fit the data (red lines).}
\label{fig:deedlshortdistcomp3d}
\end{figure}

Note that due to our interest in the high-temperature regime, we are forced to go to rather large values of the inverse gauge coupling, $\beta_g$. As discussed in \cite{Itzhaki:1998dd}, this means that the criteria for the applicability of the classical approximation for the bulk theory in which the holographic computations in Sec.~\ref{sec:holo} are carried out, are satisfied only at relatively large values of the slab-width $\ell$, compared to the lattice spacing $a$. The range of validity as function of $\beta_g$ and $\ell$ in lattice units is sketched in Fig.~\ref{fig:validityrange} for $\Nc=2$, using the inequalities from~\cite{Itzhaki:1998dd}:
\[
\frac{1}{\Nc\,g^2_c}\ll \ell_c \ll \frac{1}{\Nc^{1/5}\,g^2_c}\ ,
\]
where $g^2_c$ is the continuum theory gauge coupling. In terms of lattice quantities, using the relations $\beta_g=2\,\Nc/g^2_0$, $g^2_0=a\,g^2_c$, and $\ell_c=a\,\ell$, the inequalities take the form:
\[
\frac{\beta_g}{2\,\Nc^2} \ll \ell \ll \frac{\beta_g}{2\,\Nc^{6/5}}\ .\label{eq:holovalidityinequality}
\]

Our lattice data for the derivative of the entanglement entropy might therefore at short distances not conform with the power law behavior, expected from the discussion in Sec.~\ref{sec:holo}. 

With this in mind, we will allow for certain corrections to the expected behavior discussed in Sec.~\ref{sec:holo} when performing fits to our lattice data. An example is shown in Fig.~\ref{fig:deedlshortdistcomp3d}. From holography (cf. Eq.~\eqref{eq:holseeuvlimit}), we would expect our lattice data for $\dd S_{EE}/\dd \ell$ as function of $\ell$ to behave in the UV like a power law $\sim \ell^{-7/3}$. As our simulations are carried out at temperatures, $T$, which are 2-5 times the critical temperature $T_c$, this power law behavior is almost completely screened by the plateau within the range of validity for the holographic predictions from Eq.~\eqref{eq:holovalidityinequality} resp. Fig.~\ref{fig:validityrange}. 

However, our focus in this section is not on the $\ell$ dependency of $\dd S_{EE}/\dd \ell$ at small $\ell$, but on its $T/T_c$ dependency at large $\ell$. More precisely, we aim to verify that for $T/T_c\gg 1$ one has
\[
\lim_{\ell\to\infty}\frac{1}{\abs{\partial A}} \partial_{\ell}\,S_{EE}\of{\ell} = s_{\mathrm{th},0}\,\of{T/T_c}^{7/3}\ ,\label{eq:eeplateauthescaling}
\]
which follow from $\lim_{\ell\to\infty}\partial_{\ell}\,S_{EE}\of{\ell}\propto s_{\mathrm{th},A\of{\ell}}$ (cf. Sec.~\ref{sec:eeandterelation}), in combination with the holographic prediction that for a $\of{2+1}$-dimensional $\SU{N}$ gauge theory the thermal entropy behaves like $S_{\mathrm{th}}\propto \of{T/T_c}^{7/3}$ for $T/T_c\gg 1$.

\begin{figure}[!h]
\centering
\includegraphics[width=0.83\linewidth]{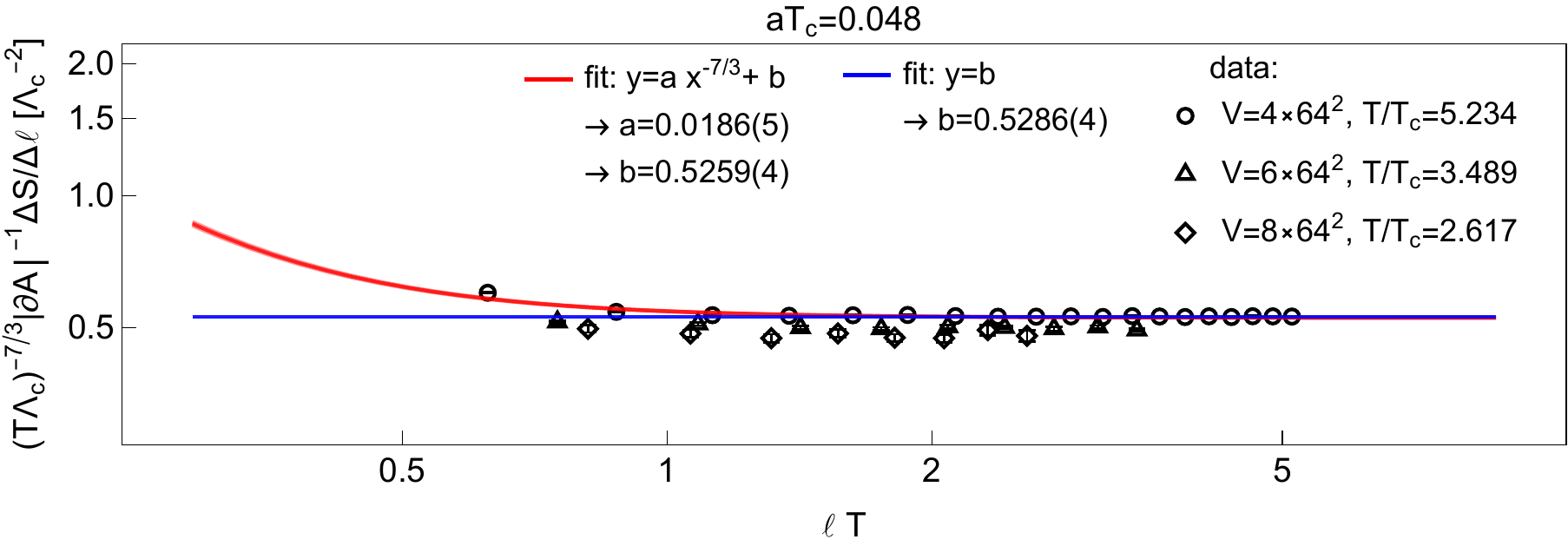}\\[2pt]
\includegraphics[width=0.83\linewidth]{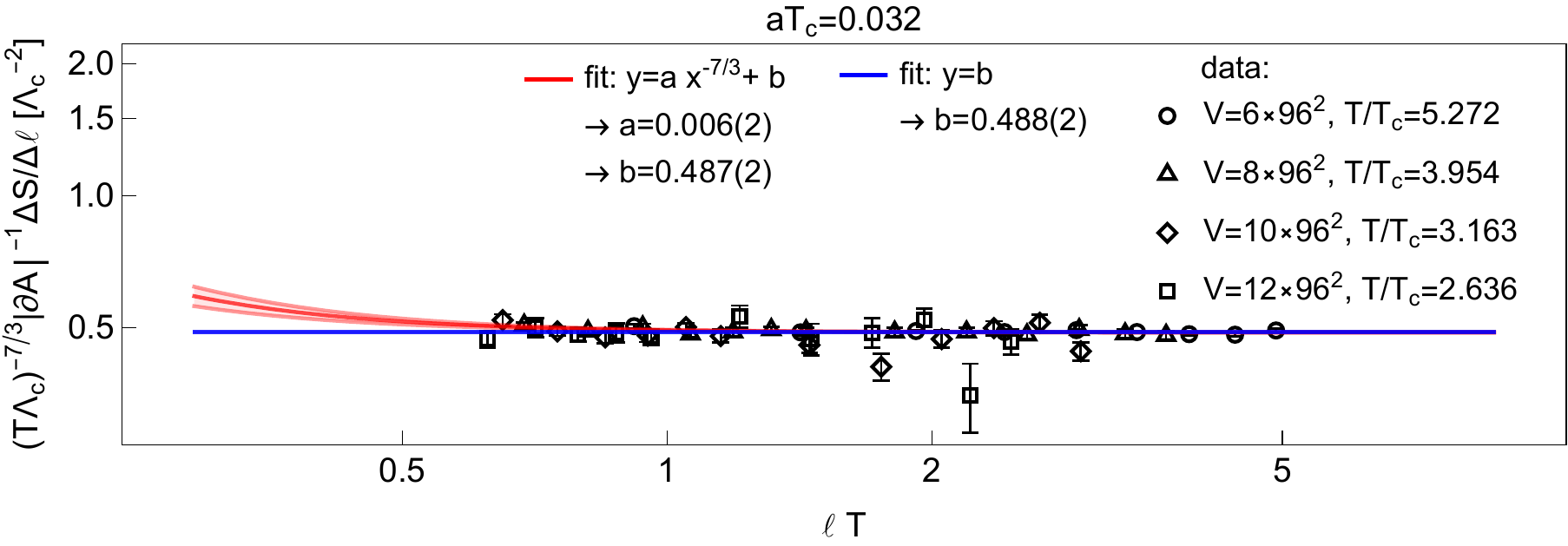}\\[2pt]
\includegraphics[width=0.83\linewidth]{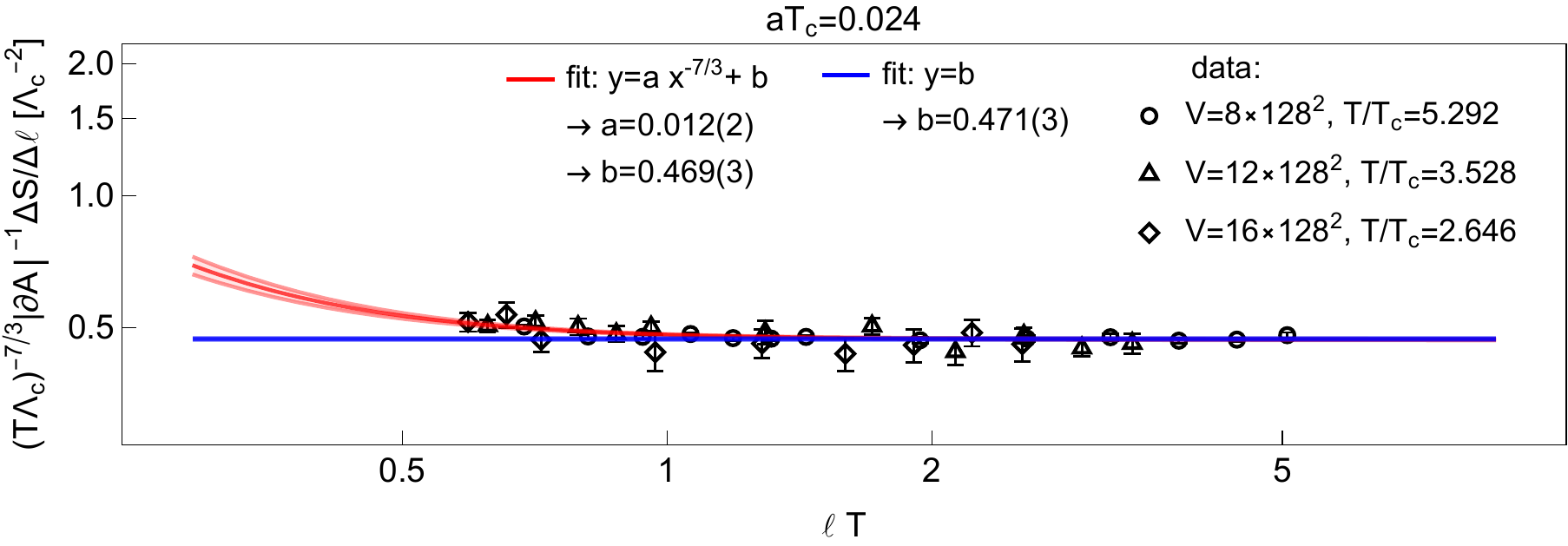}\\[6pt]
\includegraphics[width=0.83\linewidth]{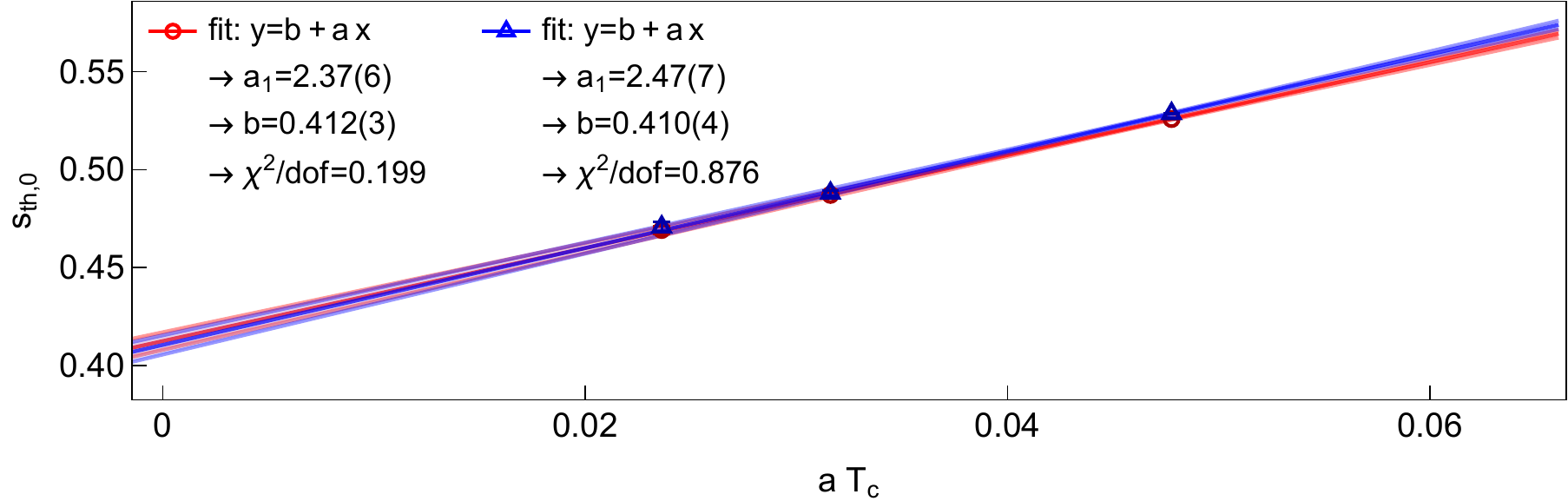}
\caption{The three upper panels show $\dd S_{EE}/\dd \ell$ for $\SU{2}$ in (2+1) dimensions, rescaled by a factor of $\of{T/T_c}^{-7/3}=\of{T\,\Lambda_{cr}}^{-7/3}$, as function of $\ell\,T$ at three different lattice spacing values, corresponding to $\beta=32$ (top), $\beta=48$ (second from top), and $\beta=64$ (third from top). Included is only data for $T/T_c>2.5$ as the scaling $\dd S_{EE}/\dd \ell \propto \of{T/T_c}^{7/3}$ is expected to hold only at large temperatures, away from the critical point. As can be seen, for the two finest lattice spacings, corresponding to $\beta=48,64$, the data collapses nicely. The plateau value is fitted by two different ansatzes, as indicated (red: $y=a\,x^{-7/3}+b$, blue: $y=b$). The bottom panel shows the continuum extrapolation of the plateau values, $b$, obtained in the previous three panels with the two different fit ansatzes in red and blue.}
\label{fig:deedlcollapses}
\end{figure}

As a first test of the scaling law from Eq.~\eqref{eq:eeplateauthescaling}, we show in the first three panels of Fig.~\ref{fig:deedlcollapses} our lattice results for $\partial_{\ell}\,S_{EE}\of{\ell}$, rescaled by $\of{T/T_c}^{-7/3}$ and as function of $\of{\ell\,T}$ for $\beta_g=32$ (top), $48$ (second from top), and $64$ (third from top). The ratio $T/T_c$ as function of $\beta_g$, $N_s$, and $N_t$ is determined as described in Appendix~\ref{ssec:scalesetting}. For each value of $\beta_g$ the plots consist of data from various temperatures. Thanks to the rescaling of $\partial_{\ell}\,S_{EE}\of{\ell}$ by $\of{T/T_c}^{-7/3}$, the plateau data corresponding to different temperatures coincides. This is in particular true for $\beta_g=48,64$, corresponding to lattice spacing values $a\,T_c\approx 0.024$, $0.032$. For $\beta_g=32$, resp. $a\,T_c=0.048$ the data does not collapse as nicely anymore, and for the even lower $\beta_g$-values, resp. larger lattice spacings, it would look even worse. The bottom panel in Fig.~\ref{fig:deedlcollapses} shows a linear extrapolation of the rescaled plateau values, which provides a first rough estimate of the continuum value of the proportionality factor, $s_{\mathrm{th},0}$ in Eq.~\eqref{eq:eeplateauthescaling}.\\ 

In order to be able to use correction terms in the continuum extrapolation of our lattice data for the plateau values $\lim_{\ell\to\infty}\partial_{\ell} S_{EE}\of{\ell}$ and refine the estimate for $s_{\mathrm{th},0}$ obtained in the bottom panel of Fig.~\ref{fig:deedlcollapses}, we extract the plateau values $\lim_{\ell\to\infty}\partial_{\ell} S_{EE}\of{\ell}$ for all simulated values of $\beta_g$ and $T/T_c=T\,\Lambda_c$ separately by fitting the ansatz
\[
y=a\,\of{x-x_0}^{-7/3}+b\ \label{eq:dseedlfitform1}
\]
to our data, where $\of{x,y}\in\cof{\of{\ell/\Lambda_c,\Lambda_c^{2}\,N_s^{-1} \dd S_{EE}/\dd \ell}\,\vert\,\forall \ell}$. The ratio $T/T_c$ is again determined as described in Appendix~\ref{ssec:scalesetting} and $a\,T_c=1/\of{N_t\,\of{T/T_c}}$. The shift $x_0$ in Eq.~\eqref{eq:dseedlfitform1} has been added to account for expected deviations from the pure $\sim\ell^{-7/3}$ power law behavior of $\dd S_{EE}/\dd \ell$, mentioned at the beginning of this section. The form of the chosen correction is not important, as long as it does not significantly affect the fit results for the parameter $b$. Alternatively, one could perform constant fits ($y=b$) on restricted data for which $\ell>T^{-1}$ (cf. Fig.~\ref{fig:deedlcollapses}).

\begin{figure}[h]
\centering
\includegraphics[width=0.8\linewidth]{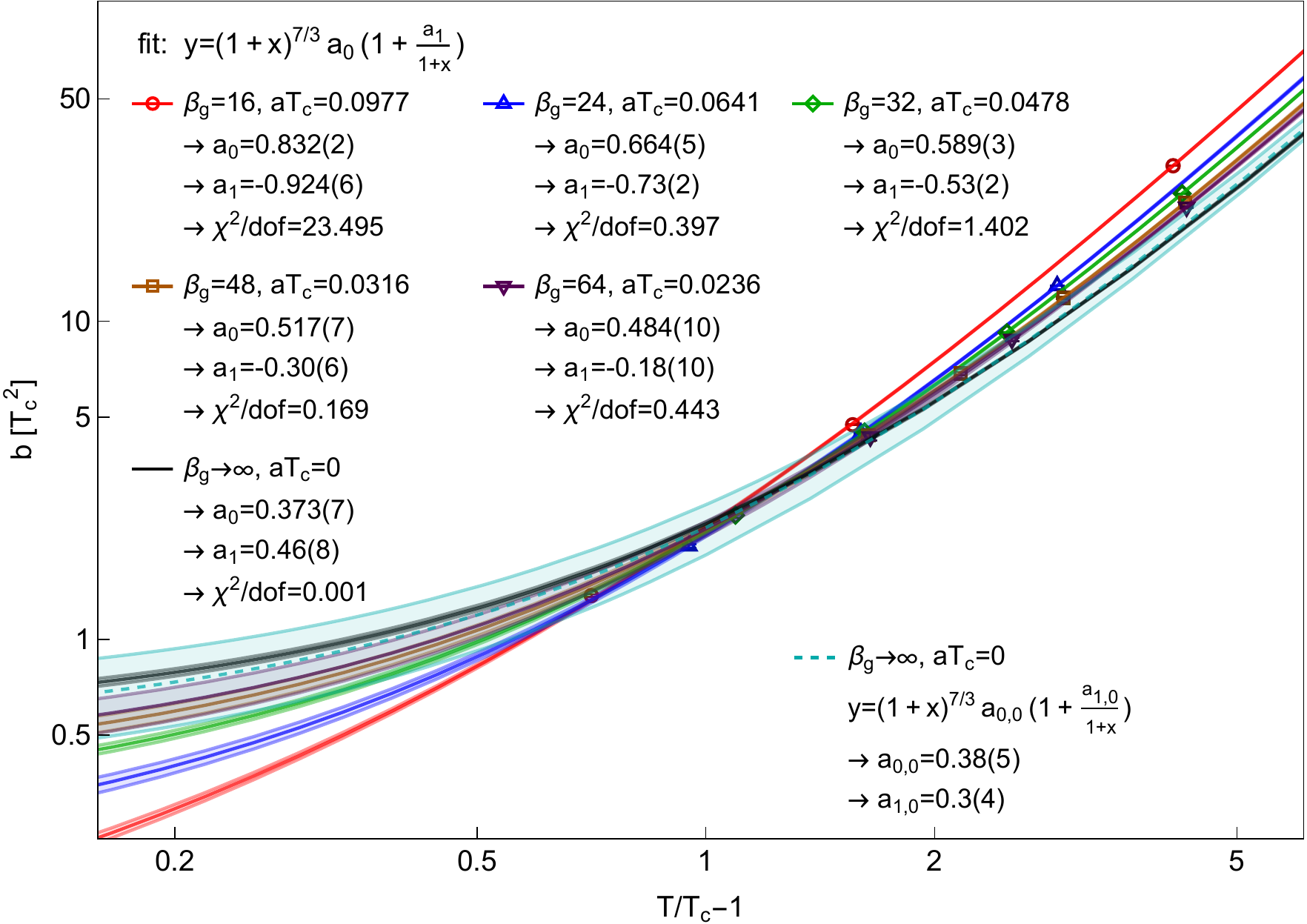}
\caption{Plateau values of $\dd S_{EE}/\dd \ell$ at large $\ell$ for $\SU{2}$ in (2+1) dimensions for various temperatures $T/T_c>1$. The different colors represent data for different lattice spacing values, $a\,T_c$ (in units of the critical temperature, $T_c$), and the corresponding fits to their temperature dependency. The black line shows the continuum extrapolation ($\beta\to \infty$).}
\label{fig:deedlplateaux3d}
\end{figure}

The so obtained plateau values, $b$, are plotted in Fig.~\ref{fig:deedlplateaux3d} as function of the reduced temperature $\tau=T/T_c-1$.  Superimposed to the data is for each value of $\beta_g$ separately a fit of the form
\[
y=a_0\,\of{1+x}^{7/3}\,\of{1+a_1\of{1+x}^{-1}}\ ,\label{eq:bvstaufit1}
\]
with $\of{x,y}\in\cof{\of{\tau,b\of{\tau}}\,\vert\,\forall \tau}$. As before, this fit ansatz is motivated by the identification of the plateau value with the thermal entropy density. The latter should, according to holography, for $T\gg \ell^{-1}$ behave like $s_{\mathrm{thermal}}\propto \of{T/T_c}^{7/3}=\of{1+\tau}^{7/3}$, and we include a leading correction term, whose relative magnitude is parametrized by $a_1$. The fits are performed on data for which $T/T_c>1.9$, as lower temperatures would require higher order corrections. The only exception is the case of $\beta_g=16$, where $T/T_c>1.7$ is used as lower bound in order to have a minimum of three data points to fit the two parameters $a_0$ and $a_1$. However, as can be seen, the quality of fit for $\beta_g=16$ is poor. This is on the one hand due to the extended lower bound of the fit-range which, as mentioned before, would require higher order corrections, and on the other hand due to the fact that the lattice is too coarse at $\beta_g=16$, so that $N_t=2$ for the highest temperature, $T/T_c\approx 5.12$. We therefore exclude the $\beta_g=16$ data from further analysis.

The data from the remaining $\beta_g$-values is used to perform a continuum extrapolation $\of{\beta_g\to\infty}$ in two different ways (black and cyan bands). 

\begin{figure}[h]
\centering
\includegraphics[width=0.83\linewidth]{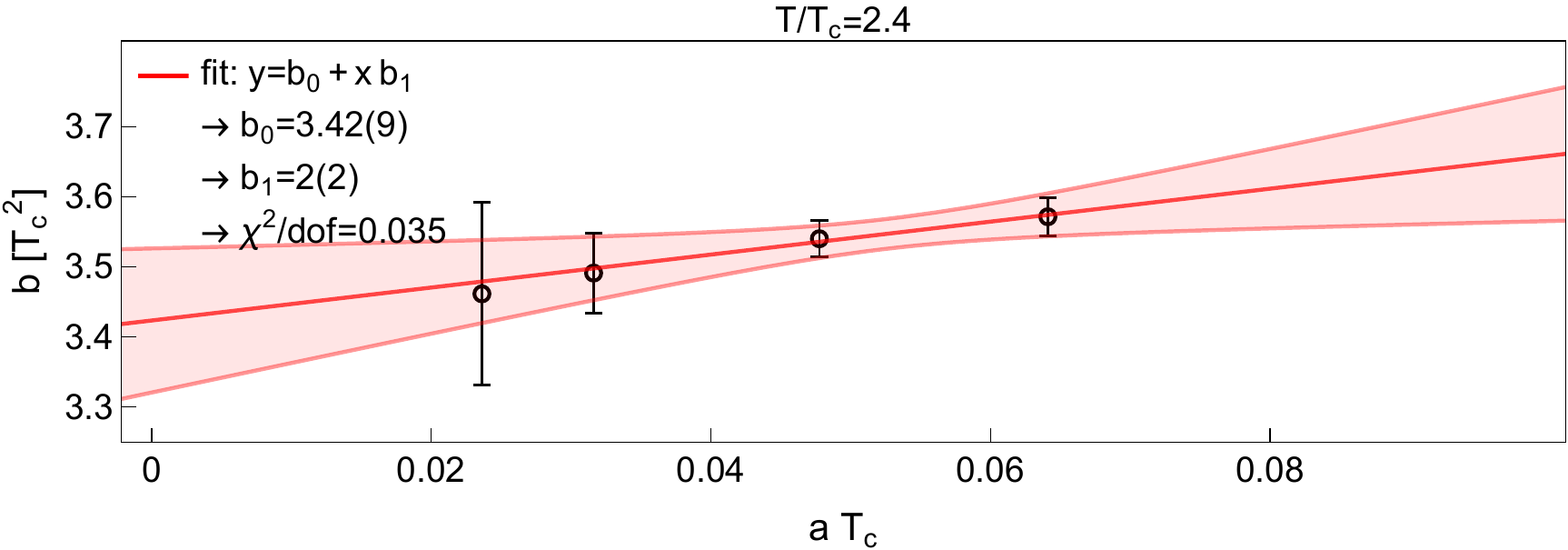}\\[5pt]
\includegraphics[width=0.83\linewidth]{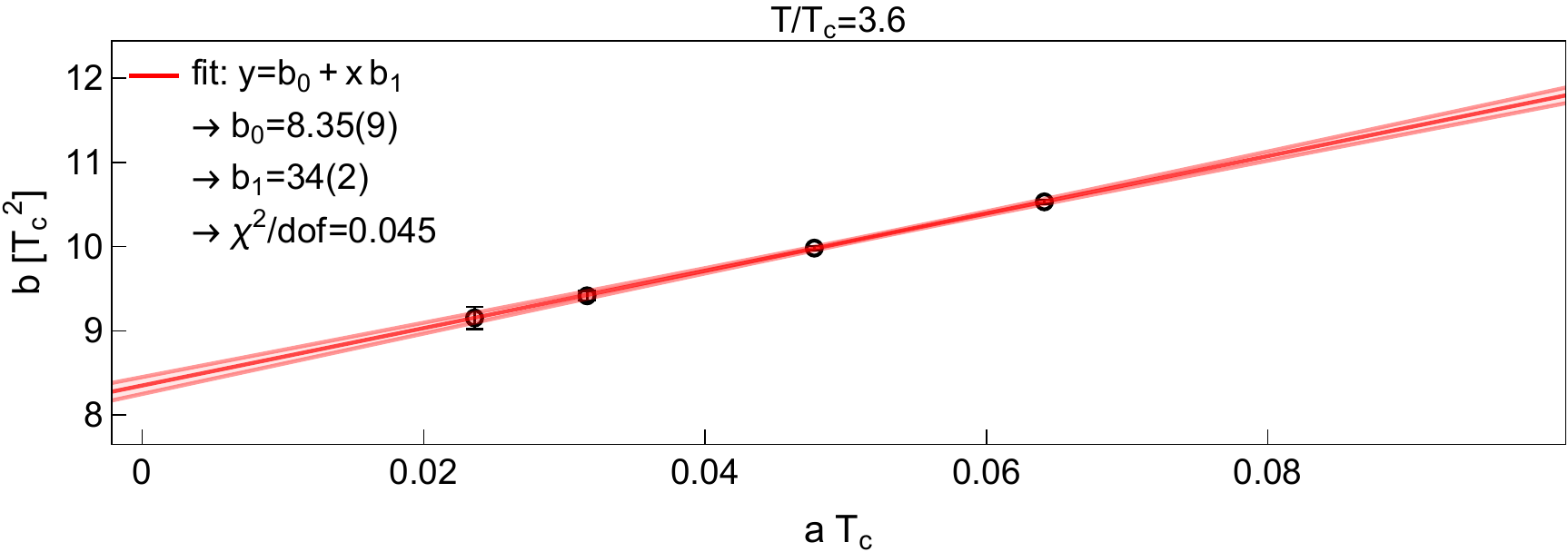}\\[5pt]
\includegraphics[width=0.83\linewidth]{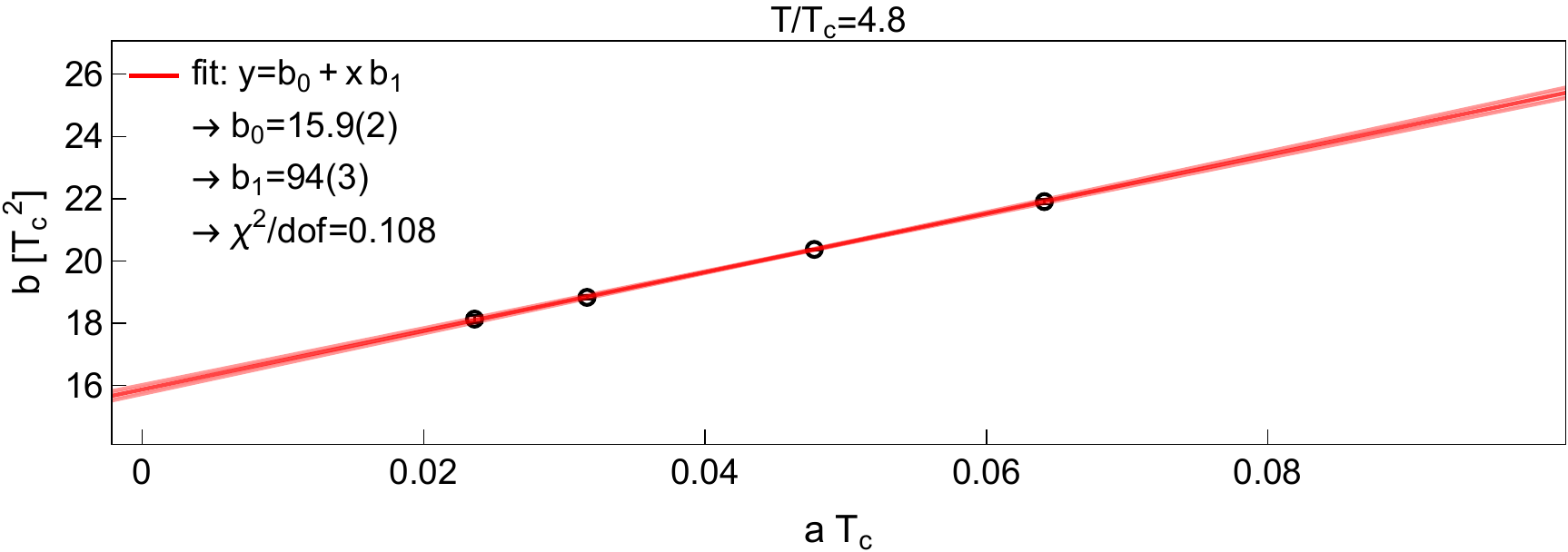}
\caption{Example of fits used to perform continuum extrapolations of the plateau value, $b$, of the entanglement entropy, as determined from the fit results shown in Fig.~\ref{fig:deedlplateaux3d} for $\beta_g=24,32,48,64$.}
\label{fig:bcontinuumexpex}
\end{figure}

The first extrapolation (black) is obtained using the fitted functions and corresponding error bands to obtain values and errors for $b$ at fixed $T/T_c$ but different lattice spacing values, $a\,T_c$, and perform a fit:
\[
y=b_0 + b_1\,x\ ,
\]
with $\of{x,y}\in\cof{\of{a\,T_c,b\of{a\,T,T/T_c}}\,\vert\,\forall\,a\,T_c}$. This is repeated for different values of $T/T_c\in\cof{2.0,2.2,\ldots,5.0}$ while keeping track of the values of $b_0$ as function of $\tau=T/T_c-1$. A few examples of such fits are shown in Fig.~\ref{fig:bcontinuumexpex}. To the so obtained data we then fit again Eq.~\eqref{eq:bvstaufit1} with $\of{x,y}\in\cof{\of{\tau, b_0\of{\tau}}\,\vert\,\forall \tau\in\cof{1.0,1.2,\ldots,4.0}}$, which yields the black curve and corresponding error band from Fig.~\ref{fig:deedlplateaux3d}.

\begin{figure}[h]
\centering
\includegraphics[width=0.83\linewidth,keepaspectratio]{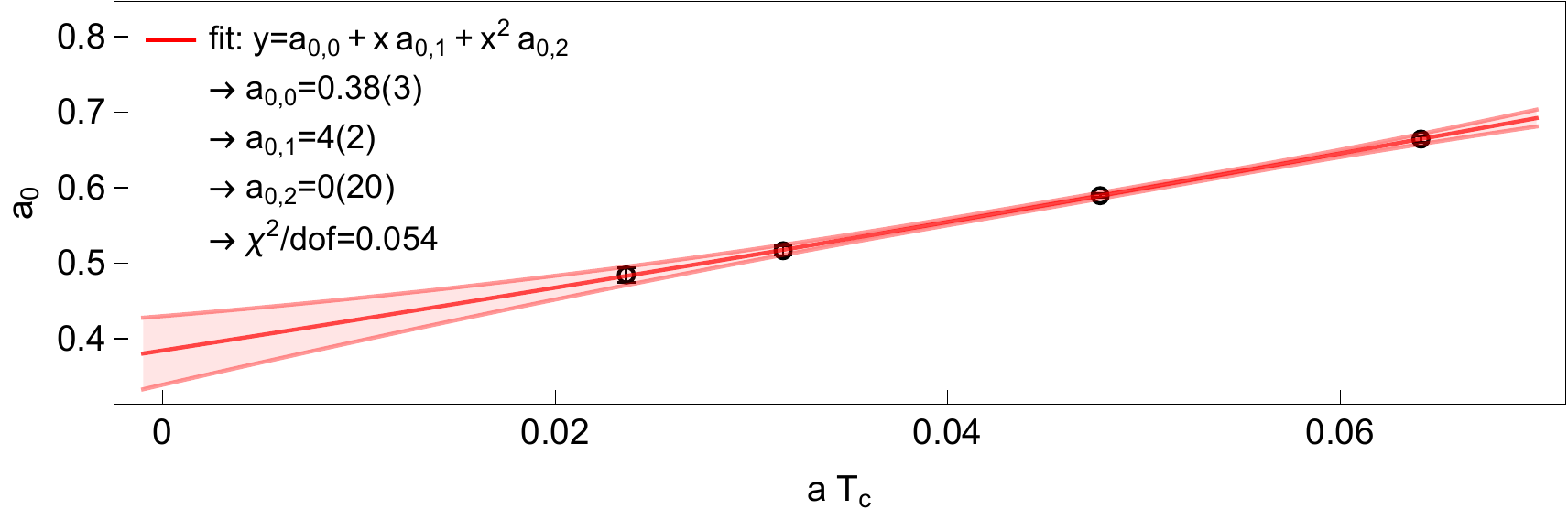}\\[10pt]
\includegraphics[width=0.83\linewidth,keepaspectratio]{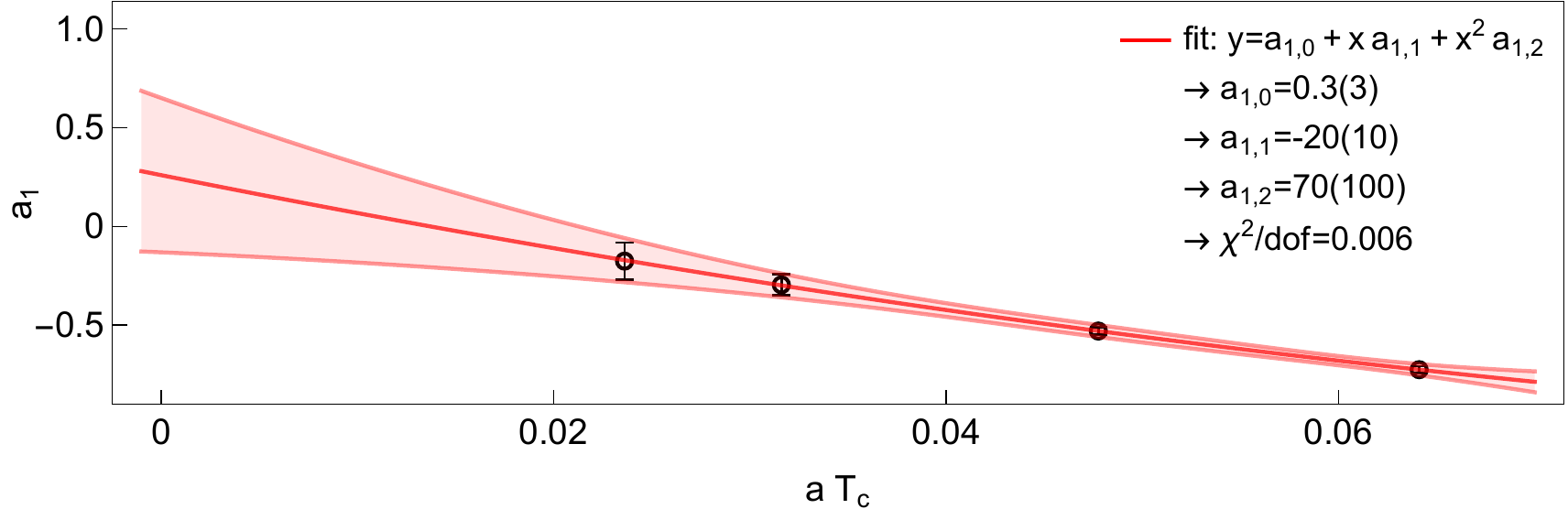}
\caption{Continuum extrapolation of the parameters $a_0$ and $a_1$, obtained from the fits in Fig.~\ref{fig:deedlplateaux3d} for $\beta_g=24,32,48,64$.}
\label{fig:bcontinuumexpex2}
\end{figure}

The second continuum extrapolation in Fig.~\ref{fig:deedlplateaux3d} (cyan) is obtained by considering directly the lattice spacing-dependency of the fit results for the parameters $a_0$ and $a_1$ for $\beta_g=24,32,48,64$. As shown in Fig.~\ref{fig:bcontinuumexpex2}, we perform quadratic fits to the data for $a_0$ and $a_1$ as functions of $a\,T_c$. The so obtained fit results for the constants $a_{0,0}$ and $a_{1,0}$ and their errors are then used to draw the cyan curve and error band in Fig.~\ref{fig:deedlplateaux3d}. As can be seen, the two continuum extrapolations are in good agreement.

\begin{figure}[h]
\centering
\includegraphics[width=0.8\linewidth]{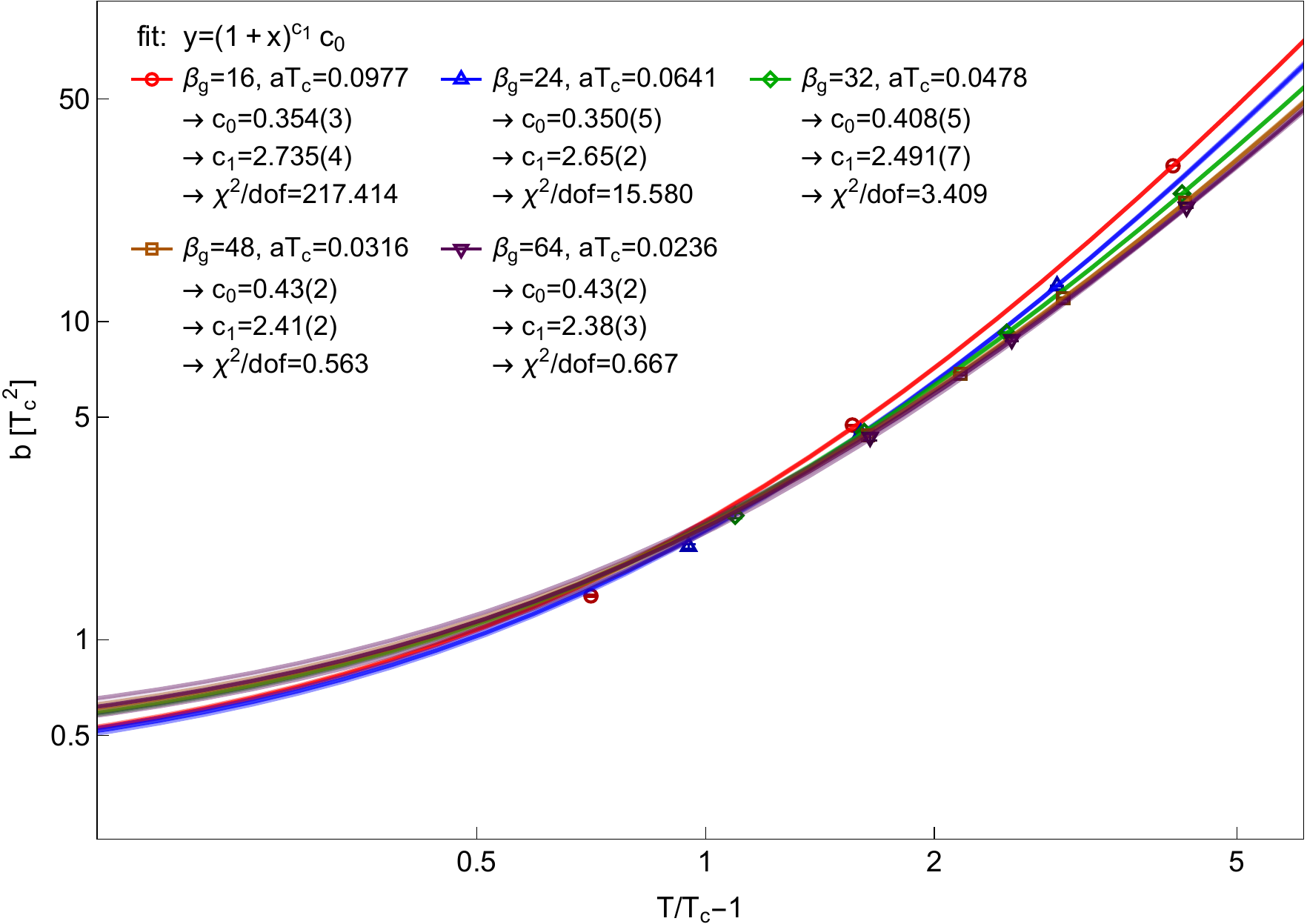}
\caption{Plateau values of $\dd S_{EE}/\dd \ell$ at large $\ell$ for $\SU{2}$ in (2+1) dimensions for various temperatures $T/T_c>1$. The different colors represent data for different lattice spacing values, $a\,T_c$ (in units of the critical temperature, $T_c$), and the corresponding fits to their temperature dependency. The black line shows the continuum extrapolation ($\beta\to \infty$).}
\label{fig:deedlplateaux3d2}
\end{figure}

Finally, let us also perform a quick check of whether in Eq.~\eqref{eq:eeplateauthescaling} a power law different from $\of{T/T_c}^{7/3}$ would be consistent with the data. To this end, we fit to the plateau data from Fig.~\ref{fig:deedlplateaux3d} for each value of $\beta_g$ separately the function form
\[
y=c_0\,\of{1+x}^{c_1}\ ,\label{eq:bvstaufitpl}
\]
where $\of{x,y}\in\cof{\of{\tau,b\of{\tau}}\vert\forall\tau}$ as before in the fits with Eq.~\eqref{eq:bvstaufit1}. As our data sets cover for some $\beta_g$ values of only three different reduced temperature values, $\tau$, we do not include in Eq.~\eqref{eq:bvstaufitpl} the leading corrections from Eq.~\eqref{eq:bvstaufit1} to keep the number of fit parameters at $2$. As can be seen in Fig.~\ref{fig:deedlplateaux3d2}, which summarizes the results of the fits of the form Eq.\eqref{eq:bvstaufitpl} to the data, without the leading correction term, $\sim a_1$, in the fit function the achievable quality of fit is generally worse than in Fig.~\ref{fig:deedlplateaux3d}. But, with increasing $\beta_g$ the fit quality improves in Fig.~\ref{fig:deedlplateaux3d2} as the correction parameter $a_1$ in the corresponding fits in Fig.~\ref{fig:deedlplateaux3d} becomes smaller in magnitude with increasing $\beta_g$. As $\beta_g$ increases, the fitted exponent $c_1$ in Fig.~\ref{fig:deedlplateaux3d2} approaches the expected value $7/3\approx 2.33$ and the value of the amplitude parameter, $c_0$, approaches the value of $a_0$ from Fig.~\ref{fig:deedlplateaux3d}.

\begin{figure}[h]
\centering
\includegraphics[width=0.83\linewidth,keepaspectratio]{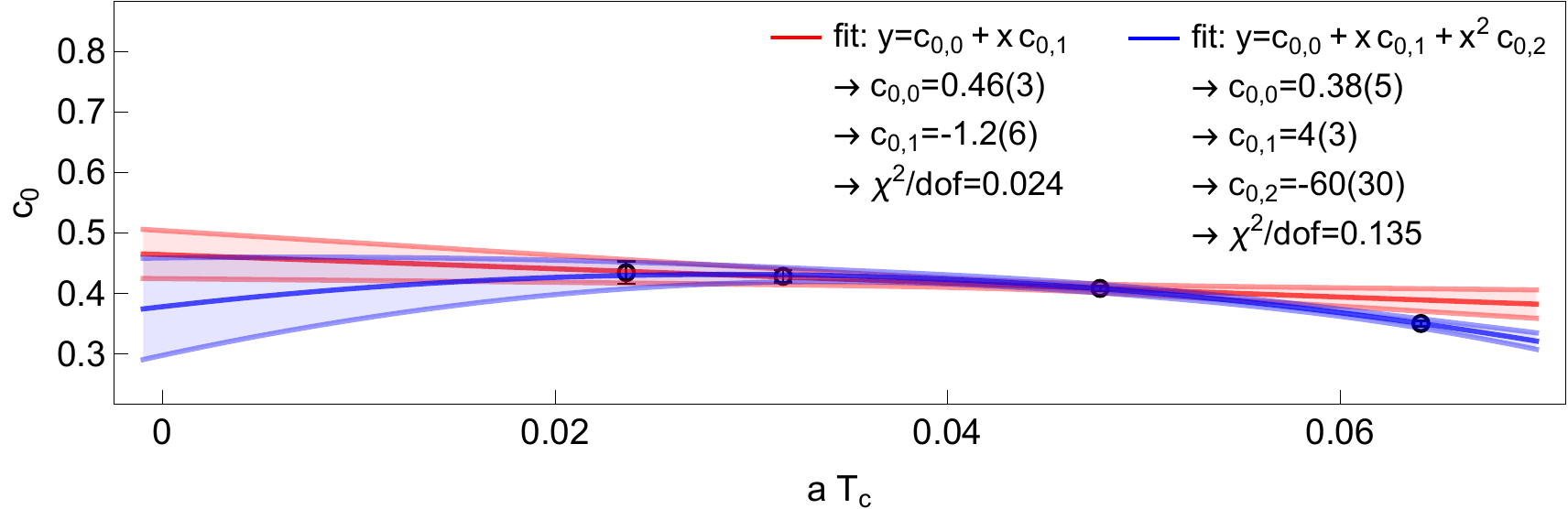}\\[10pt]
\includegraphics[width=0.83\linewidth,keepaspectratio]{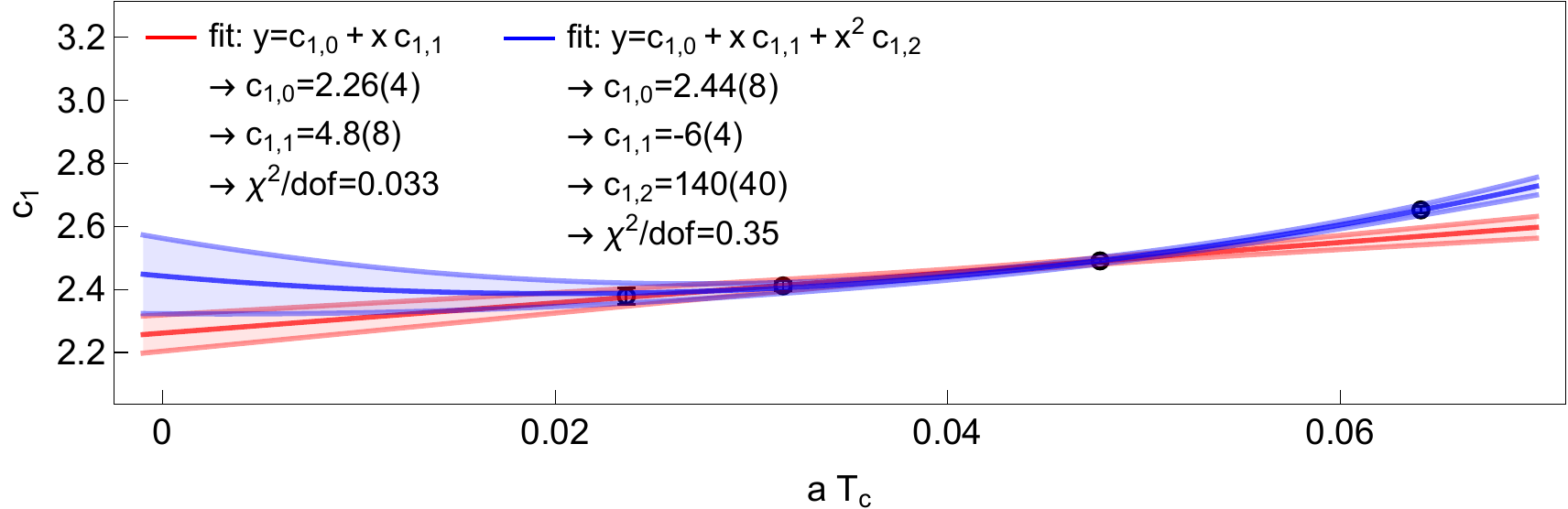}
\caption{Continuum extrapolation of the parameters $c_0$ and $c_1$, base on linear (red) and quadratic (blue) fits to the $c_0$ and $c_1$ parameters shown in Fig.~\ref{fig:deedlplateaux3d2} for the different $\beta_g$ values. The linear fits use the data from $\beta_g=32,48,64$ while the quadratic fits use also $\beta_g=24$.}
\label{fig:bcontinuumexpex3}
\end{figure}

In Fig.~\ref{fig:bcontinuumexpex3} we perform continuum extrapolations for the $c_0$ and $c_1$ parameters, based on the fit results for different $\beta_g$, shown in Fig.~\ref{fig:deedlplateaux3d2}. These extrapolations are analogous to the continuum extrapolations performed in Fig.~\ref{fig:bcontinuumexpex2} for the $a_0$ and $a_1$ parameters for different $\beta_g$ values listed in Fig.~\ref{fig:deedlplateaux3d}. However, as in Fig.~\ref{fig:deedlplateaux3d2}, the quality of fit for the $\beta_g=24$ data is much worse than was the case in Fig.~\ref{fig:deedlplateaux3d}, we show in Fig.~\ref{fig:bcontinuumexpex3} in addition to the extrapolations based on quadratic fits (blue) to the data for $\beta_g=24,32,48,64$ also extrapolations based on linear fits to the data for $\beta_g=32,48,64$. Both extrapolation methods yield within errorbars consistent continuum values for $c_0$ and $c_1$ and the results for $c_1$ are consistent with $7/3\approx 2.33$. 

\FloatBarrier
\subsection{Reconstructed bulk geometries}\label{sec:bulkreconstruction}

In Section \ref{sec:holo} we have discussed our statistical model which we use to connect the metric tensor of the holographic dual to entanglement entropy values. We will now use data of the derivative of entanglement entropy with respect to the slab width obtained from the lattice to find which metric fits these observations. More specifically, the data is for $\SU{2}$ lattice gauge theory on a $V = 128^2$ lattice at temperature $T/T_c=2.646$. Since the gauge theory is in $(2+1)$ dimensions, the appropriate holographic background is the asymptotically D2-brane geometry.

\begin{figure}[htbp]
    \centering
    \includegraphics[width=1.0\linewidth]{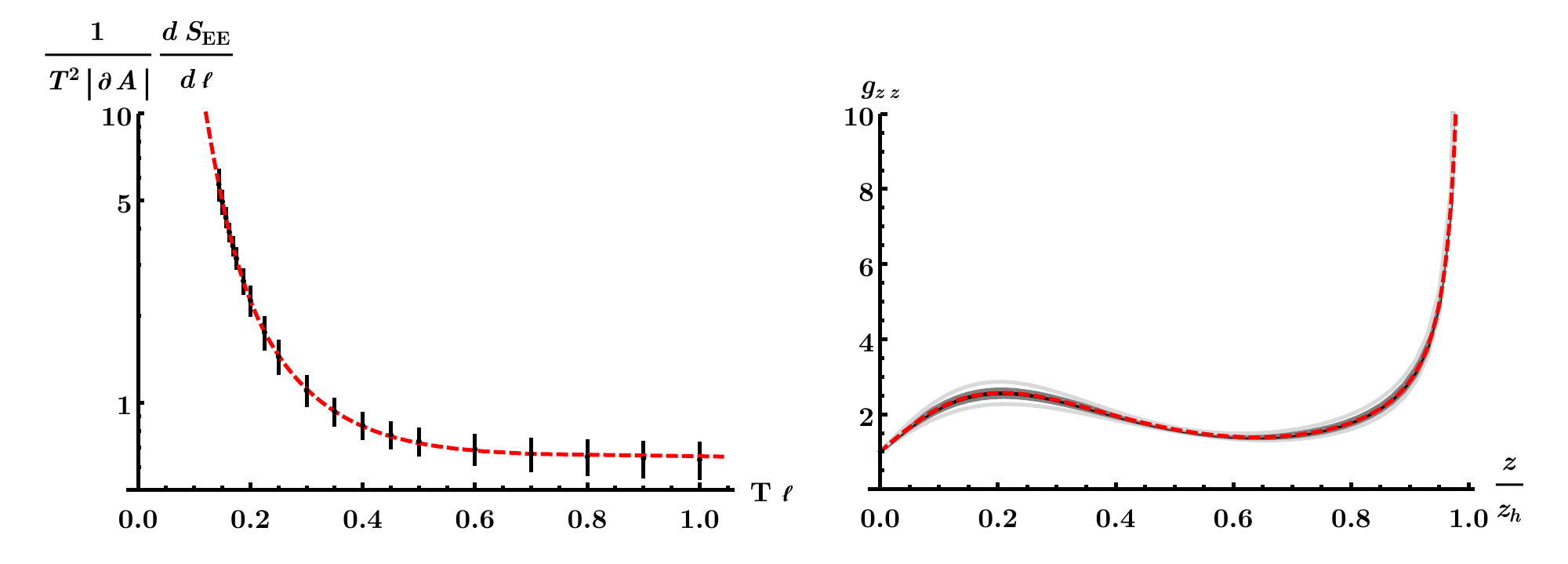}
    \caption{\textbf{Left:} Maximum likelihood fit (red dashed line) of the derivative of entanglement entropy in $\of{2+1}$ dimensions. Black dots are a subset of the interpolated lattice data for $\dd S_{EE}/\dd \ell$. \textbf{Right:} The metric function reconstructed from lattice data in $\of{2+1}$ dimensions. Red dashed line is the maximum likelihood estimate of the data. The black dashed line is the median in the distribution of $g_{zz}=a(z)^2/b(z)$. The dashed gray lines represent the $50\%$ and $95\%$ central confidence intervals of the metric. Both figures use $N_{basis}=3$. The utilized lattice data is the same as in Fig.~\ref{fig:deedlplateaux3d}.}
    \label{fig:gzz_3d}
\end{figure}

We are sampling the posterior distribution of the parameters which define the metric tensor to find metrics which would fit the lattice data well under the assumption that the holographic model represents the gauge theory well. The results are shown in Fig.~\ref{fig:gzz_3d}. There is a pronounced bump in the metric towards the UV-region of the geometry. It would be interesting to analyze the ramifications for this feature on some derived quantities.
We have also plotted the predictions for $\dd S_{EE}/\dd \ell$ using the maximum likelihood holographic geometry and find good agreement with the lattice gauge theory data.

\section{Discussion and outlook}\label{sec:conclusions}

We demonstrated that holography has entered a new phase, precision science. Gravity duals of realistic field theories can be reconstructed from underlying data that is {\emph{always}} bound to be imprecise, a fact which subsequently feeds in likelihood estimates on any holographic computations.
Systematically reducing the statistical uncertainties on the extracted quantities from the lattice leads to an increasingly accurate metric of the dual geometry. 
The current work is a proof-of-concept and we did not yet investigate much which operators drive the renormalization group flow away from the UV or what predictions the bulk reconstruction yields, other than the computation of the thermal entropy per usual identification with the Bekenstein-Hawking entropy of the dual black hole.

Given the formidable route to establishing the dual quantum geometry at finite numbers of colors, we advocate that there seems to be a (naive) way to quantify the trustworthiness of classical dual geometry. The holographic entanglement c-function has a sudden jump across the energy scale marked by the deconfinement-confinement transition. This can be associated with the  degrees of freedom scaling as ${\cal{O}}(\Nc^2)$ and ${\cal{O}}(\Nc^0)$ in the respective phases. This jump is likely an artifact of the $\Nc\to \infty$ limit, and is less drastic for finite values of $\Nc$, as first illustrated in~\cite{Itou:2015cyu} for $\Nc=3$ and later in~\cite{Rabenstein:2018bri} for $\Nc=2,3,4$. Nevertheless, if one sticks to the deconfining phase, the reconstructed dual metrics behave smoothly as a function of $\Nc$, giving one in principle means to interpolate the emerging classical geometries. There are of course many complications along the way to making this precise, especially if the entanglement entropy is being used as a tool to reconstruct the metric, since the RT formula is expected to be modified in the presence of quantum corrections \cite{Faulkner:2013ana}. It is clearly interesting to study further the entanglement entropies for higher rank Yang-Mills theories. We hope to report on entanglement c-functions in (3+1) dimensions for $\Nc>4$ in the future.

Other possible avenues for further research include the shape dependence on the entangling region. In principle, parts of this information is already encoded in our link updates and one could not only extract, say, the corner contribution \cite{Bueno:2015rda}, but globally extend the entangling region to spherical shapes. In addition to this, aspects of multipartite entanglement measures are clearly interesting targets. Simplest extension of the present work being to that of two strips (slabs) and questions related with mutual information,  entanglement negativity \cite{Vidal:2002zz}, or purification \cite{Terhal:2002}. In particular, it would be interesting to analyze if in the last case scenario one finds non-monotonous behavior as a function of the energy scale as suggested by the holographic dual quantity \cite{Jokela:2019ebz} known as the entanglement wedge cross section \cite{Takayanagi:2017knl}.

Entanglement measures are by far not the only data with which one can reconstruct the bulk geometry; work in this direction include \cite{Hammersley:2007ab,Bilson:2010ff,Spillane:2013mca,Bao:2019bib,Jokela:2020auu,Park:2022fqy}. Other quantities that have been investigated in the literature include operators \cite{Hamilton:2006az,Kabat:2011rz}, four-point correlators \cite{Caron-Huot:2022lff}, differential entropy \cite{Balasubramanian:2013lsa,Headrick:2014eia}, fidelity \cite{Kusuki:2019hcg}, complexity \cite{Hashimoto:2021umd}, light-cone cuts \cite{Engelhardt:2016wgb,Hernandez-Cuenca:2020ppu}, chiral condensate \cite{Hashimoto:2018bnb,Hashimoto:2020jug}, conductivity \cite{Li:2022zjc}, hadron spectra \cite{Akutagawa:2020yeo}, and magnetization  \cite{Hashimoto:2018ftp} with varying degree of assumptions about the dual classical bulk gravity. To all of these cases one could incorporate the Hamiltonian Monte Carlo approach discussed in Sec.~\ref{sec:bulkreconstruction} to also infer the maximum likelihood estimates of the reconstructed duals.
In addition to these examples, one can also use the Wilson/Polyakov loops to infer the bulk metric \cite{Jokela:2020auu,Hashimoto:2020mrx}, though the fundamental string breaking at finite temperature limits the most straightforward application and only part of the geometry is visible to the boundary. In an upcoming work \cite{DLWilson} the authors overcome this limitation and reconstruct the full geometry. Besides the murky status of Debye screening at strong coupling as revealed in Appendix~\ref{sec:Wilson}, the ability to use Polyakov loop data also for the reconstruction will further motivate putting extra effort in a more systematic study of quark-anti-quark potentials already in the pure Yang-Mills theory in any dimension. Having simultaneous access to bulk metric both in the string and the Einstein frame by reconstruction from $q\bar q$ potential and entanglement entropy data, respectively, would enable us to make a direct comparison with the beta function of $\SU{\Nc}$ Yang-Mills theory at strong coupling.

Finally, to address the deconfinement phase transition and lower temperatures in general in the bulk gravity side, one seeks for an alternative ansatz for the metric other than those (\ref{eq:4dmetric}), (\ref{eq:3dmetric}) where a black hole is assumed to reside at the bottom of the geometry. 


\addcontentsline{toc}{section}{Acknowledgments}
\paragraph{Acknowledgments}
We would like to thank Jacopo Ghiglieri, Marco Panero, and Javier G. Subils for useful discussions. N.~J. is supported in part by the Academy of Finland grant. no 1322307, and A.~S. and K.~R. by the Academy of Finland grants 319066, 320123 and 345070. T.~R. is supported in part by the Swiss National Science Foundation (SNSF) through the grants no. 200021\_175761 and TMPFP2\_210064. The authors wish to acknowledge CSC - IT Center for Science, Finland, for generous computational resources.


\begin{appendices}

\section{Quark-anti-quark potential}\label{sec:Wilson}

In this appendix we slightly extend the discussion in the main text to cover also the Polyakov loop which give rise to a potential between quarks and anti-quarks. This is to illustrate that the D2-brane background, which lead to the rather unconventional power law behavior for the entanglement entropy $S_{EE}\propto \ell^{-4/3}$, predicts also an unexpected power law behavior for the quark-anti-quark potential at small separation $\propto L^{-2/3}$. We will see that the lattice data that we will generate seems to conform with this and thus lends more support on the usage of the gravity dual of (2+1)-dimensional super Yang-Mills theory at largish energy scales.  In addition to this matching, we learned from EE studies that at finite temperature, the black hole present in the D2-background also explains the behavior of the entanglement entropy at large subregions sizes. It is therefore not completely unreasonable to ask if the large separation behavior of the $q\bar q$-potential be also within grasp via D2-brane black holes in this regime. Indeed, our frugal results from the lattice seem to conform with  $\propto L^{-10/3}$ behavior at large separation, extractable from the D2-brane background. If this result persists under further scrutiny, then the holographic approach predicts that the analytic continuation of the lattice results to real time cf. \cite{Burnier:2016mxc} would result in an imaginary part of the potential linear in the $q\bar q$ separation $L$.

\subsection{Static quark potential from thermal D2-brane background}\label{ssec:sqpfromthermald2brane}

To derive these $q \bar q$-potential results we start with the metric \eqref{eq:3dmetric} with $a(z) = 1$ and $b(z) = 1 - \left(\frac{z}{z_0}\right)^5$. This is a black hole geometry with a horizon at $z = z_h$. The potential is found by studying a string whose endpoints are at the boundary, separated by a distance $L$ in the $x_1$-direction. The action of such a string is
\begin{align}
    S_{NG} = \frac{1}{2\pi\alpha'} \int \sqrt{\det{g_2}} \ ,
\end{align}
where $g_2$ is the metric induced on the string worldsheet. The action can be written in terms of the profile function $x_1 = x_1(z)$ which determines the shape of the hanging string in the bulk as
\begin{align}
    S_{NG} = \frac{\tau z_p^2}{2\pi\alpha'} \int \frac{1}{z^2} \sqrt{1 + \left(\frac{z_p}{z}\right) b(z) x_1'(z)^2} \ \dd z \ , \label{eq:nambu_goto_action}
\end{align}
where $\tau$ is the time interval spanned by the string worldsheet. The $q \bar q$-potential can be found by studying the equation of motion for the string profile $x_1(z)$. The Euler-Lagrange equation is
\begin{align}
    \frac{\dd}{\dd z} \left( \frac{b(z) x_1'(z)}{z^3 \sqrt{1 + z_p b(z) x_1'(z)^2/z}} \right) = 0 \ .
\end{align}
The solution with the boundary conditions $x_1(z=0) = L/2$ and $x_1'(z \to z_*) \to -\infty$ is
\begin{align}
    x_1(z) = \frac{L}{2} - \frac{z^4}{4 z_p^{1/2} z_*^{5/2}} \sqrt{1-z_*^5/z_h^5} \ F_1\left( \frac{4}{5}; \frac{1}{2}, \frac{1}{2}; \frac{9}{5}; \frac{z^5}{z_h^5}, \frac{z^5}{z_*^5} \right) \ , \label{eq:string_profile}
\end{align}
where $F_1$ is the Appell hypergeometric function. The constants $L$ and $z_*$ are the quark separation and string turning point, respectively. They are connected by the equation $x_1(z=z_*) = L/2$, that is
\begin{align}
    L(z_*) = \frac{\sqrt\pi \Gamma\left( \frac{9}{5} \right) z_*^{3/2}}{2 \Gamma\left( \frac{13}{10} \right) z_p^{1/2}} \sqrt{1 - \frac{z_*^5}{z_h^5}} \ {}_2 F_1\left( \frac{1}{2}, \frac{4}{5}; \frac{13}{10}; \frac{z_*^5}{z_h^5} \right) \ . \label{eq:quark_separation}
\end{align}
The string action can be evaluated by plugging $x_1'(z)$ from \eqref{eq:string_profile} into \eqref{eq:nambu_goto_action}. One finds
\begin{align}
    \frac{2\pi\alpha'}{\tau z_p^2} S_{NG} = \frac{2}{\epsilon} - \frac{\sqrt\pi \Gamma\left(\frac{4}{5}\right)}{\Gamma\left( \frac{3}{10} \right) z_*}\ {}_2 F_1\left( -\frac{1}{2}, -\frac{1}{5}; \frac{3}{10}; \frac{z_*^5}{z_h^5} \right) \ , \label{eq:quark_action}
\end{align}
where $\epsilon$ is the usual UV-cutoff to regulate the otherwise divergent action. Now we can find the UV-behavior of the potential by studying \eqref{eq:quark_separation} and \eqref{eq:quark_action} in the limit $z_* \to 0$
\begin{align}
    L &= \frac{\sqrt \pi \Gamma\left(\frac{9}{5}\right)}{2 \sqrt{z_p} \Gamma\left(\frac{13}{10}\right)} z_*^{3/2} \\
    \frac{2\pi\alpha'}{\tau z_p^2} S_{NG} &= \frac{2}{\epsilon} - \frac{2\sqrt \pi \Gamma\left(\frac{4}{5}\right)}{z_* \Gamma\left(\frac{3}{10}\right)} \ .
\end{align}
Together, these formulas imply that when $z_*$ is small the action behaves as
\begin{align}
    \frac{2\pi\alpha'}{\tau z_p^2} S_{NG} &= \frac{2}{\epsilon} - \frac{2^{1/3} 4 \pi^{5/6} \Gamma\left(\frac{4}{5}\right)^{5/3}}{3^{2/3} \Gamma\left(\frac{3}{10}\right)^{5/3} z_p^{1/3} L^{2/3}} \ .
\end{align}
The $q \bar q$-potential is defined
\begin{align}
    V(L) = -\left( S_{NG}(L) - \re \left[S_{NG}(L \to \infty)\right] \right) / \tau \ . \label{eq:quark_potential_definition}
\end{align}
There are different ways to do the regularization, often the potential is computed from the difference of the string action and the action of two straight disconnected strings hanging from the boundary. This time we choose a different method, namely we find it useful to regulate so that the real part of the potential goes to zero at large separations. We will shortly show that the latter term in \eqref{eq:quark_potential_definition} is a constant. Therefore, the potential indeed is $V(L) \propto L^{-2/3}$ at small separations.

\begin{figure}
    \centering
    \includegraphics{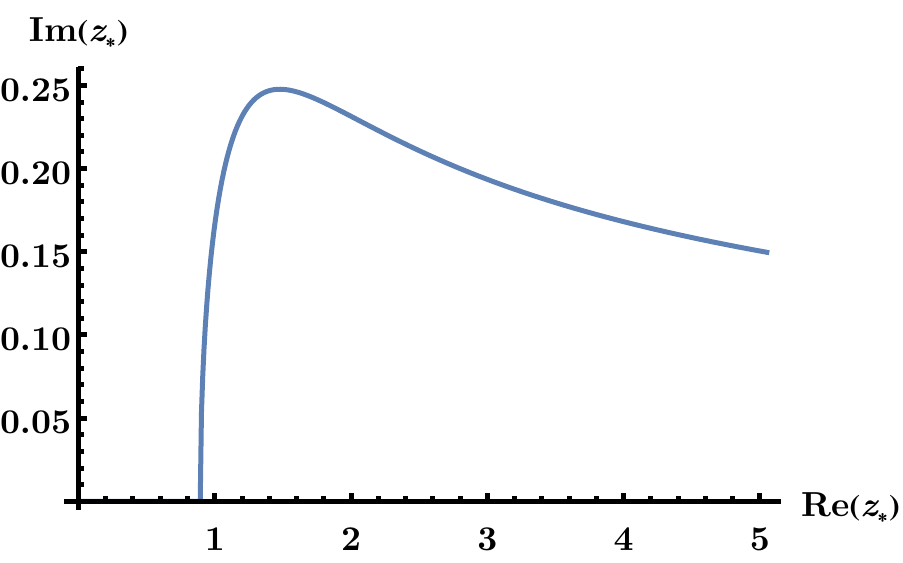}
    \caption{The curve in the complex $z_*$-plane which corresponds to positive real quark separations in \eqref{eq:quark_separation}. At quark separation $L=0$ the corresponding turning point is $z_*=0$ and $L$ increases as one moves to the right along the curve. At small separations $z_*$ is real but moves to the complex plane shortly after the separation where the string would break.}
    \label{fig:complex_zs_curve}
\end{figure}

In order to study the potential at large separations we need to analyze the string profile carefully. Usually the large separation limit is somewhat trivial as often the string is assumed to break at some critical separation making the potential constant at larger separations. It is possible, however, to extract non-trivial IR-behavior of the potential by analytically continuing $L(z_*)$ and $S_{NG}(z_*)$ to the complex $z_*$-plane \cite{Albacete:2008dz}. The quark separation $L$ is always non-negative and real but the turning point $z_*$ need not be. In Fig.~\ref{fig:complex_zs_curve} we show a curve along which $L$  is real and non-negative, corresponding to complex string configurations. The curve represents the solutions to \eqref{eq:quark_separation} when solving for $z_*$ and setting $L$ to fixed non-negative real values. At small separations $z_*$ is real as usual. The separation at which the string would break into the disconnected configuration happens in this region. Shortly after that separation the turning point becomes complex. Also the conjugate of this curve solves the equation but we choose to focus on the $\im [z_*] \geq 0$ branch.

The large separation limit can be worked out by first inverting \eqref{eq:quark_separation} at large $L$. We find
\begin{align}
    \frac{z_*}{z_h} =& \frac{5^{2/3} \Gamma\left(\frac{4}{5}\right)^{4/2}}{2^{14/15} \pi^{2/3} \Gamma\left(\frac{3}{5}\right)^{2/3}} \left(\frac{z_p L^2}{z_h^3}\right)^{1/3} \nonumber \\
    &+ \frac{4 (-1)^{3/10} 2^{1/3} (5-\sqrt 5)(5\pi)^{1/6} \Gamma\left(\frac{4}{5}\right)^{5/3}}{9 \Gamma\left(\frac{3}{10}\right)^{5/3}} \left(\frac{z_h^3}{z_p L^2}\right)^{1/6}  + \mathcal O \left( L^{-4/3} \right)\ .
\end{align}
This can be used together with \eqref{eq:quark_action} to see that the action at large separations behaves as
\begin{align}
    \frac{2\pi\alpha'}{\tau z_p^2} S_{NG}(L\to\infty) = \frac{2}{\epsilon} - i \frac{z_p^{1/2} L}{z_h^{3/2}} + \frac{2 (-1)^{4/5}}{\sqrt\pi} \Gamma\left(\frac{7}{10}\right) \Gamma\left(\frac{4}{5}\right) + \mathcal O\left(L^{-1}\right) \ . \label{eq:quark_action_IR}
\end{align}
In the definition of the $q \bar q$-potential, we can now see that the real part of the IR-action will cancel the $1/\epsilon$-divergence present in $S_{NG}(L)$. The constant term in \eqref{eq:quark_action_IR} contributes to the counter term in \eqref{eq:quark_potential_definition} and makes the real part of $V(L)$ vanish at large separations. We can also immediately see that $\im[V(L\to\infty)] = -i z_p^{1/2} L/z_h^{3/2} + \mathcal O(1)$. Further, analyzing subleading terms of \eqref{eq:quark_action} we can find that $\re[V(L\to\infty)]$ behaves as $L^{-10/3}$ like we claimed in the beginning of this section.

\subsection{Static quark potential from the lattice}

The lattice gauge theory formalism was introduced by K.~G.~Wilson in 1974 to show that quarks are confined in QCD~\cite{Wilson:1974sk}. The static quark potential was therefore from the beginning of primary interest in lattice Monte Carlo studies of $\SU{\Nc}$ gauge theories~\cite{Creutz:1980zw} and various observables have been employed to study its properties. In particular Wilson loops, which are traces over parallel transporters along closed loops, and Polyakov loops, which are Wilson loops that wrap around a compact direction (usually the Euclidean time direction), played an important role in understanding the non-perturbative properties of QCD, as, {\emph{e.g.}}, showing that $\SU{2}$ pure gauge theory is both, confining and asymptotically free~\cite{Creutz:1980wj}, and that the asymptotically free and the confining region of parameter space are separated from each other by a deconfinement transition~\cite{McLerran:1981pb}.

With increasing computer performance and the development of improved simulation techniques~\cite{Cabibbo:1982zn} these non-perturbative studies of the properties of the static quark potential have been extended first to $\SU{3}$~\cite{Barkai:1984cr,Stack:1983cw,Otto:1984qr} and then to higher $\Nc$ and spacetime dimensions different from (3+1)~\cite{Lucini:2001nv}. Results at $T=0$ in (2+1) dimensions can for $\Nc=2$, for example, be found in \cite{HariDass:2007tx,Dass:2007dv}, for $\Nc=3$ in \cite{Luscher:2002qv}, and for $\Nc=5$ in \cite{Meyer:2006gm}. Recall that the static quark potential can also be computed perturbatively~\cite{Pineda:2010mb}.

At $T=0$ the determination of the static quark potential using Polyakov loops suffers from a bad signal to noise ratio as the magnitude of Polyakov loop correlation function drops exponentially with $1/T$. At $T= 0$ the static quark potential is therefore usually determined from Wilson loops~\cite{Stack:1983cw}. As we are in the present work interested not in the zero but in high temperature properties of $\SU{\Nc}$ gauge theory, we will use Polyakov loop correlators to determine the static quark potential.    

\subsubsection{Static quarks and Polyakov loops}
In this section we illustrate the relation between Polyakov loops and static quarks. Starting point is Wilson's lattice fermion action
\[
S_F\fof{U,\bar{\psi},\psi}=\sum\limits_{x,y}\bar{\psi}\of{x}\,D_{x,y}\fof{U}\,\psi\of{y}\ ,
\]
with $\psi\of{x}$ and $\bar{\psi}\of{x}$ being Euclidean Dirac spinors appropriate for $d$ dimensions and $D_{x,y}\fof{U}$ is the corresponding Wilson-Dirac operator,
\begin{multline}
D_{x,y}\fof{U}\,=\,\delta_{x y}\id-\kappa\underbrace{\sum\limits_{\nu=1}^{d-1}\bof{\delta_{x+\hat{\nu},y}\of{\id-\gamma_{\nu}}U_{\nu}\ssof{x}+\delta_{x-\hat{\nu},y}\of{\id+\gamma_{\nu}}U^{\dagger}_{\nu}\ssof{x-\hat{\nu}}}}_{S_{x y}}\\
-\,\kappa\underbrace{\of{\delta_{x+\hat{4},y}\of{\id-\gamma_{4}}\,\e^{\mu}\,U_{4}\ssof{x}+\delta_{x-\hat{4},y}\of{\id+\gamma_{4}}\,\e^{-\mu}\,U^{\dagger}_{4}\ssof{x-\hat{4}}}}_{T_{x y}}\label{eq:wilsondiracop}\ ,
\end{multline}
with the hopping parameter, $\kappa=1/\of{2\of{m_0+d}}$, which is related to the bare quark mass, $m_0=a\,m_c$, and $\cof{\gamma_{\nu}}_{\nu=1,\ldots,d}$ are Euclidean Dirac matrices in $d$ dimensions. We have added a quark chemical potential, $\mu$, to Eq.~\eqref{eq:wilsondiracop} in order to be able to keep track of whether a terms correspond to a quark or an anti-quark in what follows. 

In the Euclidean path integral,
\[
Z=\int\DD{U,\psi,\bar{\psi}}\,\e^{-S_F\fof{U,\bar{\psi},\psi}-S_G\fof{U}}\ ,
\]
the integration over the spinor fields $\psi\of{x}$ and $\bar{\psi}\of{x}$ is a multi-dimensional Grassmann integral which upon evaluation yields a so-called fermion determinant:
\[
Z_f\fof{U}=\int\DD{\psi,\bar{\psi}}\,\e^{-S_F\fof{U,\bar{\psi},\psi}}\,=\,\Det{D\fof{U}}\ ,\label{eq:fermiondet}
\]
in terms of which:
\[
Z=\int\DD{U}\,\Det{D\fof{U}}\,\e^{-S_G\fof{U}}\ .\label{eq:fermiondetinpartf}
\]
Now, if we are interested in static quarks, {\emph{i.e.}}, in quarks that do not perform spatial hops but simply move along straight lines around the time direction, we can in Eq.~\eqref{eq:wilsondiracop} drop the spatial hopping terms, $S_{x\,y}$, and keep only the temporal hopping terms, $T_{x\,y}$. In this case, the fermion determinant in Eq.~\eqref{eq:fermiondet} simplifies significantly. If we stick for the moment to $d=4$, it takes the form:
\begin{multline}
\Det{D\fof{U}}=\Det{\id-\kappa\,T}=\\
\prod\limits_{\bar{x}}\sof{\det^2_{c}\sof{\id+\of{2\,\kappa}^{N_t}\,\e^{\mu\,N_t}\,P\of{\bar{x}}}\det^2_{c}\sof{\id+\of{2\,\kappa}^{N_t}\,\e^{-\mu\,N_t}\,P^{\dagger}\of{\bar{x}}}}\ ,\label{eq:hdfermiondet}
\end{multline}
where the subscript $c$ of the $\operatorname{det_{c}}$-operator is meant to indicate that it acts only in color-space ({\emph{i.e.}}, that the operand is a matrix with only color indices) and the product runs over all spatial locations $\bar{x}$. 

\begin{figure}[h]
\centering
\includegraphics[height=0.4\linewidth,keepaspectratio]{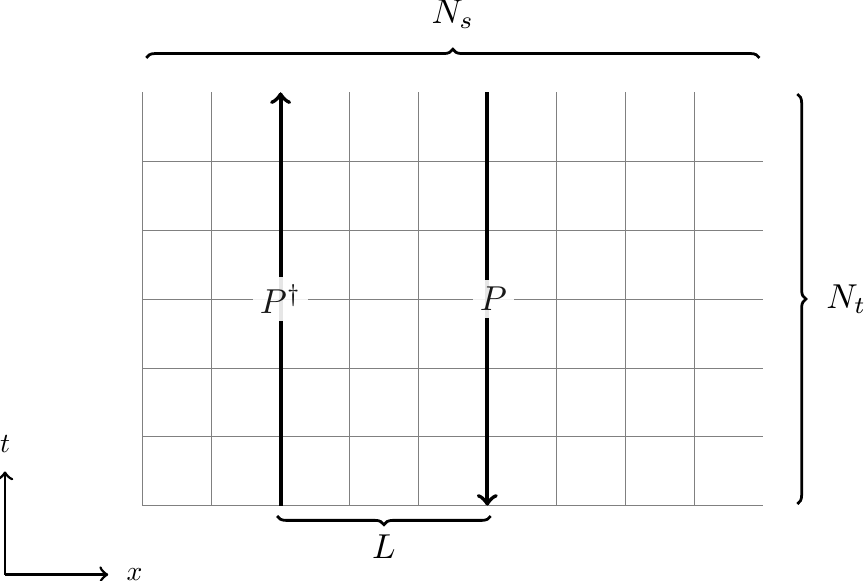}
\caption{Illustration of a pair of anti-Polyakov loop, $P^{\dagger}$, and Polyakov loop, $P$, separated by a spatial distance $L$. Both loops are wrapping around the periodic time extent of the lattice, but in opposite directions.}
\label{fig:plloops}
\end{figure}

As can be seen from Eqn.~\eqref{eq:hdfermiondet} and \eqref{eq:fermiondetinpartf}, the static quarks couple to the gauge field through the Polyakov and anti-Polyakov loops. The Polyakov loop, $P\of{\bar{x}}$ at spatial location $\bar{x}$, is given by the ordered product of all temporal link-variables (cf. Eq.~\eqref{eq:linkvar}) over $\bar{x}$, to form a parallel transporter that connects the site $x=\of{\bar{x},t=0}$ with itself by wrapping once on a straight temporal trajectory around the periodic time-direction:
\[
P\of{\bar{x}}=U_{d}\of{\bar{x},0}\,U_{d}\of{\bar{x},1}\,\ldots\,U_{d}\of{\bar{x},N_t-1}\ .
\]
Its hermitian conjugate, $P^{\dagger}\of{\bar{x}}$, does the same thing but in opposite direction, as illustrated in Fig.~\ref{fig:plloops}.

Now, as we forced the fermions in Eq.~\eqref{eq:hdfermiondet} to be static, it is reasonable to require them as well to be heavy, in which case $\kappa$ becomes very small and the color-determinants in Eq.~\eqref{eq:hdfermiondet} can be written as
\begin{subequations}\label{eq:approxcolordets}
\[
\det_{c}\sof{\id+\of{2\,\kappa}^{N_t}\e^{\mu\,N_t} P\of{\bar{x}}}=1+\of{2\,\kappa}^{N_t}\e^{\mu\,N_t}\trace\sof{P\of{\bar{x}}}+\order{\ssof{2\,\kappa}^{2\,N_t}}\ ,
\]
resp.
\[
\det_{c}\sof{\id+\of{2\,\kappa}^{N_t}\e^{-\mu\,N_t} P^{\dagger}\of{\bar{x}}}=1+\of{2\,\kappa}^{N_t}\e^{-\mu\,N_t}\trace\sof{P^{\dagger}\of{\bar{x}}}+\order{\ssof{2\,\kappa}^{2\,N_t}}\ .
\]
\end{subequations}
As can be seen from Eqn.~\eqref{eq:approxcolordets} by counting the powers of $\e^{\mu}$ resp. $\e^{-\mu}$: having a static quark or anti-quark winding around the time direction over a spatial site $\bar{x}$ is accompanied by a Polyakov loop resp. anti-Polyakov loop located on that site.

\subsubsection{Static quark potential from Polyakov loops}
We can now define the static quark potential, $V\of{L,T,\beta_g}$, in terms of the spatial correlation function between a traced Polyakov loop, $P$, and a traced anti-Polyakov loop, $P^{\dagger}$ (cf. Fig.~\ref{fig:plloops}), using the relation~\cite{Jahn:2004qr},
\[
\e^{-V\of{L,T,\beta_g}/T}=\frac{1}{\Nc^2}\avof{\trace\sof{P^{\dagger}\of{\bar{x}}} \trace\sof{P\of{\bar{y}}}}_{\beta_g}\ ,\label{eq:staticquarkpotfromcorr}
\]
with $L=\abs{\bar{y}-\bar{x}}$ and $T=1/N_t$ (in lattice units). Note that Eq.~\eqref{eq:staticquarkpotfromcorr} is particularly useful at finite temperatures, {\emph{i.e.}}, when $N_t$ is not too large. As the formula indicates, at low temperatures, the magnitude of the correlation function decays quickly with $N_t=1/T$ and develops a bad signal to noise ratio. If one is interested in the static quark potential at low or zero temperatures, one therefore typically uses a different approach which is based on so-called Wilson loops~\cite{Barkai:1984ca}.

\begin{figure}[h]
\centering
\begin{minipage}[t]{0.49\linewidth}
\centering
\hspace{40pt}{\small $a\,T_c=0.032$}\\[0pt]
\includegraphics[height=0.8\linewidth,keepaspectratio]{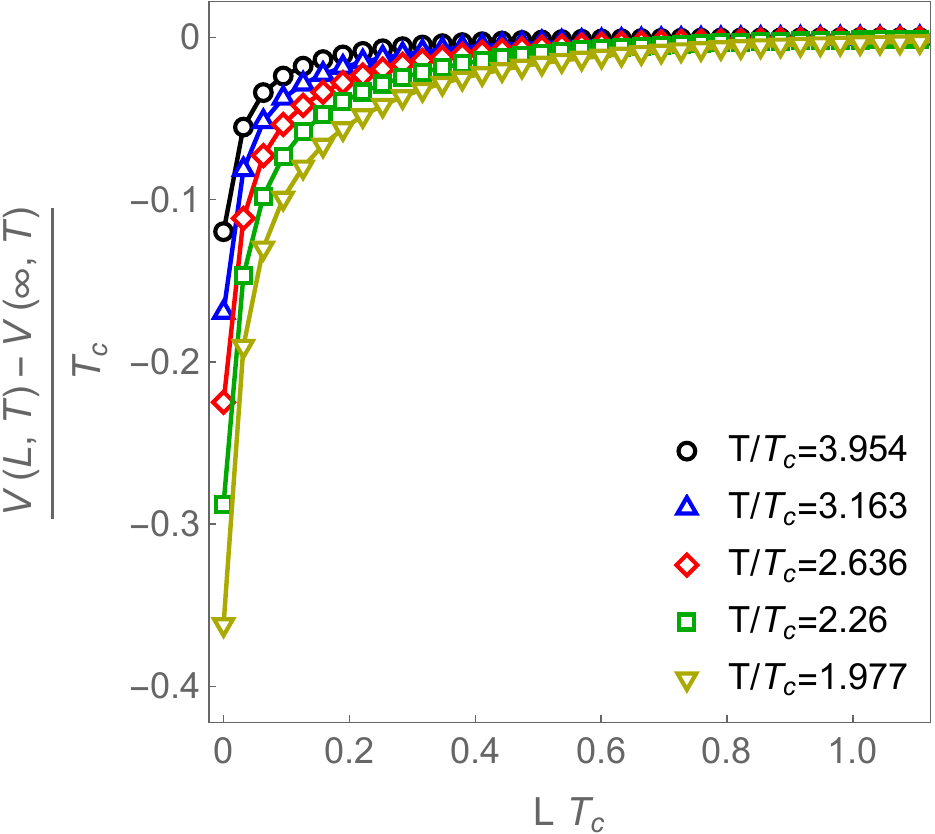}
\end{minipage}\hfill
\begin{minipage}[t]{0.49\linewidth}
\centering
\hspace{40pt}{\small $a\,T_c=0.048$}\\[0pt]
\includegraphics[height=0.8\linewidth,keepaspectratio]{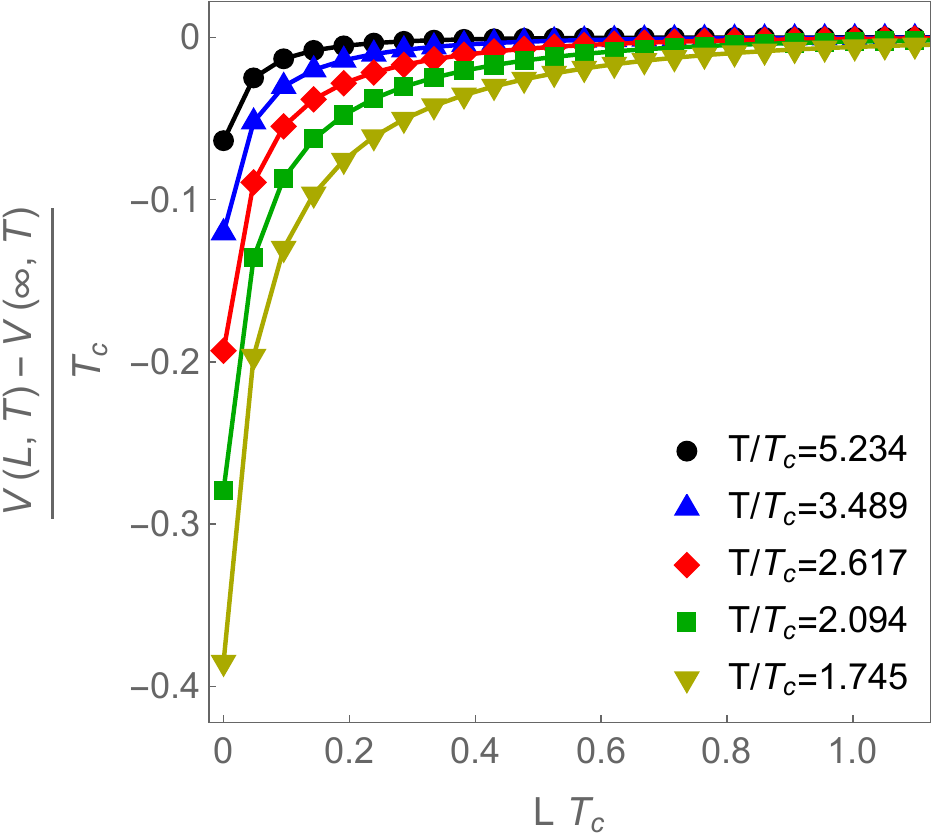}
\end{minipage}
\caption{Subtracted static quark potential from Eq.~\eqref{eq:sqpsubtracted} for $\SU{2}$ at finite temperature in (2+1) dimensions, plotted as function of distance $L$ for five different temperatures $T>T_c$. The two panels corresponds to different lattice spacings, $a\,T_c=0.032$ (left) and $a\,T_c=0.048$ (right). Error bars are smaller than the plot markers.}
\label{fig:staticpot3d0}
\end{figure}

The potential $V\of{L,T}$ in Eq.~\eqref{eq:staticquarkpotfromcorr} has an explicit temperature-dependency, which is particularly prominent for $T/T_c>1$, where the potential approaches for $L\gg T^{-1}$ a temperature-dependent plateau value. We will subtract this plateau value and consider the subtracted potential, the binding energy (in units of the critical temperature scale),
\begin{multline}
V_s\of{L,T,\beta_g}=\frac{V\of{L,T,\beta_g}-V\of{\infty,T,\beta_g}}{T_c}\\
=-\frac{T}{T_c}\log\of{\frac{\avof{\trace\sof{P^{\dagger}\of{x}} \trace\sof{P\of{y}}}_{\beta_g}}{\sabs{\avof{\trace\sof{P}}_{\beta_g}}^2}}\ ,\label{eq:sqpsubtracted}
\end{multline}
instead of $V\of{L,T,\beta_g}$ itself. Fig.~\ref{fig:staticpot3d0} shows Monte Carlo results for Eq.~\eqref{eq:sqpsubtracted} as function of $L$ in the case of $\SU{2}$ in $\of{2+1}$ dimensions at different temperatures and for two different lattice spacings, {\emph{i.e.}}, different values of $\beta_g$.

\subsubsection{Setting the scale}\label{ssec:scalesetting}
We use the critical temperature, $T_c$, as reference energy scale. To be able to do so, we need to determine the ratio $T/T_c$ as function of the inverse lattice coupling $\beta_g$ and the spatial and temporal lattice sizes $N_s$, $N_t$. We therefore want a function $C_R\of{\beta_g,N_t,N_s}=T/T_c$, which can be obtained along the lines described in~\cite[Sec.~4.3-4.4]{Edwards:2009qw}. Using the expression given in \cite{Edwards:2009qw} for the pseudo-critical lattice coupling as function of $N_t$ and $N_s$, 
\begin{multline}
\beta_{g,c}\of{N_t,N_s}=N_t\,\of{4\,c_r-d_2\of{\frac{N_t}{N_s}}^{\rho}}-\of{4\,c_1+d_1\,\of{\frac{N_t}{N_s}}^{\rho}}\\
-\frac{1}{N_t}\of{4\,\frac{c_2}{c_r}+d_0\,\of{\frac{N_t}{N_s}}^{\rho}}\ ,\label{eq:betapc}
\end{multline}
as well as the function
\[
C\of{N_t,N_s}=N_t\,\of{4\,c_r-d_2\of{\frac{N_t}{N_s}}^{\rho}}+\frac{1}{N_t}\of{4\,\frac{c_2}{c_r}+d_0\,\of{\frac{N_t}{N_s}}^{\rho}}\ ,\label{eq:tcfunc}
\]
where~\cite{Edwards:2009qw}
\[
c_1=-0.176(5)\ ,\ c_2=0.0675(5)\ ,\ c_r=\frac{T_c}{g_c^2}=0.3757(5)\ ,
\]
and
\[
d_0=-0.51(9)\ ,\ d_1=0\ ,\ d_2=0.134(4)\ ,\ \text{and}\ \rho=2.61(9)\ ,
\]
one can define a preliminary
\[
C_{R,c}\of{\beta_g,N_t,N_s}=1+\frac{\beta_g-\beta_{g,c}\of{N_t,N_s}}{C\of{N_t,N_s}}\ .\label{eq:ttcratio0}
\]
The subscript $c$ in Eq.~\eqref{eq:ttcratio0} indicates that $T/T_c\approx C_{R,c}\of{\beta_g,N_t,N_s}$ is only valid if $\beta_g$ is sufficiently close to its pseudo-critical value $\beta_{g,c}\of{N_t,N_s}$.

To obtain an expression for $C_R\of{\beta_g,N_t,N_s}$ that is valid for arbitrary values of $\beta_g$, $N_s$ and $N_t$, We note that $\of{a\,T_c}$ depends only on $\beta_g$ and $N_s$, and the lattice spacing $a$ itself is controlled merely by $\beta_g$. At constant $\beta_g$ and $N_s$ (which implies also constant $a\of{\beta_g}$), we therefore have~\cite{Edwards:2009qw}:
\[
\frac{T\of{\beta_g,N_t,N_s}}{T_c\of{\beta_g,N_s}}=\frac{T\of{\beta_g,N_t,N_s}}{T\of{\beta_g,N_t',N_s}}\frac{T\of{\beta_g,N_t',N_s}}{T_c\of{\beta_g,N_s}}=\frac{N_t'}{N_t}\,\frac{T\of{\beta_g,N_t',N_s}}{T_c\of{\beta_g,N_s}}\ ,
\]
which allows us to define the function $C_R\of{\beta_g,N_t,N_s}=T/T_c$ for arbitrary values of $\beta_g$, $N_s$ and $N_t$ by:
\[
C_R\of{\beta_g,N_t,N_s}=\frac{N_t'\of{\beta_g,N_s}}{N_t}\,C_{R,0}\of{\beta_g,N_t'\of{\beta_g,N_s},N_s}\ ,\label{eq:ttcratio}
\]
with
\[
N_t'\of{\beta_g,N_s}=\left.\floor{N_t}\right\vert_{N_t:\beta_{g,c}\of{N_t,N_s}=\beta_{g}}\ ,
\]
{\emph{i.e.}}, $N_t'\of{\beta_g,N_s}$ is the value of $N_t\in \mathbb{N}$ for which $\of{\beta_g-\beta_{g,c}\of{N_t,N_s}}$ assumes its smallest, positive value for given $\beta_g$ and $N_s$.

\subsubsection{Data and interpolation to arbitrary temperatures}
In Tab.~\ref{tab:simparamforsqp} we summarize the simulation parameters for which we have produced data. As can be seen, although the data for different lattice spacings covers a similar temperature range, the simulated temperatures are in general different. In order to be able to directly compare the static quark potentials obtained at different lattice spacings, we will have to interpolate between the simulated temperature values. 

\begin{table}[h]
\centering
\begin{tabular}{| c | c | c | c | c |} 
 \hline
 $\beta_g$ & $N_s$ & $a\,T_c$ & $N_t$ & $T/T_c$ \\[2pt] 
 \hline
 16. & 48 & 0.0976593 & 2 & 5.11984 \\
     &    &           & 4 & 2.55992 \\
     &    &           & 6 & 1.70661 \\
     &    &           & 8 & 1.27996 \\
 \hline
 32. & 64 & 0.0477628 & 4 & 5.2342 \\
     &    &           & 6 & 3.48947 \\
     &    &           & 8 & 2.6171 \\
     &    &           & 10 & 2.09368 \\
     &    &           & 12 & 1.74473 \\
 \hline
 48. & 96 & 0.0316115 & 8 & 3.95426 \\
     &    &           & 10 & 3.16341 \\
     &    &           & 12 & 2.63617 \\
     &    &           & 14 & 2.25958 \\
     &    &           & 16 & 1.97713 \\
 \hline
\end{tabular}
\caption{Simulation parameters and corresponding lattice spacing- and $T/T_c$-values for $\SU{2}$ pure gauge theory on a (2+1)-dimensional lattice.}
\label{tab:simparamforsqp}
\end{table}

\begin{figure}[h]
\centering
\begin{minipage}[t]{0.49\linewidth}
\centering
\includegraphics[height=0.85\linewidth,keepaspectratio]{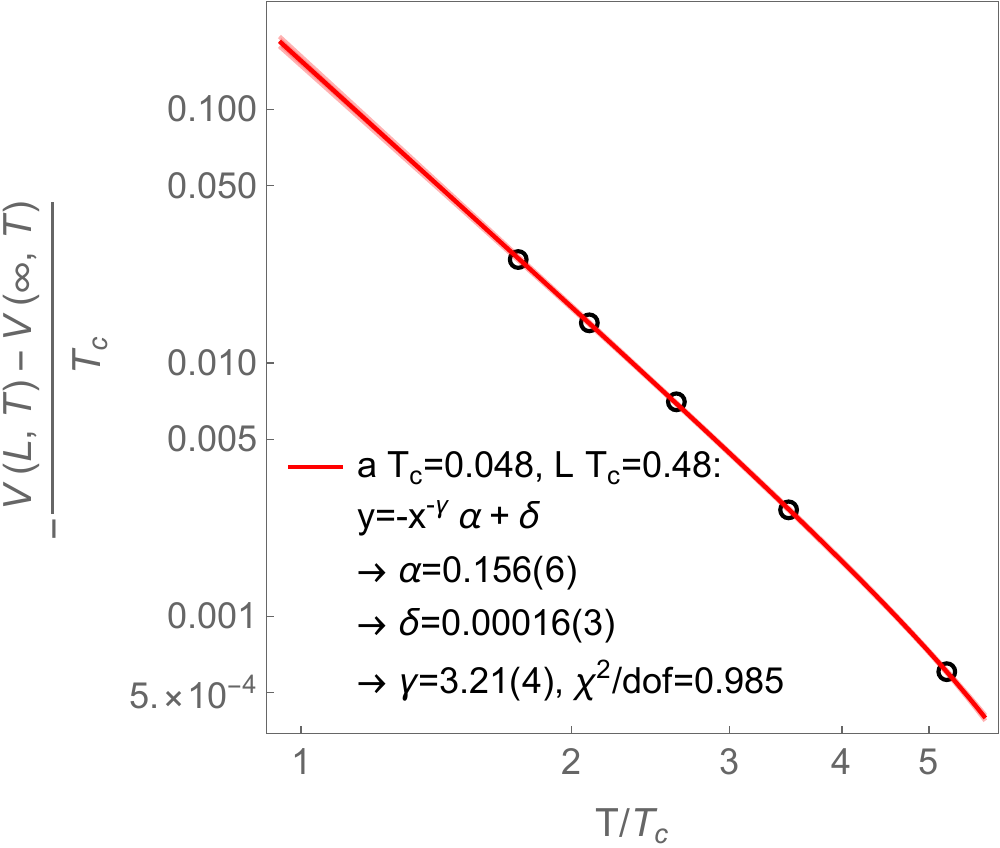}
\end{minipage}\hfill
\begin{minipage}[t]{0.49\linewidth}
\centering
\includegraphics[height=0.85\linewidth,keepaspectratio]{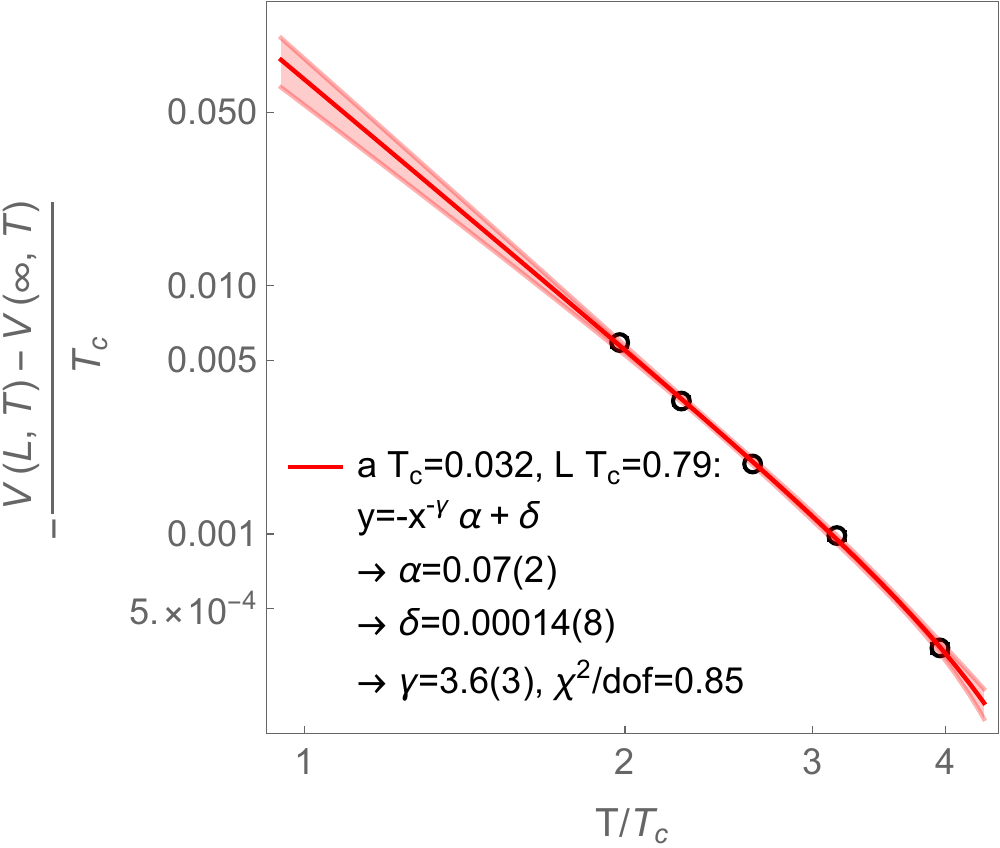}
\end{minipage}
\caption{Two examples of fits of the form \eqref{eq:vsfitfunc} to the subtracted static quark potential from Eq.~\eqref{eq:sqpsubtracted} for $\SU{2}$ in (2+1) dimensions, considered as function of $T/T_c$ at fixed $\of{\beta_g,L}$, resp. $\of{a\,T_c,L\,T_c}$. The error bands correspond to the 65\% confidence band.}
\label{fig:sqptempfitexamples}
\end{figure}

\begin{figure}[h]
\centering
\begin{minipage}[t]{0.33\linewidth}
\centering
\includegraphics[height=0.95\linewidth,keepaspectratio]{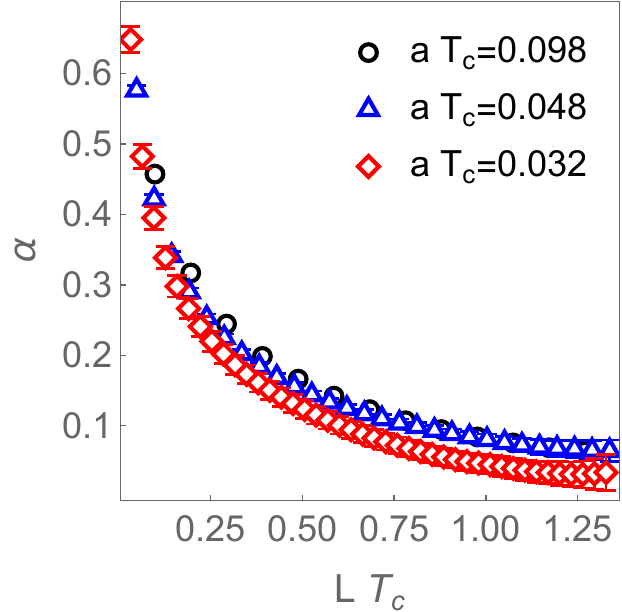}
\end{minipage}\hfill
\begin{minipage}[t]{0.33\linewidth}
\centering
\includegraphics[height=0.95\linewidth,keepaspectratio]{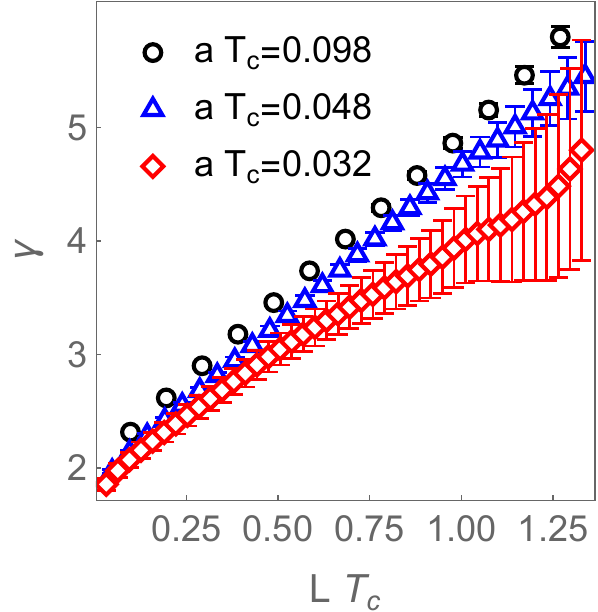}
\end{minipage}\hfill
\begin{minipage}[t]{0.33\linewidth}
\centering
\includegraphics[height=0.95\linewidth,keepaspectratio]{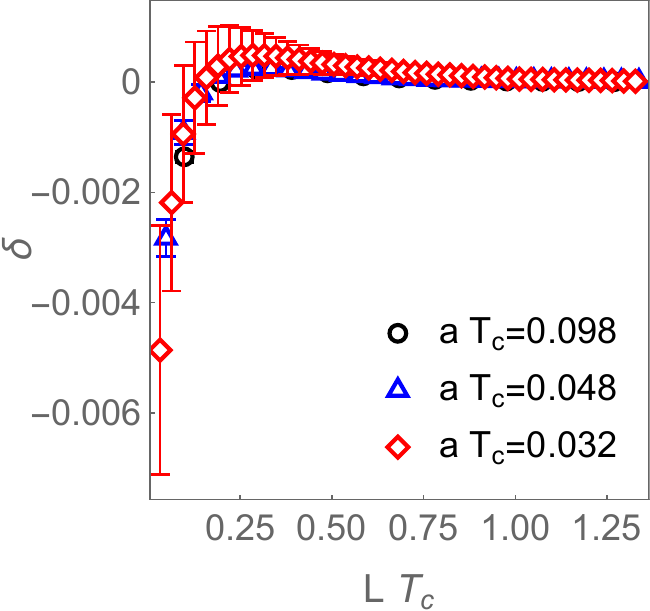}
\end{minipage}
\caption{The panels show the values of $\alpha_{\of{L,\beta_g}}$ (left), $\gamma_{\of{L,\beta_g}}$ (middle), and $\delta_{\of{L,\beta_g}}$ (right), used to parametrize Eq.~\eqref{eq:fittedvsoft}. The parameters are plotted as functions of $L\,T_c$ for three different lattice spacings, corresponding to $\beta_g=16$ (black circles), $32$ (blue triangles), and $48$ (red diamonds).}
\label{fig:sqptempfitparamres}
\end{figure}

We perform this interpolation as follows: we pick one of the simulated $\beta_g$ and a value of $L$, and fit for this choice the form
\[
y=\alpha\,x^{\gamma}+\delta\ \label{eq:vsfitfunc}
\]
to the data $\of{x,y}\in\cof{\of{T/T_c,V_s\of{L,T,\beta_g}}}_{T}$, with $V_s\of{L,T,\beta_g}$ from Eq.~\eqref{eq:sqpsubtracted} considered as function of $T$. This fit yields a function
\[
\tilde{V}_{s,\of{L,\beta_g}}\of{T}=\alpha_{\of{L,\beta_g}}\,\of{\frac{T}{T_c}}^{\gamma_{\of{L,\beta_g}}}+\delta_{\of{L,\beta_g}}\ ,\label{eq:fittedvsoft}
\]
which allows us to obtain for the given values of $\beta_g$ and $L$ the subtracted potential at arbitrary $T$. This procedure is then repeated for all available $L$ and $\beta_g$. The error in the value for the subtracted potential, given by Eq.~\eqref{eq:fittedvsoft} at arbitrary $T$, is obtained by determining the 65\% confidence band for Eq.~\eqref{eq:fittedvsoft}. Some fit examples are shown in Fig.~\ref{fig:sqptempfitexamples} and Fig.~\ref{fig:sqptempfitparamres} provides an overview over the fitted parameter values at different lattice spacings $a$ and different distances $L$.

Note that the chosen form of the fit function, Eq.~\eqref{eq:vsfitfunc}, is essentially a power law with a small additive correction, $\delta$, to account for the fact that we subtract in Eq.~\eqref{eq:sqpsubtracted} effectively the log of $\avof{\ssabs{P}}^2_{\beta_g}$ instead of $\ssabs{\avof{P}_{\beta_g}}^2$, in order to avoid cancellations between contributions to the Polyakov loop observable coming from the two degenerate vacua that exist in the deconfined phase for $\SU{2}$, where $\Repart{P}$ is either positive or negative. Provided that the ansatz \eqref{eq:vsfitfunc} works well and assuming that the parameter $\delta$ indeed just accounts for the systematic error caused by using $\avof{\ssabs{P}}^2_{\beta_g}$ instead of $\ssabs{\avof{P}_{\beta_g}}^2$ in \eqref{eq:sqpsubtracted}, we could in principle subtract $\delta_{\of{L,\beta_g}}$ from Eq.~\eqref{eq:fittedvsoft} in order to get rid of this systematic error. Instead of Eq.~\eqref{eq:fittedvsoft} we would therefore use,
\[
\tilde{V}_{s,\mathrm{imp},\of{L,\beta_g}}\of{T}=\alpha_{\of{L,\beta_g}}\,\of{\frac{T}{T_c}}^{\gamma_{\of{L,\beta_g}}}\ ,\label{eq:fittedvsoftimp}
\]
with the same values of $\alpha_{\of{L,\beta_g}}$ and $\gamma_{\of{L,\beta_g}}$. However, as our spatial lattices are relatively large, $\delta_{\of{L,\beta_g}}$ is typically very small (cf. Fig.~\ref{fig:sqptempfitparamres}, right) and whether one uses Eq.~\eqref{eq:fittedvsoftimp} or sticks with Eq.~\eqref{eq:fittedvsoft}, does not give rise to significantly (within error bars) different results.

\subsubsection{Scaling of deconfined potential at short and large distances}
As discussed in Sec.~\ref{ssec:sqpfromthermald2brane} holographic considerations suggest that the static quark potential in the deconfined phase of a strongly coupled gauge theory in $\of{2+1}$ dimensions should follow different power laws at short and long distances, namely $V_{\of{L\,T\ll 1}}\of{T,L}\propto L^{-2/3}$ and $V_{\of{L\,T\gg 1}}\of{T,L}\propto L^{-10/3}$.

\begin{figure}[h]
\centering
\begin{minipage}[t]{0.49\linewidth}
\centering
\includegraphics[height=1.19\linewidth,keepaspectratio]{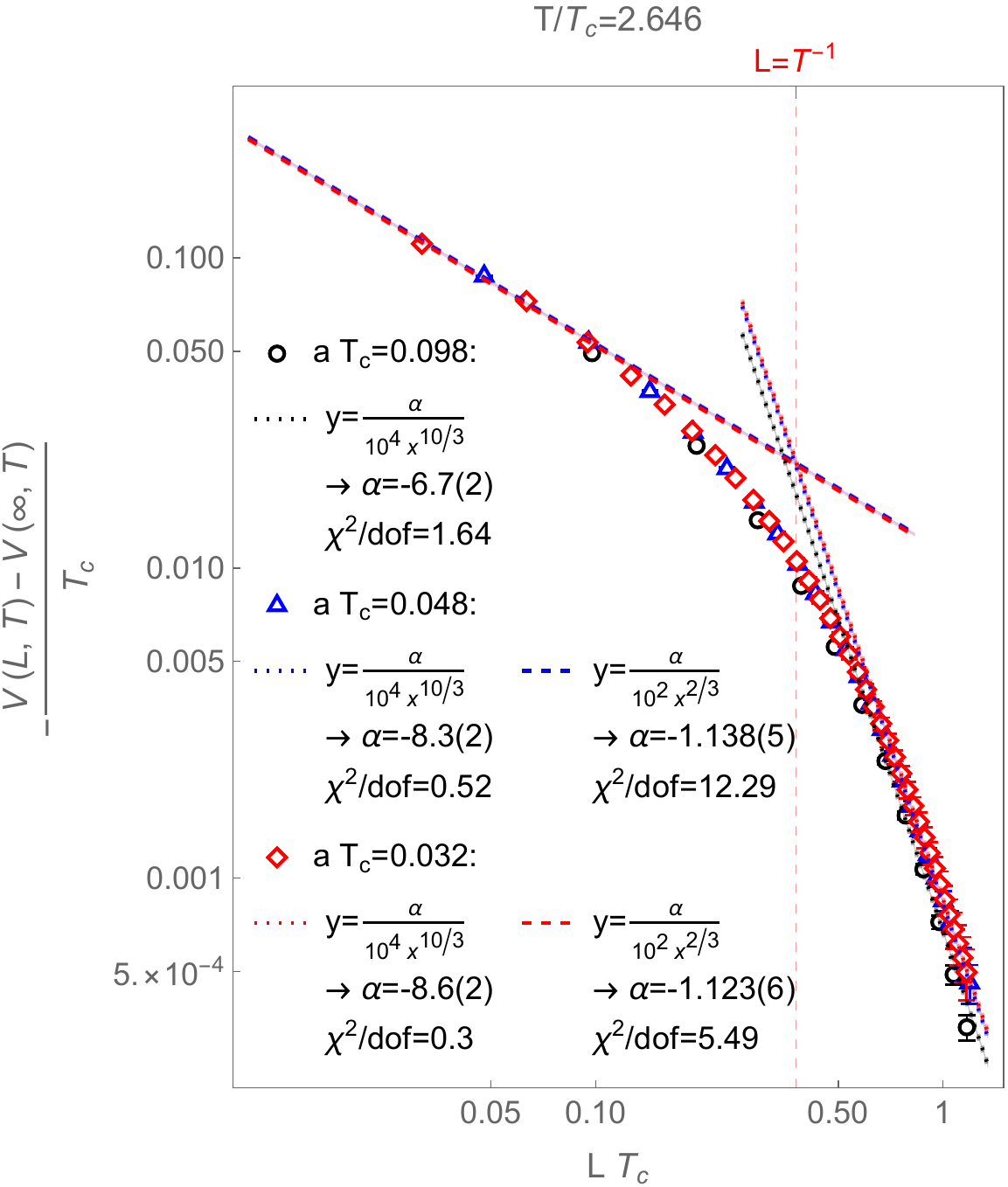}
\end{minipage}\hfill
\begin{minipage}[t]{0.49\linewidth}
\centering
\includegraphics[height=1.19\linewidth,keepaspectratio]{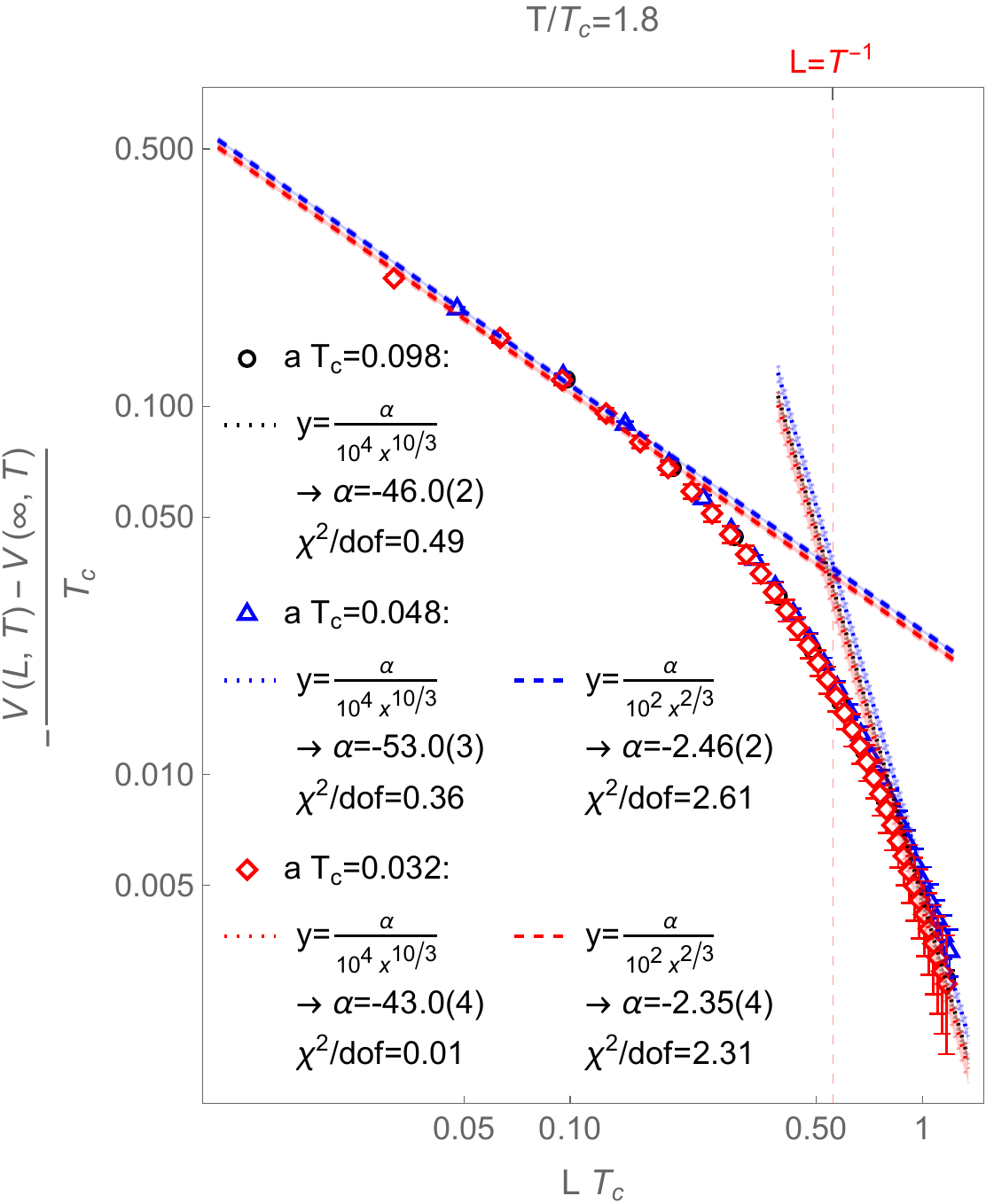}
\end{minipage}
\caption{Static quark potential for $\SU{2}$ in (2+1) dimensions as function of distance $L$ at finite temperature $T=2.646\,T_c$ (left) and $1.8\,T_c$ (right) for three different lattice spacings $a=$0.098$\,T_c^{-1}$ (black), 0.048$\,T_c^{-1}$ (blue), and 0.032$\,T_c^{-1}$ (red). At short distances, well below $L\sim T^{-1}$, the behavior of the potential appears compatible with an asymptotic power-law, $\sim L^{-2/3}$ (dashed lines). The displayed fits were performed on data for $L<0.3\,T^{-1}$ (where available). At distances above $L\sim T^{-1}$, the potential appears to behave like $\sim L^{-10/3}$ (dotted lines). Here the fits were obtained on data for $L>1.7\,T^{-1}$.}
\label{fig:staticpot3d_2}
\end{figure}

In Fig.~\ref{fig:staticpot3d_2} we fit these power law ansatzes to the short and long distance pieces of the interpolated potential \eqref{eq:fittedvsoft} (as function of $L\,T_c$) at fixed $T/T_c=2.646$ (left) and fixed $T/T_c=1.8$ (right) for different lattice spacing values. In order to stay away from the region around $L\,T\approx 1$, where the log-log plots in Fig.~\ref{fig:staticpot3d_2} have clearly visible curvature, we included for the short distance fits only data with $L\,T<0.3$, and for the long distance fits only data with $L\,T>1.7$. The short distance fit could therefore only be performed for the two smaller lattice spacing values, as for the largest one with $a\,T_c=0.98$, there is only one point satisfying the criterion $L\,T<0.3$ . 

At the current stage, these fits should of course not be taken too seriously, as the available data is of limited statistics and does also not sufficiently cover the asymptotic short- and long-distance regimes. In both regimes, we see at most the onset of a possible linear behavior of the subtracted potential in the log-log plots in Fig.~\ref{fig:staticpot3d_2}. It should also be mentioned that at short distances, the displayed data contains also the points corresponding to $L=1\,a$, which can be expected to suffer from relevant finite-lattice spacing effects. At large distances one can infer that the data does not cover a sufficiently large range to decide whether the potential really approaches a power law and starts to follow a straight line in the log-log plots in Fig.~\ref{fig:staticpot3d_2}, or whether its slope will continue to become even steeper with increasing $L$.\\

\begin{figure}[h]
\centering
\begin{minipage}[t]{0.49\linewidth}
\centering
\includegraphics[height=1.19\linewidth,keepaspectratio]{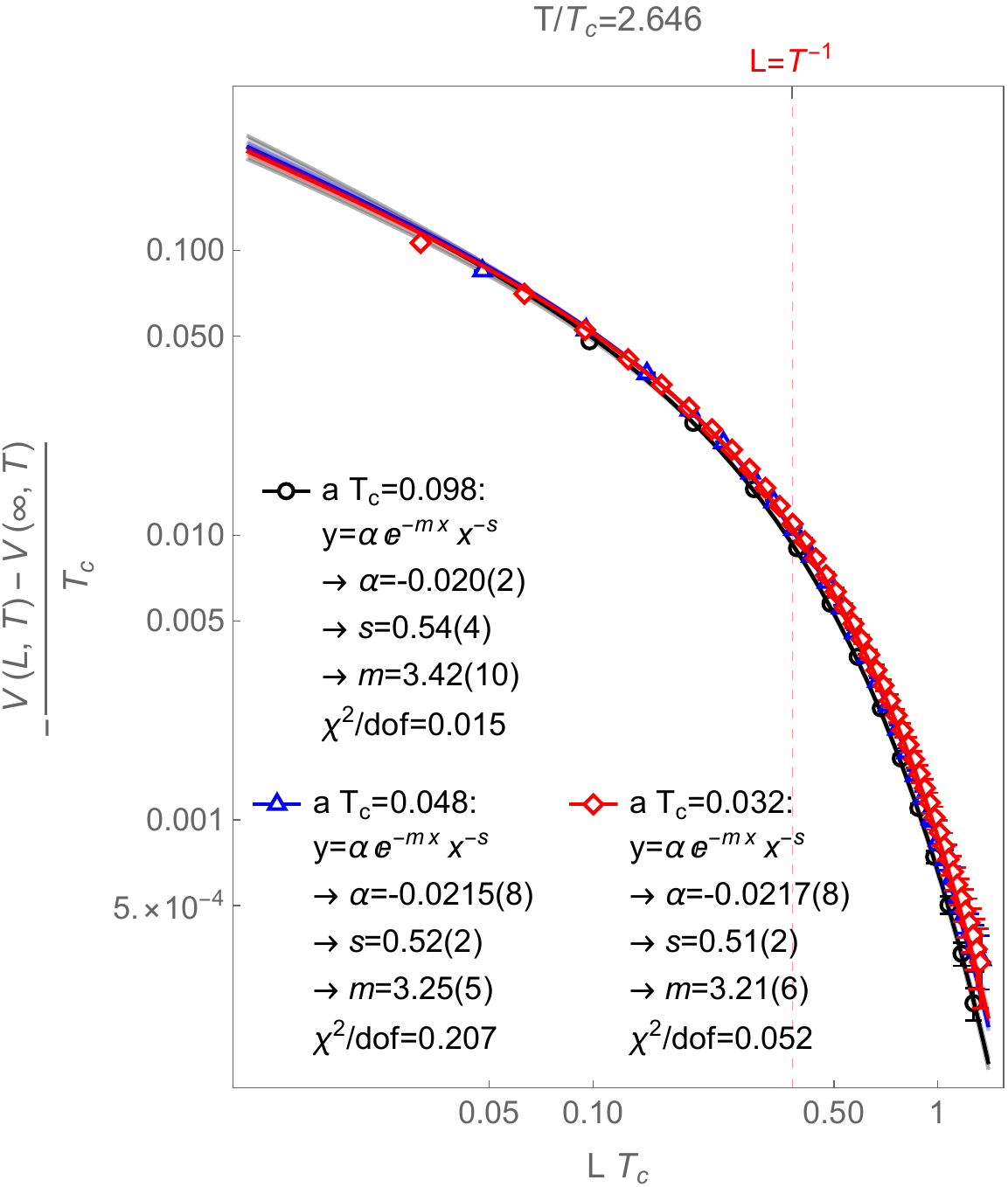}
\end{minipage}\hfill
\begin{minipage}[t]{0.49\linewidth}
\centering
\includegraphics[height=1.19\linewidth,keepaspectratio]{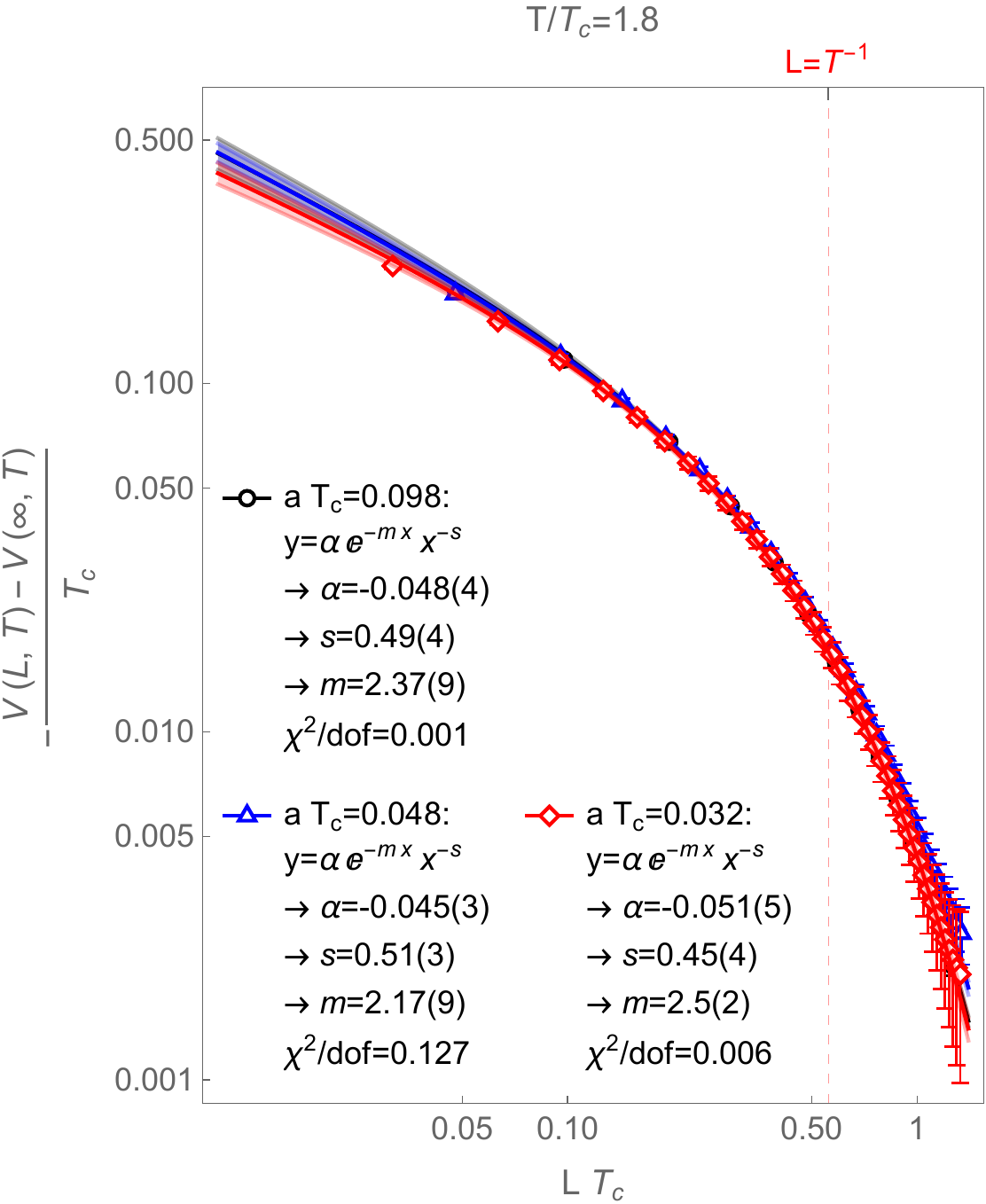}
\end{minipage}
\caption{Same data as in Fig.~\ref{fig:staticpot3d_2}, but fitted with a global ({\emph{i.e.}}, all $L$-values with $L>a$) screened power-law ansatz.}
\label{fig:staticpot3d_3}
\end{figure}

To this end, we attempted to fit also a screened power law, {\emph{i.e.}}, the form
\[
y=x^{-s}\,\e^{-m\,x}\ ,
\]
to the same data, where $\of{x,y}\in \scof{\sof{L\,T_c,\tilde{V}_{s,\ssof{L,\beta_g}}\of{T}}\vert L/a>1\  ,L\,T_c<1.35}$, with the interpolating potential from Eq.~\eqref{eq:fittedvsoft}. The results from these fits are summarized in Fig.~\ref{fig:staticpot3d_3}. Note that these are global fits in $L$, where we have excluded only the points at $L/a\leq 1$ to avoid the most-severe finite lattice spacing effects, and the points with $L\,T_c>1.35$ where for the two smaller lattice spacings the finite lattice volume would start to become visible. As can be seen, these fits worked out quite well for the given quality of the data. 

\begin{figure}[h]
\centering
\begin{minipage}[t]{0.49\linewidth}
\centering
\includegraphics[height=1.19\linewidth,keepaspectratio]{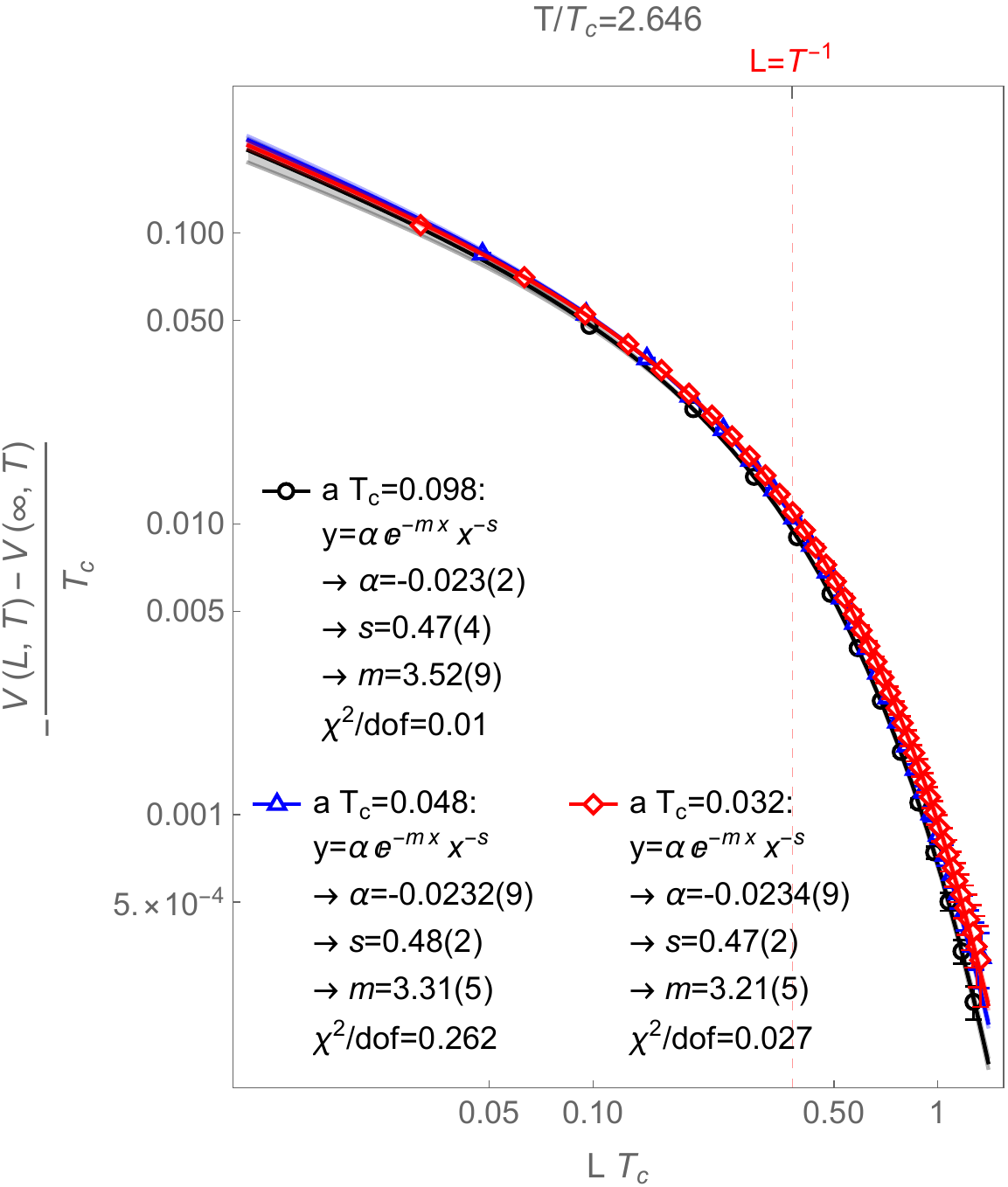}
\end{minipage}\hfill
\begin{minipage}[t]{0.49\linewidth}
\centering
\includegraphics[height=1.19\linewidth,keepaspectratio]{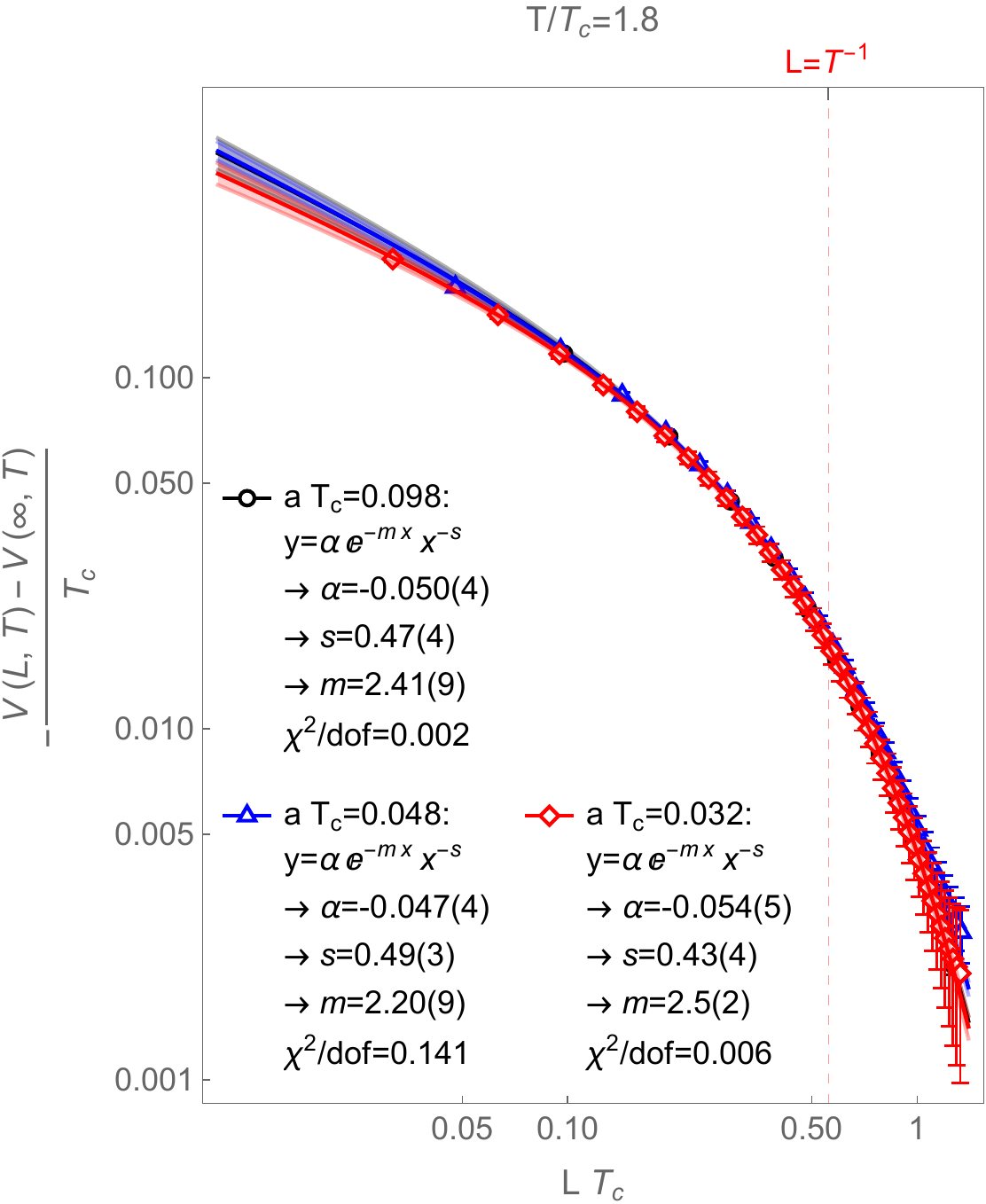}
\end{minipage}
\caption{Same data as in Fig.~\ref{fig:staticpot3d_3}, but using Eq.~\eqref{eq:fittedvsoftimp} instead of Eq.~\eqref{eq:fittedvsoft} to obtain the values of the subtracted potential at the specified values of $T/T_c$.}
\label{fig:staticpot3d_4}
\end{figure}

In order to confirm that the use of Eq.~\eqref{eq:fittedvsoftimp} instead of Eq.~\eqref{eq:fittedvsoft} to interpolate our data for the subtracted static quark potential between the simulated temperature values, would not significantly affect the fit results, we show in Fig.~\ref{fig:staticpot3d_4} the analogous plots to those in Fig.~\ref{fig:staticpot3d_3}, using the improved interpolating potential form Eq.~\eqref{eq:fittedvsoftimp}. These latter fits might appear slightly cleaner than the ones in Fig.~\ref{fig:staticpot3d_3}, in regards of the lattice spacing dependency of the fitted parameters. But, the fitted parameters in Fig.~\ref{fig:staticpot3d_3} and Fig.~\ref{fig:staticpot3d_4} are mutually compatible within error bars.

\section{Change of temporal boundary conditions and gauge invariance}\label{sec:changeoftempbc}

Recall from Sec.~\ref{ssec:imprlatEEmethod} that when updating the boundary of region $A$ by changing the temporal boundary conditions over spatial links, so that they change from region $B$ to region $A$ or vice versa, we undergo transitions between different partition functions $\cof{Z_i}_{i=0,\ldots}$, as introduced above Eq.~\eqref{eq:imprfreeenergydiff}. We are interested in the free energy difference between subsequent $Z_i$, which we determine numerically by probing the overlap between the distributions of gauge field configurations of neighboring $Z_i$.

\begin{figure}[h]
  \centering
  \begin{minipage}[t]{0.49\linewidth}
    \centering
    \includegraphics[width=0.8\linewidth]{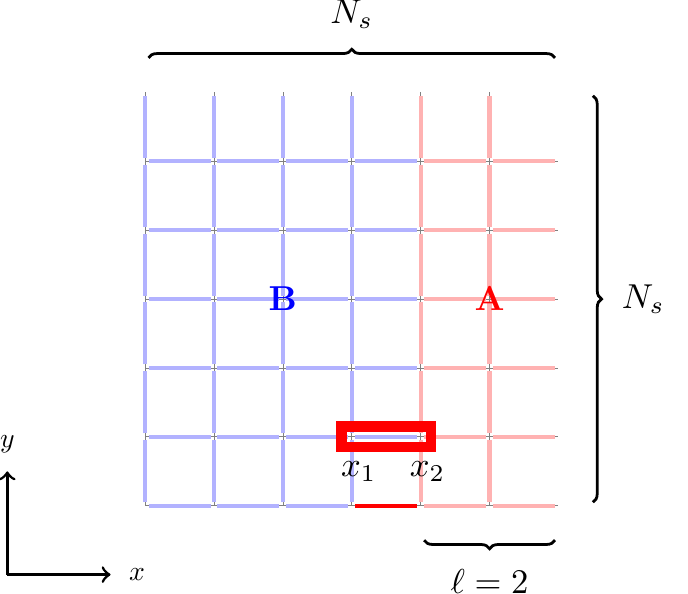}
  \end{minipage}\hfill
  \begin{minipage}[t]{0.49\linewidth}
    \centering
    \includegraphics[width=0.8\linewidth]{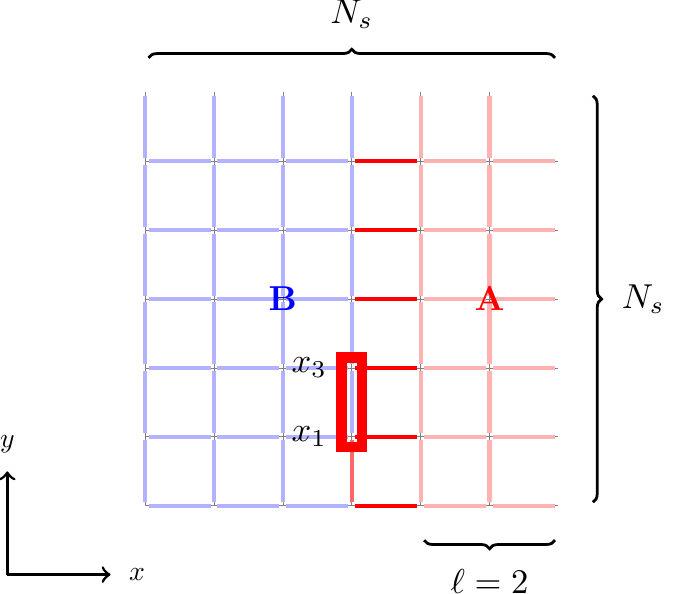}
  \end{minipage}\\[0pt]
  \caption{The two panels show the spatial links corresponding to regions $A$ (red) and $B$ (blue) at two intermediate stages during the process of interpolating between the situations depicted on the left- resp. right-hand side of Fig.~\ref{fig:bdcond}. The spatial link that is highlighted by a red frame in the left-hand panel is at the moment blue and belongs therefore to region $B$. The spatial site $x_1$ is touched only by spatial links belonging to region $B$, while the spatial site $x_2$ is touched only by spatial links belonging to region $A$, except for the highlighted spatial link. If the highlighted link changes from region $B$ to region $A$, the spatial site $x_2$ will be surrounded merrily by spatial links from region $A$, while the spatial site $x_1$ will now also be touched by one link from region $A$. However, before and after the change of the highlighted spatial link from region $B$ to region $A$, one of its ends is connected to spatial site that is touched by spatial links of only one region. In contrast: the spatial sites at both ends of the highlighted spatial link in the right-hand panel are always touched by spatial sites from both regions, $A$ and $B$, no matter whether the highlighted link is still in region $B$ or changes to region $A$.}
  \label{fig:bdupdategaugeinvb}
\end{figure}

When changing from one $Z_i$ to a neighboring one, it turns out that it makes a huge difference whether the spatial link over which the temporal boundary conditions are changed has before and after the update at least one of its ends connect to a spatial site that is either completely in region $A$ or completely in region $B$ (cf. Fig.~\ref{fig:bdupdategaugeinvb}, left), or the spatial link touches before and/or after the update at both ends a spatial site at which spatial links from both regions meet (cf. Fig.~\ref{fig:bdupdategaugeinvb}, right).    

In order to illustrate why this is the case, we look at a transition between some $Z_i$ and corresponding neighboring $Z_{i+1}$, which will require a spatial link to change from region $B$ to region $A$. We recall from Fig.~\ref{fig:bdcond} that when a spatial link changes from region $B$ to region $A$ or vice versa, two temporal plaquettes over that spatial link, denoted in Fig.~\ref{fig:bdcond} by $P_1$ and $P_2$, switch their top links, due to the change of temporal boundary conditions. Such an update is in general gauge-dependent: if it were, {\emph{e.g.}}, possible to use a local gauge transformation to make the to-be-swapped links the same before doing the change of temporal boundary conditions, then the Euclidean action would remain unchanged when changing form $Z_i$ to $Z_{i+1}$. Such a gauge-dependency is okay here as we are sampling the overlap between gauge configuration distributions of different systems with different temporal boundary conditions and therefore different sites identified with each other. The latter, {\emph{i.e.}}, the fact that in the two systems different sites are identified, affects locally the amount of gauge-freedom the system has as one cannot perform independent local gauge-transformations on identified sites. The overlap between the gauge configuration distributions of neighboring $Z_i$ has therefore an entropic component coming from the change in the amount of gauge freedom due to the identification of different numbers of sites.

\begin{figure}[h]
  \centering
  \begin{minipage}[t]{0.49\linewidth}
    \centering
    \includegraphics[width=0.8\linewidth]{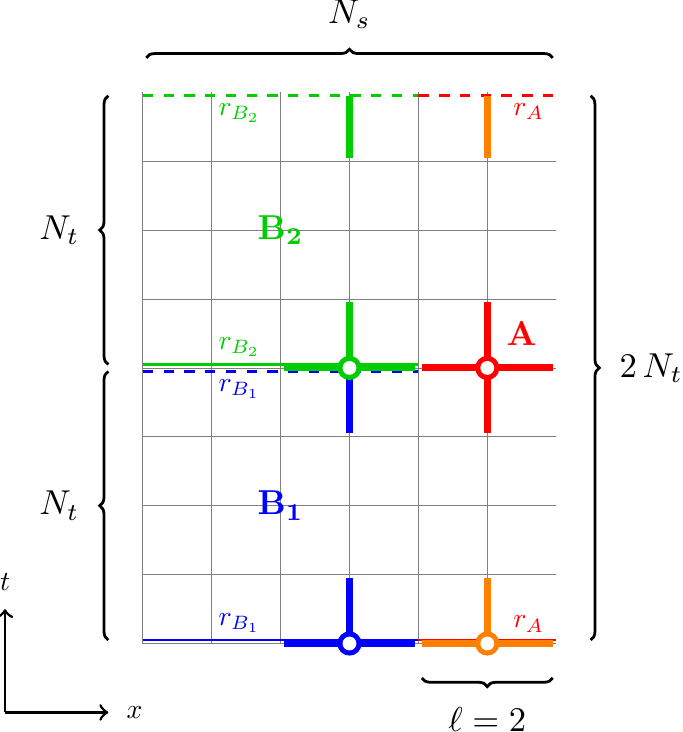}
  \end{minipage}\hfill
  \begin{minipage}[t]{0.49\linewidth}
    \centering
    \includegraphics[width=0.8\linewidth]{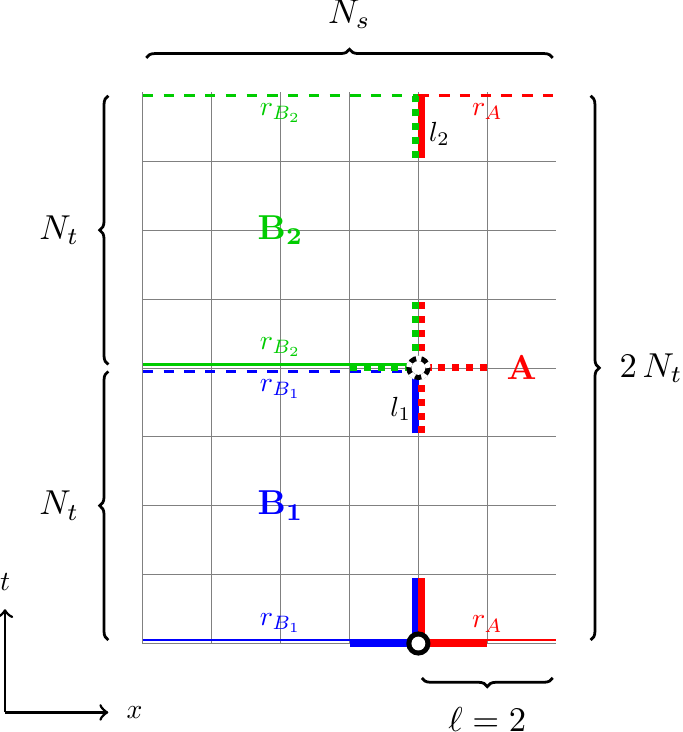}
  \end{minipage}\\[0pt]
  \caption{Left: the four sites highlighted by a red, orange, blue and green circle are fully inside of one of the two regions $A$ (red,orange) or region $B$ (blue,green) and therefore touched only by links that are all subject to the same type of temporal boundary conditions: either those of region $A$, or those of region $B$. If is therefore possible to apply independent local gauge transformations at the green and blue, as well as the red and orange sites, which will affect the correspondingly colored links.\\
Right: if one applies a local gauge transformation to the site highlighted by a solid-lined black circle, which is touched by links from both regions, $A$ and $B$, the situation is more complicated. In region $B$, a gauge transformation at this site changes the link $l_1$. From the point of view from region $A$, however, the link $l_1$ is connected to the site highlighted by the dashed black circle, on which we therefore have to apply the same local gauge transformation as on the site with the solid black circle in order to leave the gauge action unchanged, which will affect all the links highlighted by dashed lines. The same conclusion is reached by focusing on the link $l_2$ instead of $l_1$. The gauge transformation on the site with the black circle changes in region $A$ directly the link $l_2$. In region $B$, the link $l_2$ is connected to the site with the dashed black circle, on which we therefore have to apply the same gauge transformation as on the site with the solid-lined black circle.}
  \label{fig:bdupdategaugeinv}
\end{figure}

Consider as a concrete example the situation depicted on the left-hand side of Fig.~\ref{fig:bdupdategaugeinvb}: if the indicated link between $x_1$ and $x_2$ belongs to region $B$ (blue), we are in the system described by $Z_1$, and if the link changes to region $A$ (red), we would be in the system described by $Z_2$. If the link under consideration is currently part of region $B$, {\emph{i.e.}}, we are in the $Z_1$-system, the spatial site $x_1$ is touched only by spatial links that belong to region $B$, and as explained in the left-hand panel of Fig.~\ref{fig:bdupdategaugeinv}, we could therefore apply independent gauge transformations over $x_1$ to make the two links the same that would be swapped between the plaquettes $P_1$ and $P_2$ from Fig.~\ref{fig:bdcond} when the temporal boundary conditions are changed to those from region $A$. The same is true for the opposite move, where the marked spatial link from the left-hand panel of Fig.~\ref{fig:bdupdategaugeinvb} between the spatial sites $x_1$ and $x_2$ is initially part of region $A$. In this case the site $x_2$ is touched only by links from region $A$ and would allow for the required local gauge transformations to make the to-be-swapped top-links of plaquettes $P_1$ and $P_2$ identical. We can therefore define a Metropolis update that connects a gauge configuration of $Z_1$ to a gauge configuration of $Z_2$: if the system currently belongs to $Z_1$, the update proposes to pick a local gauge transformation over $x_1$ that makes the link-swap trivial, to then perform the link swap, and to finally do an inverse gauge transformation over $x_2$; if the system belongs currently to $Z_2$, the Metropolis update would do the same thing in the opposite direction, {\emph{i.e.}}, with initial local gauge transformation over $x_2$ and inverse gauge transformation over $x_1$ after the swap. As the freedom in choosing the appropriate gauge transformations is for both directions the same, and the action does not change during the move, the corresponding transition probability is always $1$. Although the way in which sites are identified changes during such an update due to the change of temporal boundary conditions, the number of identified points remains the same. Note also that, although we can achieve that the swapped links themselves are the same before and after such a move, the gauge field around them gets changed, as the local gauge transformations that are applied at the beginning and end of the move are applied at different spatial locations. 

As a second concrete example consider now the situation depicted in the right-hand panel of Fig.~\ref{fig:bdupdategaugeinvb}. The spatial link emphasized there by the red frame connects the sites $x_1$ and $x_3$. If the link belongs to region $B$, the system is described by $Z_7$ and if the link belongs to region $A$ by $Z_8$. Both spatial sites, $x_1$ and $x_3$, are touched by links from both regions, $A$ and $B$. As explained in the right-hand panel of Fig.~\ref{fig:bdupdategaugeinv}, this means that it is not possible to find a local gauge transformation that makes the links, that are swapped between $P_1$ and $P_2$ in Fig.~\ref{fig:bdcond} when the temporal boundary conditions change, the same. Spatial sites that are touched by spatial links from both regions, $A$ and $B$, are (indirectly) subject to both temporal boundary conditions simultaneously, meaning that the corresponding site at time $t=0$ over such a spatial site is not just identified with either the site at the same spatial location at time $t=N_t$ (as inside region $B$) or time $t=2\,N_t$ (as inside region $A$) but with both.

Note that in $\of{1+1}$ dimensions, the situation from the second example cannot occur and one is always in the analogous situation to the first example. In $\of{1+1}$ dimensions, the boundary update of region $A$ is always trivial and the derivative of the entanglement entropy  with respect to the width, $\ell$, of region $A$ has therefore to be identically zero.  

\end{appendices}

\printbibliography

\end{document}